\documentclass[graybox]{svmult}

\usepackage{amsfonts}
\usepackage{amsmath}

\usepackage{url}
\usepackage{float}
\usepackage{amsmath}
\usepackage{amssymb}
\usepackage{mathptmx}       
\usepackage{helvet}         
\usepackage{courier}        
\usepackage{type1cm}        
\usepackage{makeidx}         
\usepackage{graphicx}        
\usepackage{multicol}        
\usepackage[bottom]{footmisc}

\usepackage{comment}
\usepackage{hyperref}

\makeindex             

\begin{document}


\title*{Spectral and transport properties from lattice QCD}
\titlerunning{Spectral and transport properties from lattice QCD}
\author{Olaf Kaczmarek and Hai-Tao Shu}
\institute{Fakult\"at f\"ur Physik, Universit\"at Bielefeld, D-33165 Bielefeld, Germany\\
\email{okacz@physik.uni-bielefeld.de}\\
\email{htshu@physik.uni-bielefeld.de}}

\maketitle

\abstract{In these lecture notes
we will discuss recent progress in extracting spectral and transport properties from lattice QCD.
We will focus on results of probes of the 
thermal QCD medium as well as transport coefficients which are important ingredients for hydrodynamic and transport models that describe the evolution of the produced medium. These include electromagnetic 
probes, like the rates of emitted photons and dileptons, quarkonium spectral functions as 
well as transport coefficients, like the electrical conductivity or heavy flavor diffusion coefficients, of the quark gluon plasma (QGP).
All these real time quantities are encoded in the vector meson spectral function. A direct determination of the spectral functions is not possible in Euclidean lattice QCD calculations. Fortunately the spectral functions can be analytically continued from imaginary to real time, i.e. they are even equivalent in real and imaginary time. Therefore it is possible to relate the spectral function to the corresponding Euclidean correlation functions, although this requires a spectral reconstruction to obtain it from the corresponding correlation functions. In the following sections we will discuss the procedure to determine the required correlation functions and the extraction of the spectral functions from lattice QCD correlators. 
We will illustrate the concepts and methods to obtain spectral functions and related physical observables from continuum extrapolated correlation functions.  
We will focus here on results obtained from continuum extrapolated lattice correlation functions, which requires large and fine lattices, which so far was only possible to obtain in the so-called quenched approximation, where the effect of dynamical degrees of freedom in the medium are neglected.  
We will only give a brief introduction to lattice QCD and refer to the textbooks 
\cite{Gattringer:2010zz,Montvay:1994cy,Rothe:1992nt,DeGrand:2006zz}
and lecture notes \cite{Karsch:2003jg}
for more detailed introductions to lattice field theory. For the topics addressed in this lecture note we also like to refer to the overview articles
on QCD thermodynamics and the QCD phase transition\index{QCD!phase transition}
\cite{Karsch:2003jg, Ding:2015ona, Guenther:2020vqg}
and quarkonium in extreme conditions \cite{Rothkopf:2019ipj}.
}

\section{Motivation}
\label{sec:1}

\begin{figure}[htbp]
\centering{
\includegraphics[width=0.46\textwidth]{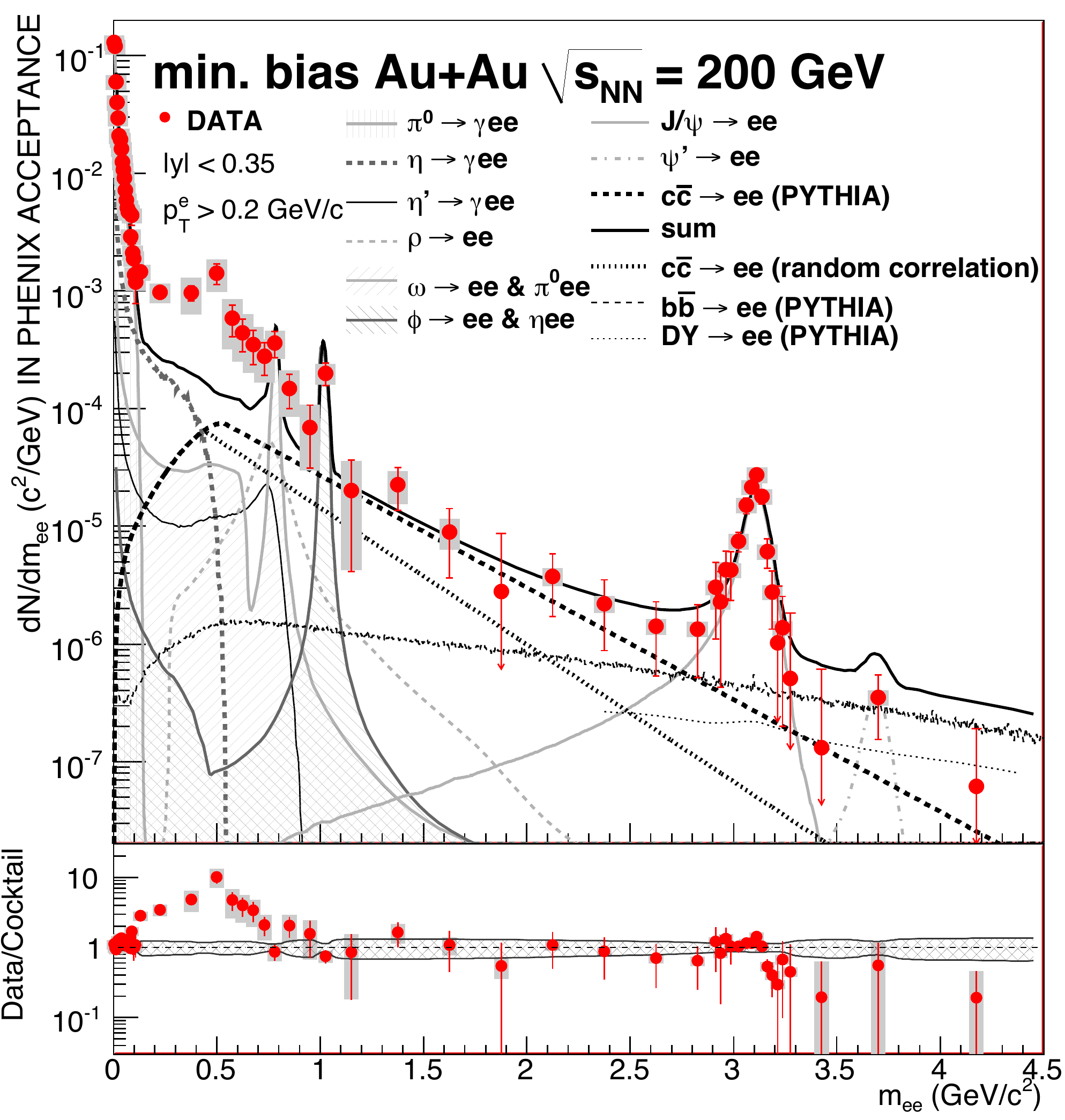}
\includegraphics[width=0.49\textwidth]{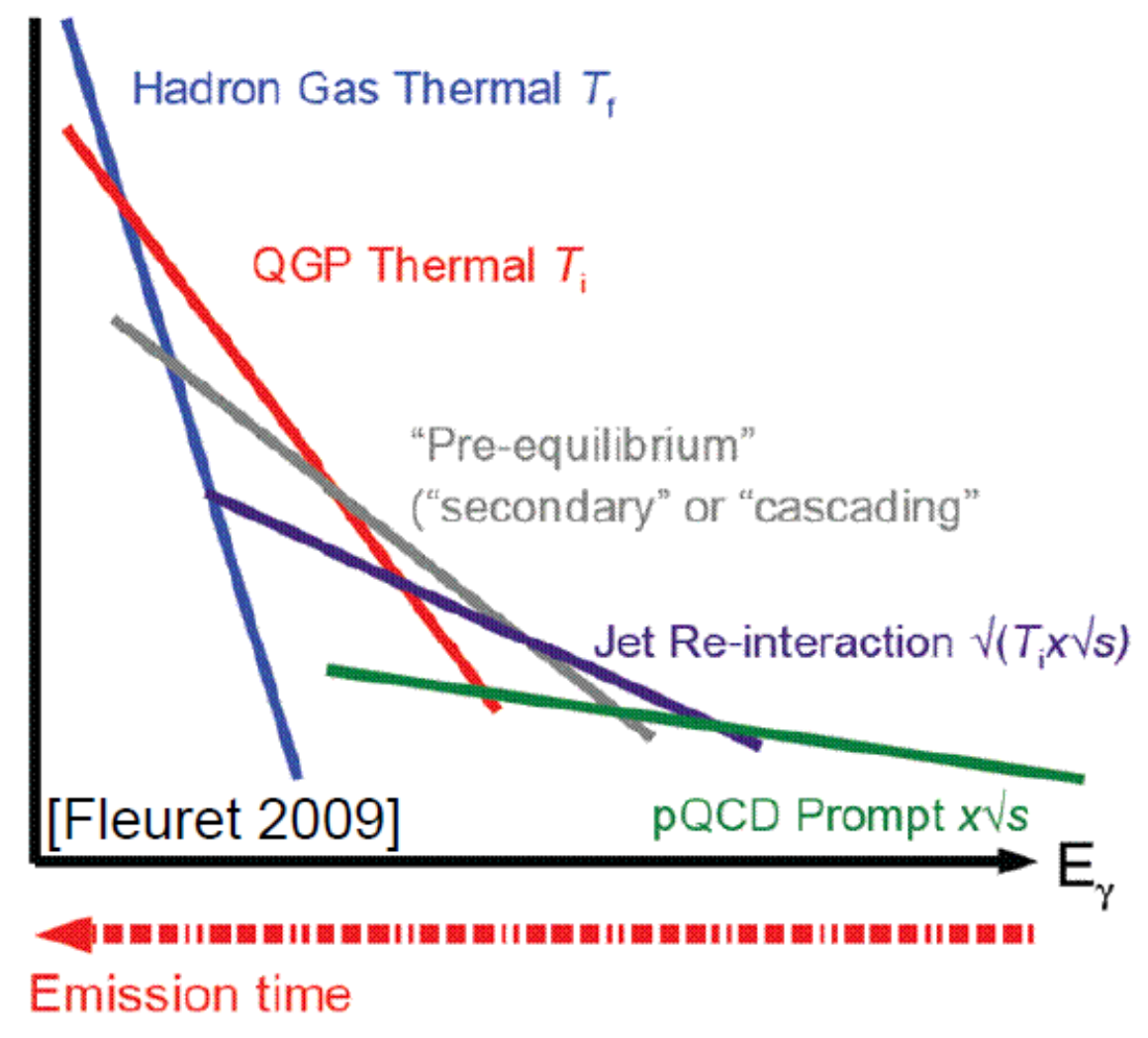}
}
\caption{Thermal dilepton rates from the PHENIX experiment (left), from \cite{Adare:2009qk}. A sketch of different sources of photons (right), from \cite{Fleuret:2009zza}.}
\label{fig8a}       
\end{figure}

Direct photons and leptons are produced in all stages of the heavy-ion collision and the subsequent evolution of the produced medium. Therefore they are good probes that carry information of the evolving medium, including information on the quark gluon plasma phase. As a matter of fact, these two objects form the basis for the electroweak interaction. Since their coupling to the QGP\index{QGP} constituents is weak \cite{David:2006sr}, once produced they escape from the interaction region hardly changed. The experimental quantities characterizing the thermal production rate\index{thermal!production rate} of these two objects are called thermal dilepton rates and thermal photon rates.
Fig.~\ref{fig8a}~(left) shows an example of experimentally measured dilepton yields, in this case from electron-positron pairs detected in at the PHENIX\index{PHENIX} experiment \cite{Adare:2009qk}, compared to the expected contributions from various hadronic decays. The excess in the low-mass region below 1~GeV$/c^2$ is expected to originate from the in-medium modification of the $\rho$-meson may be related the restoration of \index{chiral symmetry!restoration} chiral symmetry.
Fig.~\ref{fig8a}~(right) shows a sketch \cite{Fleuret:2009zza} of the expected photon rates from different stages of the medium produced in heavy-ion collisions. At intermediate photon energies thermal radiation from the QGP may be a dominant source which allows the study the thermal medium using photons.


As these electromagnetic probes are produced in all stages of the evolution of the system, the experimentally measurable dilepton and photon rates are integrated over the whole life-time of the dynamical evolving system. In lattice QCD\index{QCD!lattice} calculations contributions can be extracted for individual (equilibrium) stages of the thermal medium\index{thermal!medium}. This can provide important input for model calculations of the evolution of the system and the resulting integrated rates.   

The dilepton and photon rates could be determined from the vector meson spectral functions. For the thermal dilepton rate\index{thermal!dilepton rate} the relation is \cite{Ghiglieri:2016tvj}
\begin{equation}
 \frac{{\rm d} \Gamma_{\ell^-\ell^+}(\omega,\vec{k})}
   {{\rm d}\omega\, {\rm d}^3\vec{k} }  =  
 \frac{ 2 e^4 \sum_f Q_f^2 \theta(M^2 - 4 m_\ell^2) }{3 (2\pi)^5 M^2} 
 \biggl( 1 + \frac{2 m_\ell^2}{M^2}
 \biggr)
 \biggl(
 1 - \frac{4 m_\ell^2}{M^2} 
 \biggr)^{\frac{1}{2}} n^{ }_{B}(\omega) 
 \rho_{V}^{ }(\omega,\vec{k}) ,
\label{eq:LeptonRate}
\end{equation}
while for the thermal photon rate\index{thermal!photon rate} the formula reads
\begin{equation}
  \frac{{\rm d}\Gamma_\gamma(\vec{k})}{{\rm d}^3\vec{k}} = \frac{e^2 \sum_{f=1}^{N_f} Q_f^2 }{(2\pi)^3 k} n^{}_B(\omega)\rho_{V}^{ }(\omega=k,\vec{k}).
\label{eq:PhotonRata}
\end{equation}
Here $Q_f$ is the electric charge of flavour $f$ in units of the elementary charge $e$ and $M$ is the invariant mass $M^2=\omega^2-k^2$, $k=|\vec{k}|$. Note the vector channel spectral function in our notation is defined as                                          
\begin{equation}
\begin{split}
 \rho_{V}^{ }(\omega,\vec{k}) 
 \; &\equiv \rho^{ii}(\omega,\vec{k}) -\rho^{00}(\omega,\vec{k}) \\
 &\equiv \;
 \int {\rm d^4}x\  
  e^{i(\omega t - \vec{k}\cdot\vec{x})}
 \Big\langle 
    \frac{1}{2} \bigl[
     V^{i}_{ } (t,\vec{x}) \, , \,
     V^{i}_{ } (0)
    \bigr]
 -
    \frac{1}{2} \bigl[
     V^{0}_{ } (t,\vec{x}) \, , \,
     V^{0}_{ } (0) 
    \bigr]
 \Big\rangle^{ }_{c}
 \;, 
 \end{split}
\label{rhoV}
\end{equation}
where $\langle...\rangle_{c}$ means only the connected part is considered and $V_f^i\equiv \bar{\psi}_f\gamma^{i}\psi_f$.\\ 

In addition to the information on dilepton and photon rates,
the vector meson spectral function discussed has another contribution which allows to extract transport coefficients, i.e. diffusion coefficients and the electrical conductivity. The lattice determination of the transport coefficients relies on a Kubo formula\index{Kubo formula}, which establishes the connection between the low frequency regime of the spectral function and transport properties of the system from linear response theory. The derivation of Kubo formula can be found in common textbooks~\cite{Bellac:2011kqa,Yagi:2005yb,Kapusta:2006pm} and review papers~\cite{Meyer:2011gj,Hong:2010at,Petreczky:2005nh}. 

As an example the flavor diffusion
coefficient can be obtained from the frequency to zero limit of the slope of the vector meson correlation functions,
\begin{equation}
D = \frac{1}{3\chi_q}\lim_{\omega\rightarrow0^+}\sum_{i=1}^3\frac{\rho^{ii}(\omega,\vec{0})}{\omega}\;
\label{eq:D}
\end{equation}
from which the electric conductivity can also be obtained as
\begin{equation}
\sigma = e^2\sum_{f=1}^{N_f}Q^2_f\chi_qD\;,
\label{eq:sigma2}
\end{equation}
where $\chi_q$ is the quark number susceptibility defined by
the temporal component of the vector meson correlator,
\begin{equation}
\chi_q = \int_0^\beta d\tau\int d^3x\langle V^0(\tau,\vec{x})V^0(0)\rangle\;.
\label{eq:chiq}
\end{equation}
Other transport coefficients could be expressed in similar ways from different correlation functions. Transport coefficients are important ingredients for hydrodynamic and transport model descriptions of the evolution of the system.
Examples are the heavy quark diffusion constant $D$ \cite{Ding:2012sp},
the heavy-quark momentum diffusion coefficient $\kappa$ \cite{Francis:2015daa}
and for light quarks the electric conductivity \cite{Ding:2010ga,Ding:2016hua}.

In the following we will discuss how to calculate the corresponding hadronic correlation functions on the lattice and how to determine spectral functions from these. This will include a discussion on spectral reconstruction methods, i.e. inversion methods, as well as perturbative constraints on the spectral that allow to improve the extraction of spectral properties from lattice correlation functions.

\section{Hadronic correlation functions}
\label{sec:2}
The correlation function of interest is defined as the expectation value of a two point function of the meson current
\begin{equation}
G_H(\tau,\vec{x}) = \langle J_H(\tau,\vec{x}) J_H^\dagger(0,\vec{0})\rangle \; ,
\label{eq:Gdef}
\end{equation}
with the meson current defined as
\begin{equation}
J(\tau,\vec{x}) = 2\kappa Z_H\bar{\psi}(\tau,\vec{x})\Gamma_H\psi(\tau,\vec{x})\;.
\label{eq:Jdef}
\end{equation}
The matrix $\Gamma_H=1,\gamma_5,\gamma_\mu,\gamma_5\gamma_\mu$ determines
the scalar, pseudoscalar, vector, and axial vector mesons respectively. $\kappa$ is called $hopping\ parameter$ responsible for the quark mass
\begin{equation}
    \kappa=\frac{1}{2(am_q+4)}.
\end{equation}
$Z_H$ are the renormalization constants that can be found in \cite{Skouroupathis:2007dk,Skouroupathis:2008mf,Gockeler:2010yr,Luscher:1996jn}. The quark mass should be tuned in lattice calculations to set the hadron mass to the desired value. 
Making use of the integral property of $Grassmann$ $number$ Eq.~(\ref{eq:Gdef}) can be rewritten as
\begin{equation}
\begin{split}
    &\langle G_H(m|n)\rangle=-\frac{1}{\mathcal{Z}}\int \mathcal{D}[U]e^{-S_G[U]}\prod_{f'}\det [D_{f'}]\times\big{\{}\mathrm{Tr}[\Gamma_HD_f^{-1}(m|n)\Gamma_HD_f^{-1}(m|n)]\\
    &\ \ \ \ \ \ \ \ \ \ \ \ \ \ \ \ \ \ \ \ \ \ \ \ \ \ \ \ \ \ \ \ \ \ \ \ \ \ \ \ \ \ \ \ \ \ \ \ \ \ \ \ \ \ \ \ \ \ \ \ \ \ \ \ \ \ \ -2\mathrm{Tr}[\Gamma_HD^{-1}_f(n|n)]\mathrm{Tr}[\Gamma_HD^{-1}_f(m|m)]\big{\}},\\
    &\mathcal{Z}=\int \mathcal{D}[U]e^{-S_G[U]}\prod_{f'}\det [D_{f'}],\\
    \end{split}
    \label{full_corr}
\end{equation}
where $m$ and $n$ are the positions of the two currents and $f$, $f'$ denote the flavor of quarks. $U$ are gauge links constructed from the gauge fields.

When calculating mesonic correlators Eq.~(\ref{full_corr}) on the lattice, one usually only considers the connected part, i.e. the first line of Eq.~(\ref{full_corr}). The disconnected parts are expensive to calculate, and according to OZI suppression their contribution is small so are neglected in most calculations. Furthermore they do not contribute to iso-vector mesons.
Since the lattice simulations are carried out at finite lattice spacings, there are lattice cutoff effects in the calculated quantities. To eliminate the cutoff effects one usually simulates the same physics at different lattice spacings (and lattice extensions), and then performs the so-called continuum extrapolations to reach the continuum limit. The Ans\"{a}tz used in the extrapolation must respect the order of errors in the observables considered.

In this lecture we will focus mainly on vector mesons. Performing
the Fourier transform along the spatial directions we obtain the correlator
in a mixed representation
\begin{equation}
G(\tau,\vec{p}) = \int {\rm d^3}x e^{-i \vec{p}\vec{x}} \langle J(\tau,\vec{x}) J^\dagger(0,\vec{0})\rangle .
\label{eq:FourierG}
\end{equation}
 Spectral functions are related to the corresponding correlators 
by an integral equation \footnote{There are different conventions in the definition of $\rho$ in the literature (also in these lecture notes from chapter to chapter), for instance sometimes $G(\tau,\vec{p}) = \int_0^\infty \frac{{\rm d}\omega}{\pi}\rho(\omega,\vec{p},T)K(\tau,\omega,T)$ is used. In this case the definition of transport coefficient via spectral function also changes accordingly but the physics remains unchanged.}
\begin{equation}
G(\tau,\vec{p}) = \int_0^\infty \frac{{\rm d}\omega}{2\pi}\rho(\omega,\vec{p},T)K(\tau,\omega,T)\;,
\label{eq:IntRelation}
\end{equation} 
with the kernel function
\begin{equation}
K(\tau,\omega,T) =\frac{\cosh(\omega(\tau-\frac{1}{2T}))}{\sinh(\frac{\omega}{2T})}\; 
\label{eq:Kernel}
\end{equation}
respecting the periodic boundary condition.

As mentioned above, coefficients like thermal dilepton/photon production rate, electric conductivity and heavy quark diffusion coefficient,  all can be obtained from the vector current-current correlators. But one should be clear in mind what valence quark should be used when calculating the vector correlators. For thermal dilepton/photon production rate\index{thermal!production rate}, one should use light quarks. As for the electric conductivity, in principle it should be a summation of all possible flavors (but usually only $u$, $d$ and $s$ are considered). And for heavy quark diffusion, obviously only heavy flavors, i.e. charm and beauty will be considered. And in different scenarios, different prior information can be obtained from perturbative theories for the spectral function. These information impose strong constraint on the shape or structure of the spectral function and makes the spectral reconstruction possible. In the following sections after a short introduction to the reconstruction methods we will show some lattice results to the fore-mentioned transport coefficients case by case based on different spectral reconstruction strategies.

\section{Inversion methods}\label{Sec:inverse}
To infer the spectral function within lattice QCD methods
one needs to invert Eq.~\eqref{eq:IntRelation}. Since numerical
data is only available for a finite number of points this 
procedure cannot be unique, i.e. there are infinitely many spectral
functions leading to the same correlator. This is the famous known ill-posed problem. And to solve this problem, there are many different methods on the market. In this section we give a brief introduction to some of them. 

First we introduce a very commonly used method: Maximum Entropy Method (MEM) \cite{Jarrell:1996rrw}. It is based on Bayes' theorem and the solution is obtained by maximizing the probability of having the solution $\rho$ from the given lattice correlators $G$ and the prior information known about the solution $D$
\begin{equation}
\label{posterior_probability}
P[\rho|G,D,\alpha] = \frac{P[G|\rho,D,\alpha]P[\rho|D,\alpha]}{P[G|D,\alpha]}.
\end{equation}
Here $P[G|\rho,D,\alpha]$ and $P[\rho|D,\alpha]$ are the likelihood function and the prior probability, respectively. The likelihood function can be obtained purely from our lattice data
\begin{equation}
P[G|\rho,D,\alpha]\propto \exp(-\chi^2/2)
\end{equation}
based on the Gaussian distribution of our lattice data. While the prior probability is given by
\begin{equation}
P[\rho|D,\alpha]\propto \exp(\alpha S)
\end{equation}
where $S$ is the Shannon-Jaynes entropy\index{entropy} defined as
\begin{equation}
\label{entropy}
S[\rho] = \int_0^\infty\frac{\mathrm{d}\omega}{2\pi}\;[\rho(\omega)-D(\omega)-\rho(\omega)\;\ln \frac{\rho(\omega)}{D(\omega)}],
\end{equation}
where $\alpha$ is a regularization parameter that controls contributions to the reconstructed image from the prior information relative to the data. The final solution is obtained by integrating out $\alpha$ with special solution at each $\alpha$ and weights constructed from the above ingredients. Lattice studies using this method can be found in \cite{Ding:2012sp,Asakawa:2000tr,Aarts:2007pk,Ikeda:2016czj,Ding:2018uhl}. And later on we will show the charmonia and bottomonia spectral functions obtained from this method. Recently a novel Bayesian approach was proposed by replacing the entropy\index{entropy} in MEM by a different term in \cite{Burnier:2013nla}
\begin{equation}
S_{MEM}[\rho] \Longrightarrow S_{BR}[\rho]\equiv \int_0^\infty\frac{\mathrm{d}\omega}{2\pi}\;[1-\frac{\rho(\omega)}{D(\omega)}+\ln \frac{\rho(\omega)}{D(\omega)}].
\end{equation}
Studies using this method can be found in \cite{Rothkopf:2013kya,Burnier:2015tda,Burnier:2015nsa,Burnier:2014ssa}. 

Recently two approaches based on stochastic sampling of spectral images were introduced to extract the spectral functions from the correlators \cite{Ding:2017std}, namely stochastic optimization method (SOM) and stochastic analytical inference (SAI).  SOM has the advantage that no default model is needed while  SAI is proven to be a generalized of MEM and reduces to MEM in mean field approximation. And for a specific choice of the prior SAI can also proven to be equivalent to SOM. Other available methods include Backus Gilbert method  \cite{Francis:2015daa,Brandt:2015aqk}, which manipulates in the local vicinity of some frequency range in a model-independent way,  Tikhonov method with Morozov discrepancy principle~\cite{Dudal:2013yva} and sparse modeling \cite{Itou:2020azb}. Apart from all these methods, a classic approach that is widely used is $\chi^2$-fitting. In this method one usually needs extra physical constraints imposed on the spectral function to guarantee meaningful results. And a few fit parameters are introduced to account for the uncertainties when adapting perturbative based Ans\"{a}tz to non-perturbative lattice data. In the following sections we will show results based on some of the above methods.

\section{Thermal dilepton rate and electrical conductivity}\label{sec:dilep}
In this section we will describe the procedure to determine estimates for the thermal dilepton rate and electrical conductivity of the QGP\index{QGP} using light vector meson correlation functions calculated on the lattice. We will first discuss the comparison to the correlators in the free non-interacting limit, which is also used to normalize the lattice correlators. Continuum extrapolated vector meson correlation functions will then provide the basis for the spectral reconstruction, where a perturbatively inspired Ans\"{a}tz in the UV part and a transport peak in the IR part of the spectral function is used to fit the continuum vector meson correlation function. 
As this study required large and fine lattices together with a subsequent continuum extrapolation, the lattice calculations were done in the quenched approximation, where dynamical fermionic degrees of freedom have been neglected, i.e. in a deconfined gluonic medium.
The lattice setup and other details for this study can be found in ~\cite{Ding:2016hua}. 
For a recent review on the determination of the electrical conductivity from lattice QCD\index{QCD!lattice} see \cite{Aarts:2020dda}.

\subsection{Continuum extrapolated vector meson correlation functions}

\begin{figure}[htbp]
\centering{
\includegraphics[scale=.55]{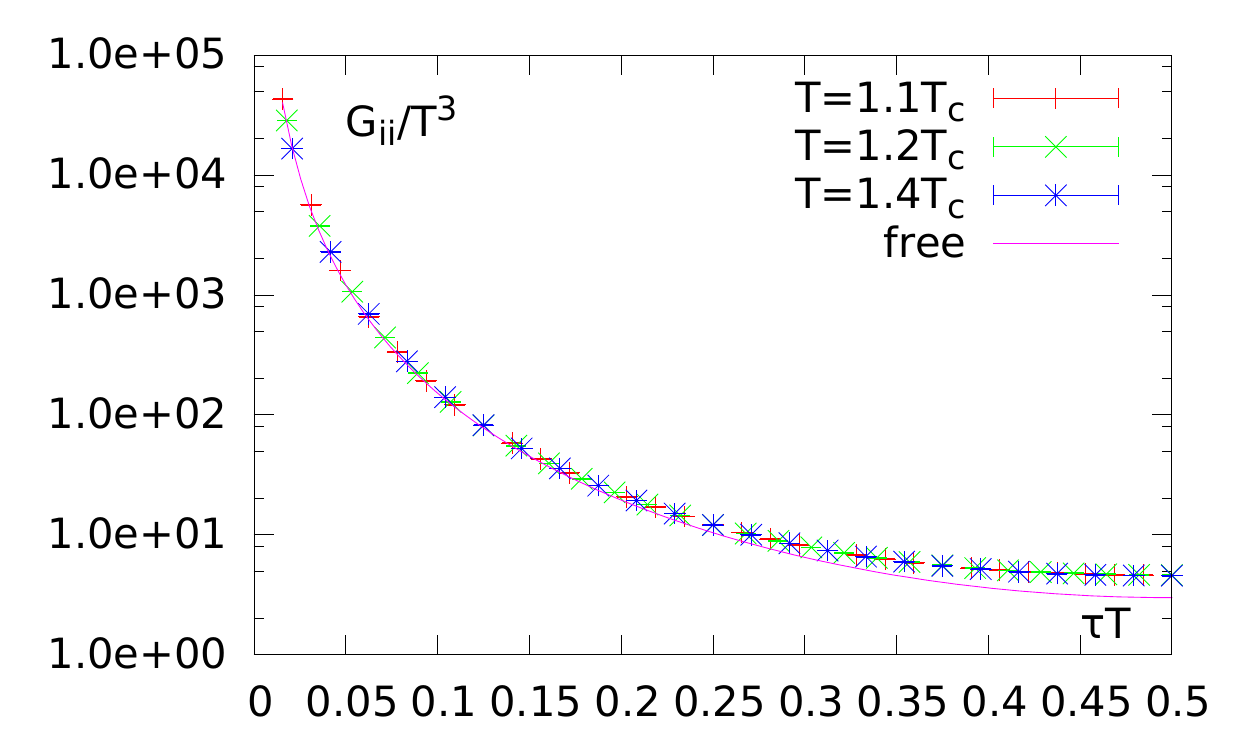}}
\caption{Lattice results of the vector meson correlation functions for three different temperatures compared to the free continuum correlator, from \cite{Ding:2014dua}.}
\label{fig9}
\end{figure}

As an example for light quark meson correlators in the vector channel we show results on finite lattices at some lattice spacing $a$ with lattices extents $N_t$ corresponding to three different temperatures above the critical temperature\index{critical!temperature} in Fig.~\ref{fig9}. The correlation functions show exponential damping behavior up to the midpoint of the lattice. Note that the Euclidean correlation functions are periodic around the midpoint, $\tau T=0.5$, where the temperature is given by $T=1/aN_{\tau}$ and $\tau a$ is the separation in the temporal direction. On this scale no temperature effects are visible and the correlators are close to the non-interacting free correlator. 

To see the thermal effects we normalize the correlation functions by the massless non-interacting (free) correlator,
\begin{equation}
G_V^{\rm free}(\tau,\omega,T) = 6T^3\left(\pi(1-2\tau T )\frac{1+\cos^2(2\pi \tau T)}{\sin^3(2\pi \tau T)}+2\frac{\cos(2\pi \tau T)}{\sin^2(2\pi \tau T)}\right)\;.
\label{eq:Gfree}
\end{equation}
which is also shown in Fig.~\ref{fig9}. In general the correlators calculated on finite lattices need to be renormalized before extrapolating to the continuum. For the vector channel it is more suitable to use renormalization independent ratios by dividing by the quark number susceptibility $\chi_q/T^2=-G_{00}/T^3$ which is obtained from the (constant) temporal component of the vector meson correlator and shares the same renormalization as the spatial component. In this way any 
ambiguities stemming from renormalization are absent.

\begin{figure}[htbp]
\centering{
\includegraphics[width=0.48\textwidth]{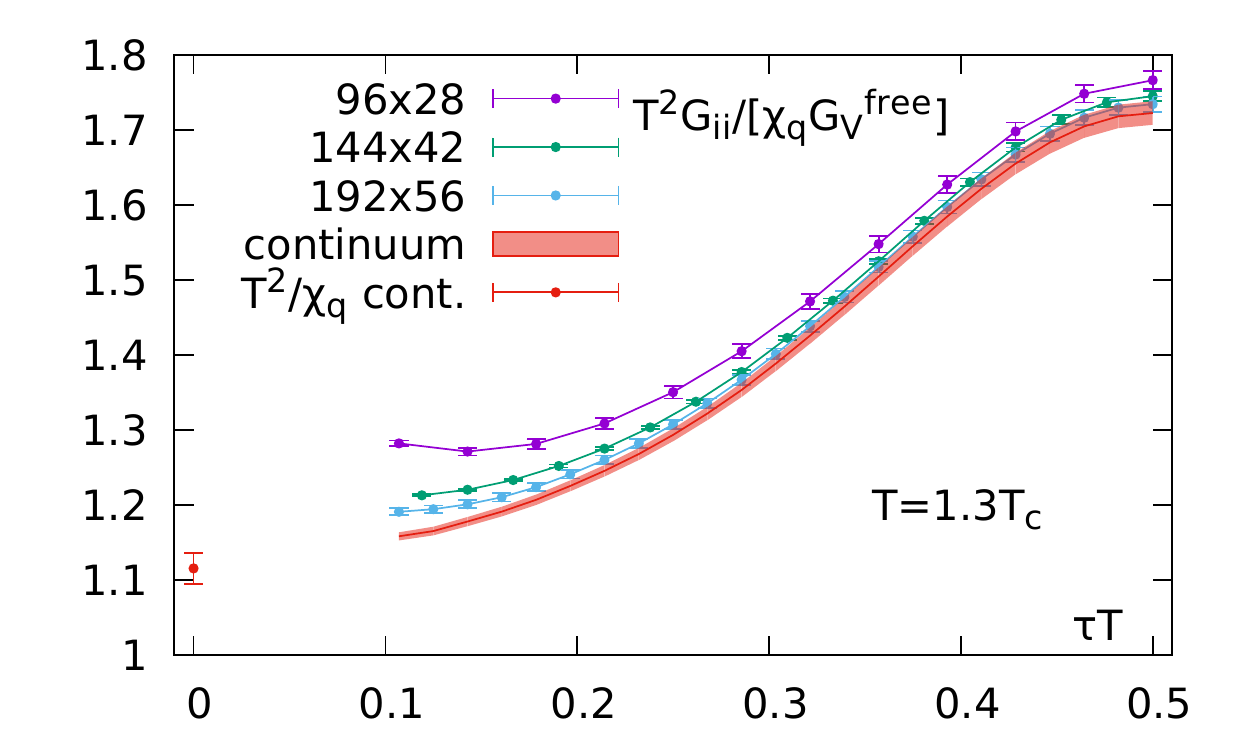}
\includegraphics[width=0.48\textwidth]{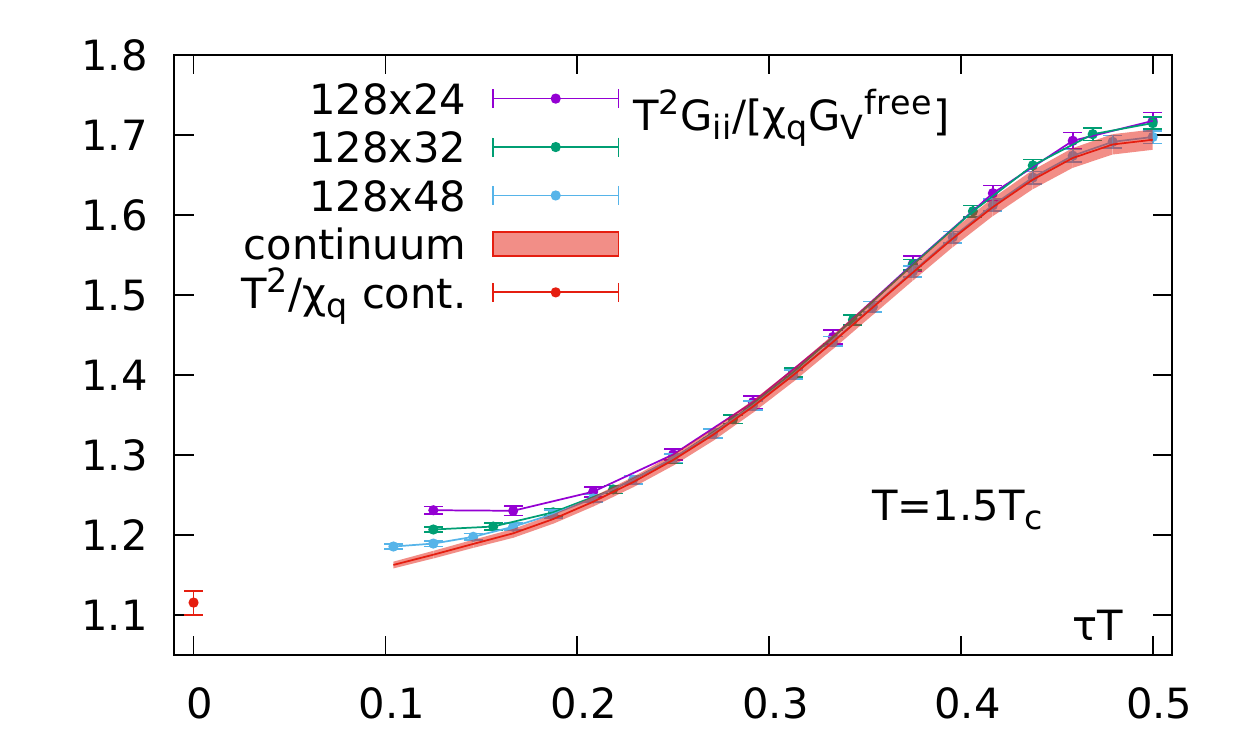}}
\centering{
\includegraphics[width=0.48\textwidth]{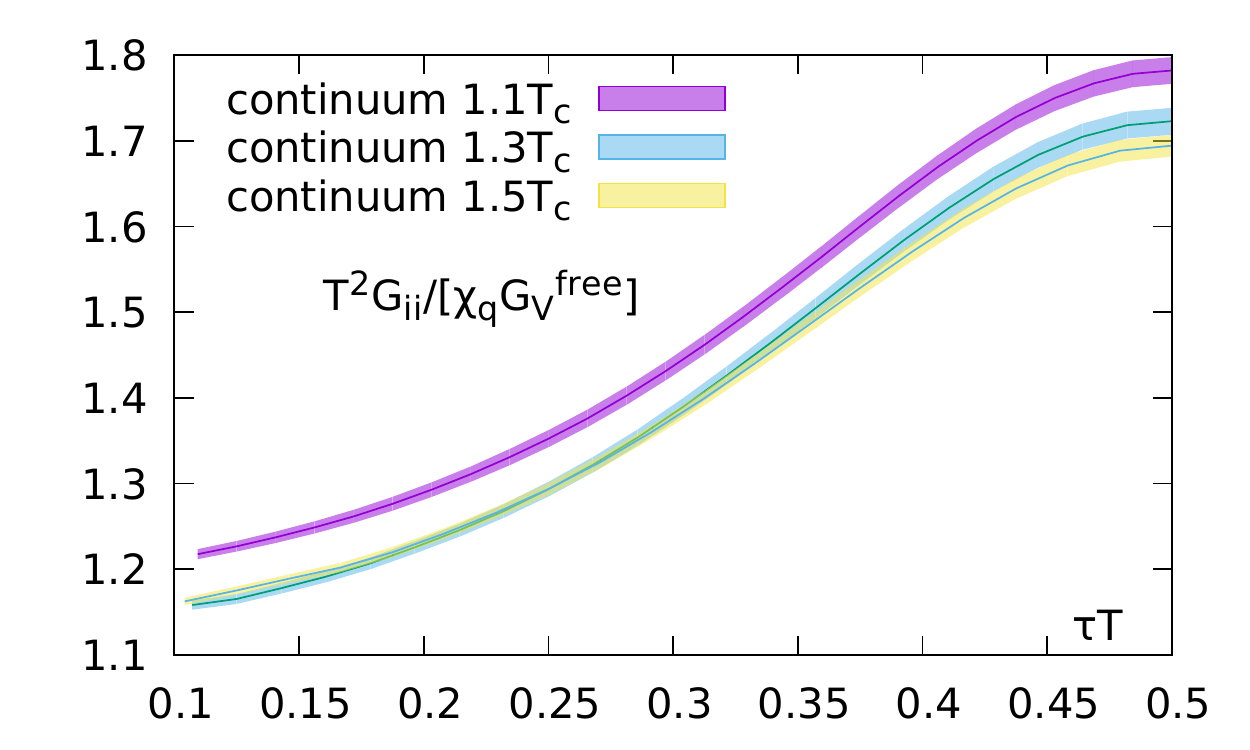}
\includegraphics[width=0.48\textwidth]{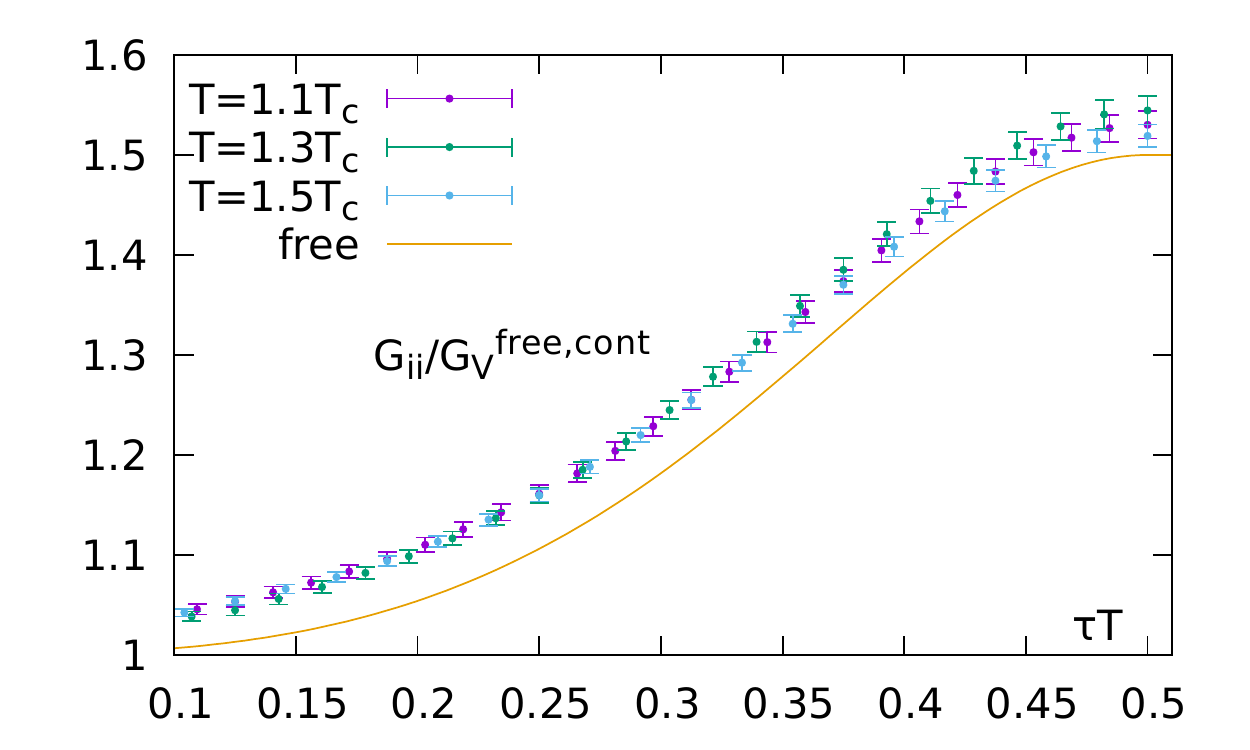}
}
	\caption{
	Normalized lattice correlators and the resulting continuum extrapolated correlators in the vector channel. Taken from \cite{Ding:2016hua}.}
\label{fig10}         
\end{figure}

Two examples of the lattice correlation functions normalized in this way are shown in Fig.~\ref{fig10}(top). When comparing the results for the three different lattices strong cut-off effects are visible especially at small $\tau T$. To remove the cut-off effects a continuum extrapolation was performed in $1/N_{\tau}^2$ for all distances $\tau T$ down to around $\tau T\simeq 0.1$. The results of the continuum extrapolated vector meson correlation functions for three temperatures in the QGP\index{QGP} are shown in Fig.~\ref{fig10}(bottom). For the bottom right plot the continuum extrapolated values for $\chi_q$ were used to remove the dependence on the quark number susceptibility.
The temperature effects observed in the lower left plot are mainly caused by the normalization with $\chi_q/T^2$ and disappear in the lower right plot.


 Note that in the lower-right panel of Fig.~\ref{fig10}, for $\tau\rightarrow0$, the ratio $G_{ii}/G_{V}^{\rm free}$ approaches unity,
 which manifests that at small distance the correlator is dominated by the UV part of the spectral function, which should approach to the free spectral function according to asymptotic freedom\index{asymptotic freedom}. At large $\tau$ one sees substantial deviations from the free case reaching up to around $50\%$. This is already an indication for non-perturbative contributions and has an impact on the determination of the transport coefficients and thermal dilepton rates as will be discussed in the following section. 
In this temperature window the only scale is the temperature and no resonance contributions, e.g. a rho meson, are expected in the gluon plasma\index{gluon!plasma} at these temperatures. Therefore the weak temperature dependence already indicates that temperature effects in the temperature scaled dilepton rates and the electrical conductivities will be rather small.


\subsection{Lattice estimate on the thermal dilepton rate and electrical conductivity}

We will now discuss the spectral reconstruction in terms of a phenomenologically inspired Ans\"{a}tz for the spectral function 
and show results for the thermal dilepton rate\index{thermal!dilepton rate} and electrical conductivity based on the continuum extrapolated vector meson correlation functions discussed in the previous section.
These results obtained from continuum correlators are only available in the quenched approximation, as only in this case the required large temporal extents of the lattice could be achieved achieved so far. Nevertheless, this already allows to study the these observables in a gluonic medium and provides a methodology that can be extended to a more realistic medium including light dynamical degrees of freedom in the future. 


\begin{figure}[thbp]
\centering{ \includegraphics[width=0.46\textwidth]{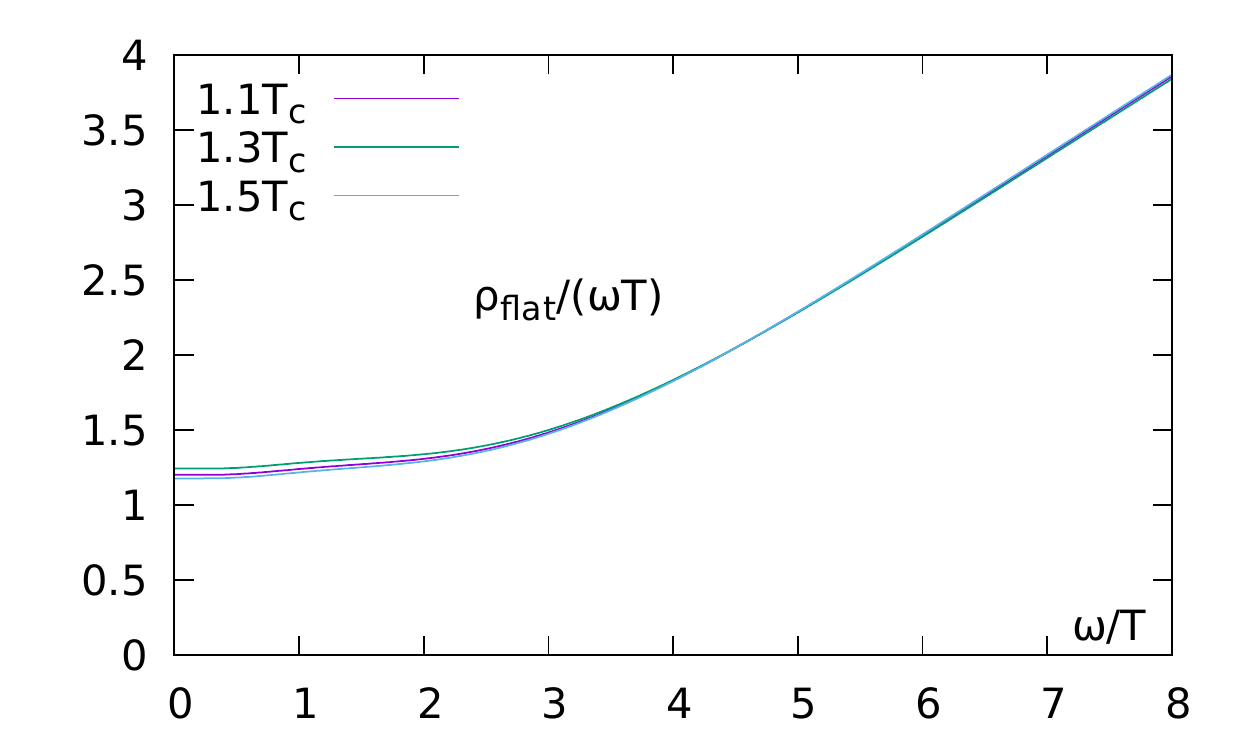}
 \hfill
 \includegraphics[width=0.46\textwidth]{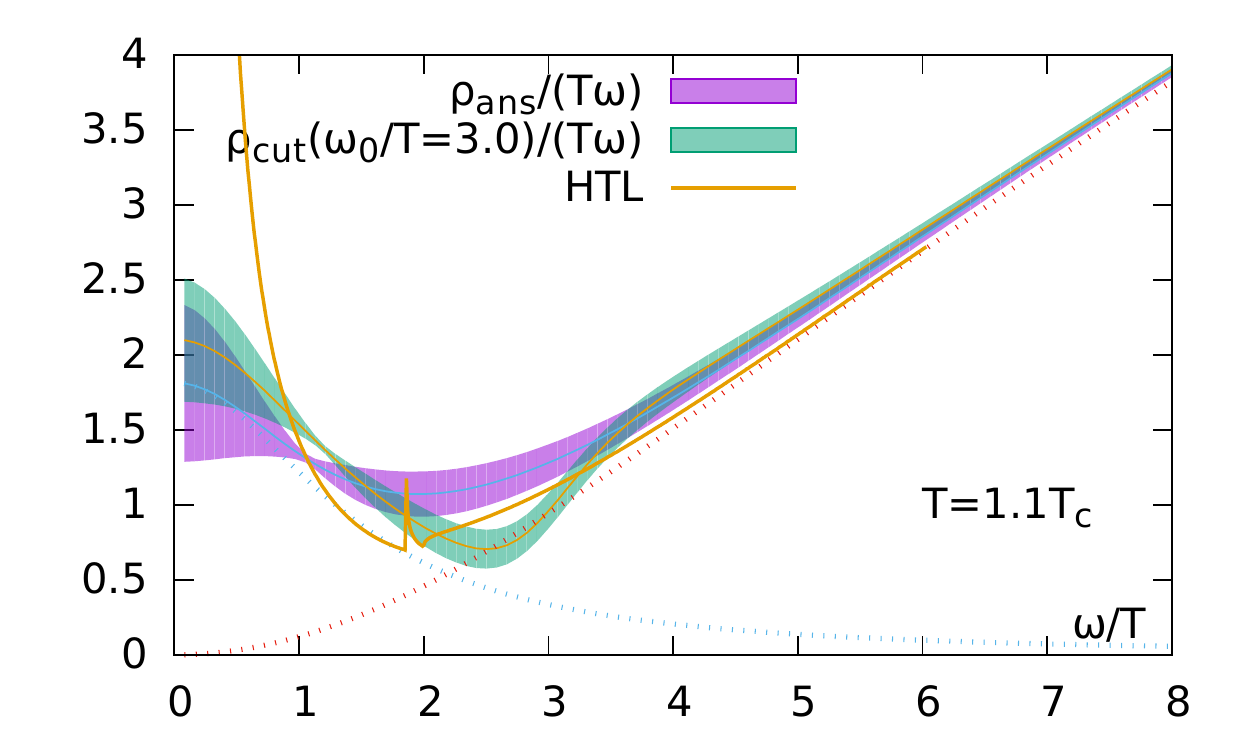}
 }
 \centering{ \includegraphics[width=0.46\textwidth]{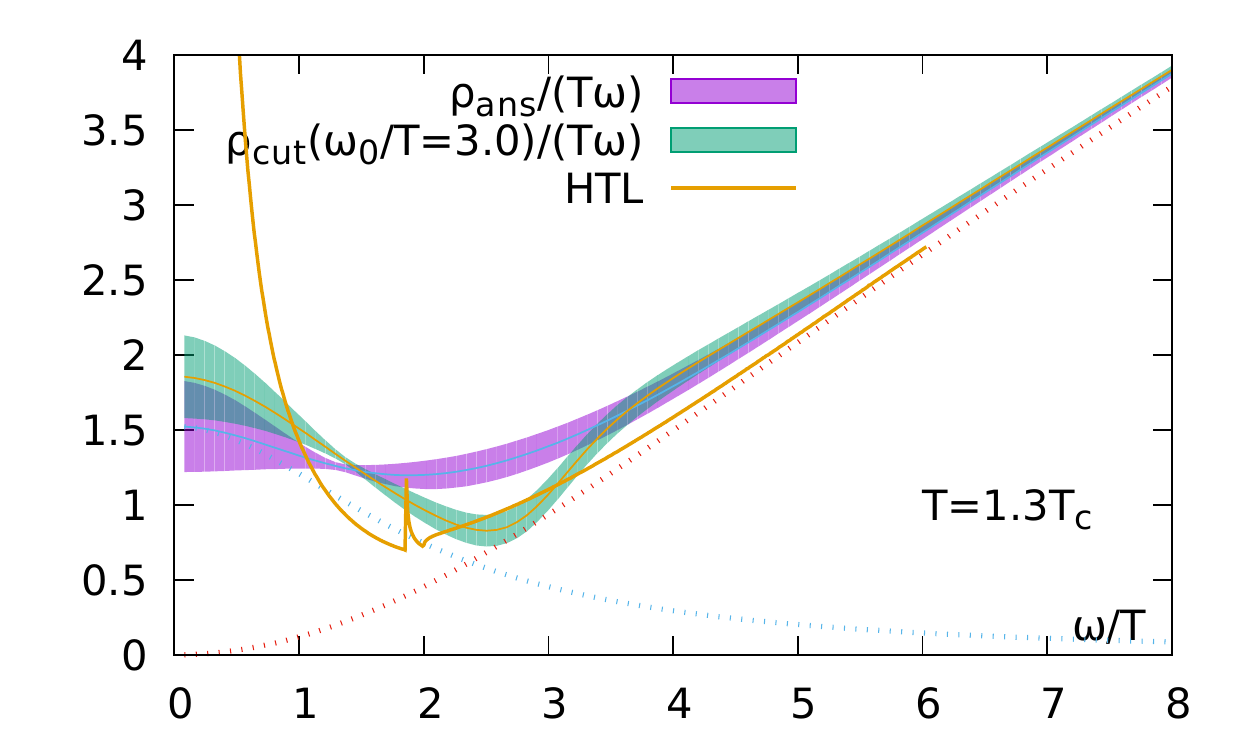}
 \hfill
 \includegraphics[width=0.46\textwidth]{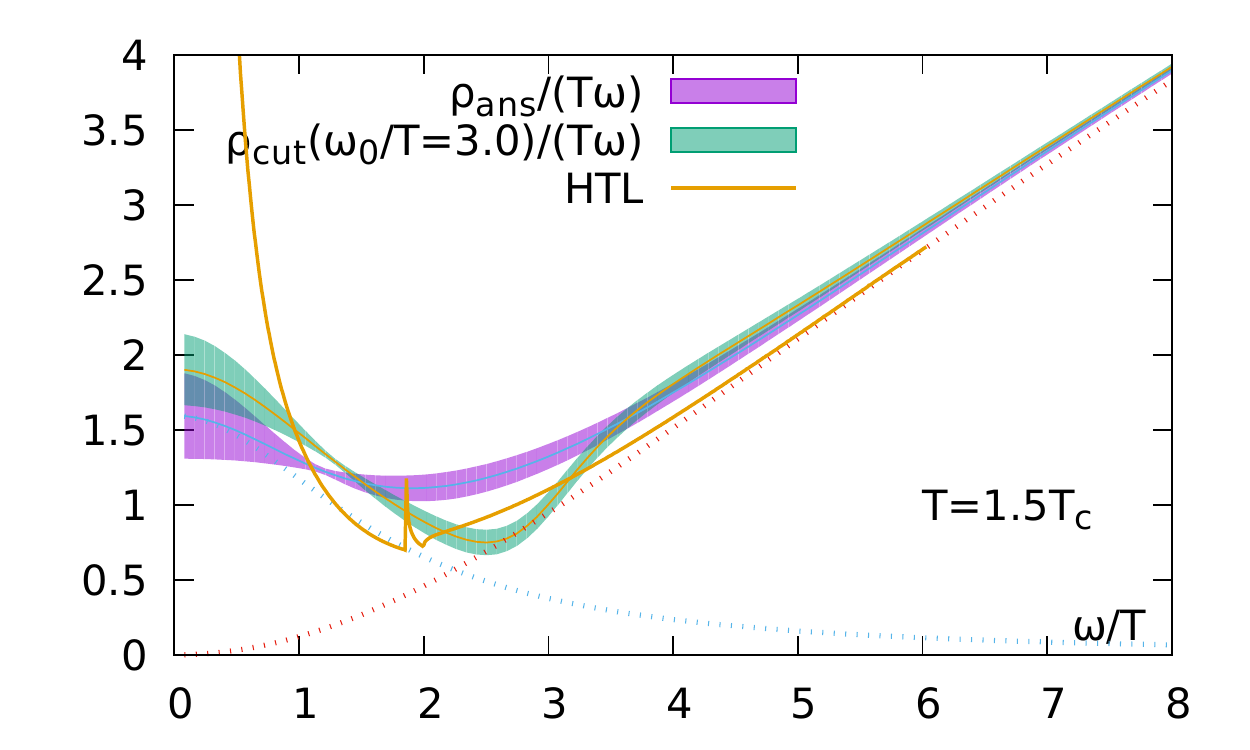}
 }
	\caption{Fitted shape of the vector spectral function to the continuum lattice correlation function. Upper-left panel shows results for Eq.~(\ref{eq:rhoflat}). The other panels show results for Eq.~(\ref{eq:rho}). Also shown is the hard thermal loop\index{thermal!loop} (HTL) result. Taken from \cite{Ding:2016hua}.}
\label{fig12}   
\end{figure}


In principle the spectral function can be split into different parts. Due to asymptotic freedom\index{asymptotic freedom}, at asymptotically large frequencies the correlation functions should approach their free continuum limit, which at large frequencies gets perturbative corrections.
At intermediate frequencies possible contributions from bound states may arise at small temperatures, but are not expected in the temperature region discussed here, as already indicated by the discussion on the correlator level in the previous section.
Since in this study all correlators are obtained at above $T_c$ temperature, even the lightest meson, e.g. the $\rho$ meson in the vector channel, can be neglected. At very low frequencies transport contributions, i.e. a transport peak, is expected in the vector channel. 

Due to asymptotic freedom the high temperature
limit of the spectral function should approach
the free particle limit. The latter can be computed analytically \cite{Karsch:2003wy,Aarts:2005hg},
\begin{equation}
\rho_{00}^{\rm free}(\omega) = 2\pi\omega T^2 \delta(\omega)\; ,
\label{eq:rho00}
\end{equation}
\begin{equation}
\rho_{ii}^{\rm free}(\omega) = 2\pi\omega T^2 \delta(\omega) + \frac{3}{2\pi}\omega^2\tanh(\omega/(4 T))\; . 
\label{eq:rhoij}
\end{equation}
and eventually $\rho_V(\omega) = \rho_{ii}(\omega)-\rho_{00}(\omega)$ so that delta 
contributions cancel.
When one includes interactions the delta function in the
temporal component is protected but the one in the spatial components gets smeared out due to interactions leading to a transport peak at low frequencies. This motivates the following Ans\"{a}tz 
\begin{equation}
\rho_{00}(\omega) = 2\pi\chi_q\omega\delta(\omega)\; ,
\label{eq:rho001}
\end{equation}
\begin{equation}
\rho_{ii}(\omega) = 2\pi\chi_q c_{\rm BW}\frac{\omega\Gamma/2}{\omega^2+(\Gamma/2)^2} + \frac{3}{2\pi}(1+\kappa)\omega^2\tanh(\omega/(4T))\;,
\label{eq:rho}
\end{equation}
with the quark number susceptibility $\chi_q$, two parameters for the transport peak, $c_{\rm BW}$ and $\Gamma$ and a coefficient $\kappa$, which attributes to perturbative corrections of the continuum part of the spectral function.
At the leading order in perturbation theory $\kappa=\alpha_s/\pi$.
The electric conductivity, $\sigma$, is given by the frequency to zero limit (Kubo formula\index{Kubo formula}) and can be expressed by the parameters introduced in the Ans\"{a}tz as 
\begin{equation}
\frac{\sigma}{T} = \frac{C_{em}}{6}\lim_{\omega\rightarrow0}\frac{\rho_{ii}(\omega,\vec{p}=0,T)}{\omega T} = \frac{2}{3}\frac{\chi_q}{T^2}\frac{T}{\Gamma} c_{BW}C_{em}\;,
\label{eq:Cem}
\end{equation}
where $C_{em}=\sum_f Q_f^2$ is the sum of
the square of the elementary charges of the quark flavors f.
This rather simple Ans\"{a}tz for the spectral function can be fitted to the continuum extrapolated vector meson correlator and already describes the data rather well \cite{Ding:2016hua}.
Nevertheless to study systematic uncertainties of the resulting spectral functions, 
one can further modify this Ans\"{a}tz accounting for effects in the frequency regions where the transport peak and the continuum part overlap. As the continuum part contributes also in the low frequency regime we modify this part in such a way that the small $\omega$ contribution is smoothly removed, 
\begin{equation}
\rho_{ii}(\omega) = 2\pi\chi_q c_{\rm BW}\frac{\omega\Gamma/2}{\omega^2+(\Gamma/2)^2} + \frac{3}{2\pi}(1+\kappa)\omega^2\tanh(\omega/(4T))\Theta(\omega_0,\Delta_\omega)\;,
\label{eq:rhoii}
\end{equation}
with
\begin{equation}
\Theta(\omega,\omega_i,\Delta_i) = \left(1+\exp(\frac{\omega_i^2-\omega^2}{\omega\Delta_i})\right)^{-1}\;.
\label{eq:tilderTheta}
\end{equation}
The effect of this modification can be seen in the right panel of Fig.~\ref{fig12}, which shows the corresponding result of the fit.

\begin{figure}[thbp]
\centering{
\includegraphics[height=.45\textwidth]{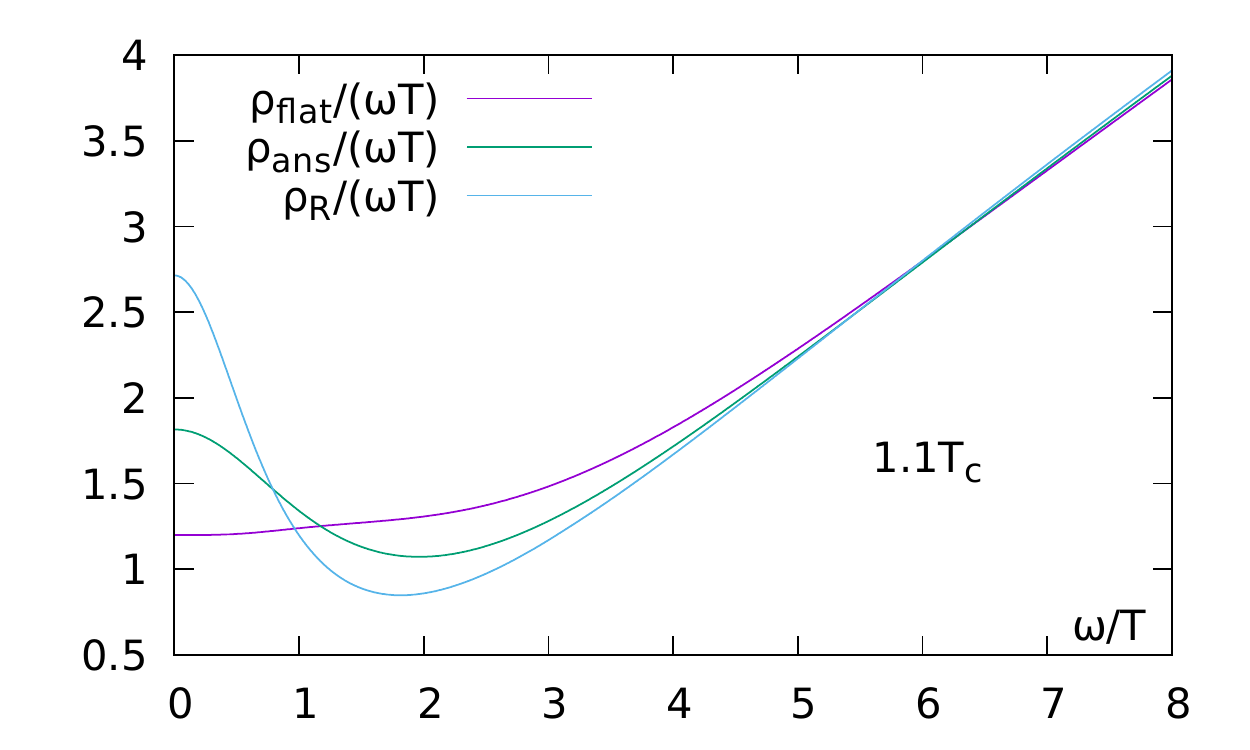}
}
\caption{Result for the spectral function fitted to the continuum extrapolated correlator at $1.1~T_c$ for the three Ans\"{a}tz of the spectral function \cite{Ding:2016hua}.}
\label{fig13}       
\end{figure}


As the Ans\"{a}tz \ref{eq:rho} and the modified Ans\"{a}tz \ref{eq:rhoii} contain a real transport peak, the corresponding fits are not sensitive to a behavior where the small frequency part is rather flat behavior. Therefore an additional
Ans\"{a}tz explores the small frequency behavior to examine
the appearance of a flat frequency to zero limit rather than a real transport peak,
\begin{equation}
\rho_{ii}^{\rm flat}(\omega) = a\chi_q\omega(1-\Theta(\omega,\omega_0,\Delta_0)) + (1+k)\rho_{\rm free}(\omega)\Theta(\omega,\omega_1,\Delta_1)\;.
\label{eq:rhoflat}
\end{equation}

\begin{figure}[thbp]
\centering{
\includegraphics[width=0.49\textwidth]{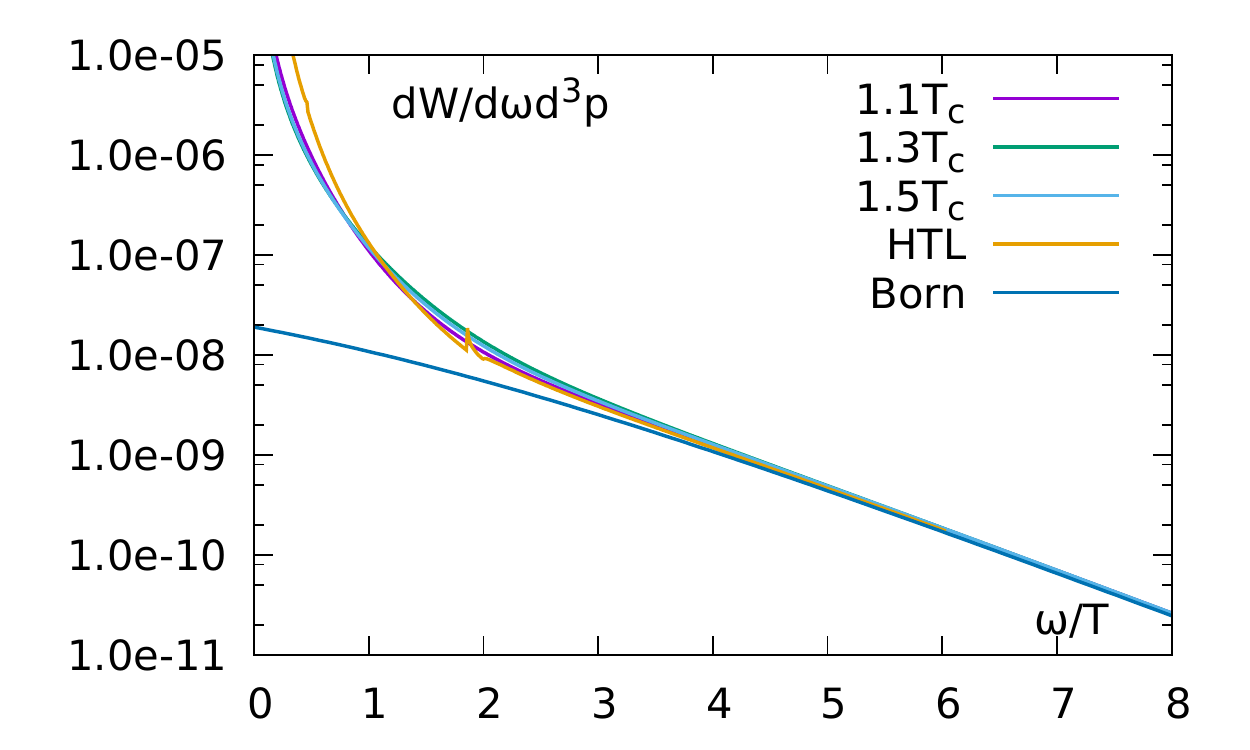}}
\includegraphics[width=0.49\textwidth]{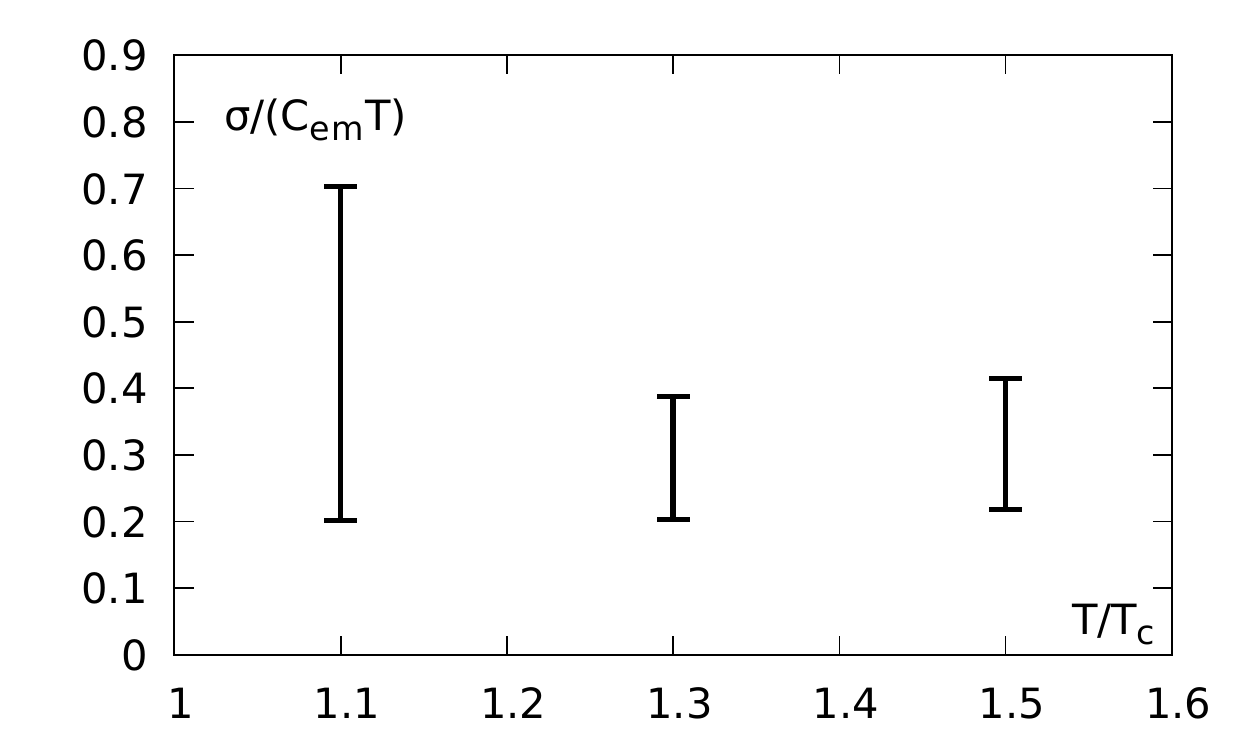}
\caption{
Left: The thermal dilepton rate obtained from $\rho_{\text{R}}$ 
		as a function of $\omega/T$. Also shown is the hard thermal loop (HTL rate) and the non interacting Born rate.
		Right: The final results for the electrical conductivity incorporating the
		the full systematic uncertainties, i.e. the minimum and maximum conductivities, 
	respectively, of $\rho_{\text{ans}}$ and $\rho_{\text{R}}$.
Taken from \cite{Ding:2016hua}.}
\label{fig14}       
\end{figure}

This is motivated by the strongly coupled results of the AdS/CFT correspondence
where a rather featureless and flat low energy behavior is generic \cite{Teaney:2006nc}.
The fitted shape of Eq.~(\ref{eq:rhoflat}) to lattice data is presented in left panel of Fig.~\ref{fig12}.

Another region of changing or improving the Ans\"{a}tz is in the UV behavior of the spectral function. So far we have used the leading order perturbative correction to the free behavior incorporated by constant, i.e. not running coupling constant.
At very high frequencies the spectral function can be determined from perturbation theory and furthermore thermal effects should be suppressed
at sufficiently large frequencies. Therefore vacuum perturbation theory is expected to become reliable and can be used to improve the model in the high frequency regime.
The five loops vacuum spectral function \cite{Baikov:2008jh,Baikov:2009uw} reads
\begin{equation}
\rho_V(\omega) = \frac{3\omega^2}{4\pi}R(\omega^2)\;,
\label{eq:rhoV}
\end{equation}
where 
\begin{eqnarray}
R(\omega^2) &=& r_{0,0} + r_{1,0}\alpha_s + (r_{2,0}+r_{2,1}l)\alpha_s^2 + \\
\nonumber &+& (r_{3,0} + r_{3,1}l + r_{3,2}l^2)\alpha_s^3+\\
\nonumber &+& (r_{4,0} + r_{4,1}l + r_{4,2}l^2 + r_{4,3}l^3)\alpha_s^4 + O(\alpha_s^5)\;,
\label{eq:Romega}
\end{eqnarray}
Using 3-loop running coupling\index{running coupling} $\alpha_s$ and $l=\log(\mu^2/\omega^2)$
with  $\mu=1.5\max(\pi T,\omega)$.
Taking into account the leading effects of the temperature this perturbative vacuum behavior can be incorporate in the Ans\"{a}tz \cite{Burnier:2012ts} for the spectral function and one obtains
\begin{equation}
\rho_{ii}^{(T)}(\omega) = \frac{3\omega^2}{4\pi}\left[1 - 2n_F(\omega/2)\right]R(\omega^2)+\pi\chi_q^{\rm free}\omega\delta(\omega)\;.
\label{eq:rhoiiT}
\end{equation}
Using this as an Ans\"{a}tz for the high frequency part together with 
the Breit-Wigner description of the non-perturbative, low frequency 
part one gets an excellent description of the lattice data
\begin{equation}
\rho_R(\omega) = \rho_{\rm BW}(\omega) + \frac{3\omega^2}{4\pi}\left[1 - 2n_F(\omega/2)\right]R(\omega^2)\;,
\label{eq:rhoR}
\end{equation}
where
\begin{equation}
\rho_{\rm BW} = 2\pi\chi_q c_{\rm BW}\frac{\omega\Gamma/2}{\omega^2+(\Gamma/2)^2}\;.
\label{eq:rhoBQ}
\end{equation}

The fit strategy using these three Ans\"{a}tz for the spectral function and the results of the fits to the continuum extrapolated vector meson correlation functions are explained in \cite{Ding:2016hua} in more detail.
All three Ans\"{a}tz of the spectral functions reproduce the continuum extrapolated lattice data at the three temperatures $T=1.1, 1.3$ and $1.5 T_c$ with high accuracy and $\chi^2/\mathrm{dof}$ of order unity.
While all spectral functions converge at high frequencies, they differ substantially in the infrared region as can be seen in Fig.~\ref{fig13}.
This indicates the general difficulties in the extraction of transport coefficients from lattice QCD\index{QCD!lattice}. 
Comparing the three Ans\"{a}tz, we see that the area under each peak of the
spectral functions is very similar, but the sensitivity to the shape of the spectral in this region is limited. Nevertheless these results allow for an estimate of the electric conductivity and 
the discrepancy in the zero frequency limit acts as a measure of the systematic uncertainty in the results for the electrical conductivity, shown in Fig.~\ref{fig14}~(right), which is estimated by the respective minimum and maximum of the two Ans\"{a}tz $\rho_{ans}$ and $\rho_R$.

The resulting thermal dilepton rates, obtained from the spectral function
$\rho_{\text{R}}$ inserted in Eq.~(\ref{eq:LeptonRate})
are shown in Fig.~\ref{fig14}~(left) for all three temperatures 
and a sum of squared charges of $C_{em}=\sum_f q_f^2=5/9$, corresponding to 
two valence quark flavors $u$ and $d$. 
The results are qualitatively comparable to the rate obtained by an hard thermal loop\index{thermal!loop} (HTL) calculation
\cite{Braaten:1989mz} in the large frequency region, as well as to the 
leading order (Born) rate. However, compared to the HTL computation, 
the lattice results show an enhancement in the intermediate region $\omega/T\sim2$ and a qualitatively different behavior for small frequencies.

\begin{figure}[htbp]
\centering{
\includegraphics[width=0.49\textwidth]{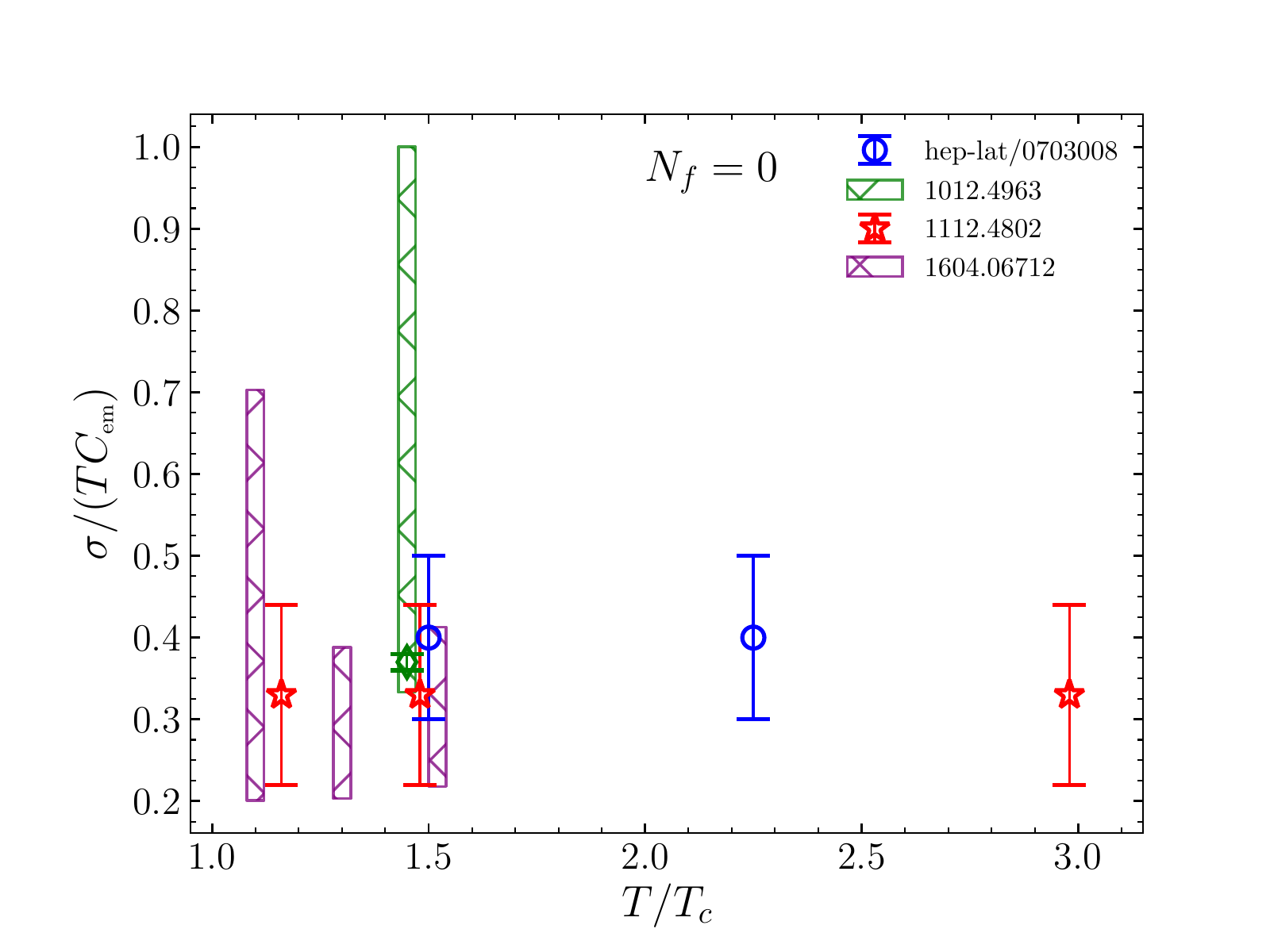}
\includegraphics[width=0.49\textwidth]{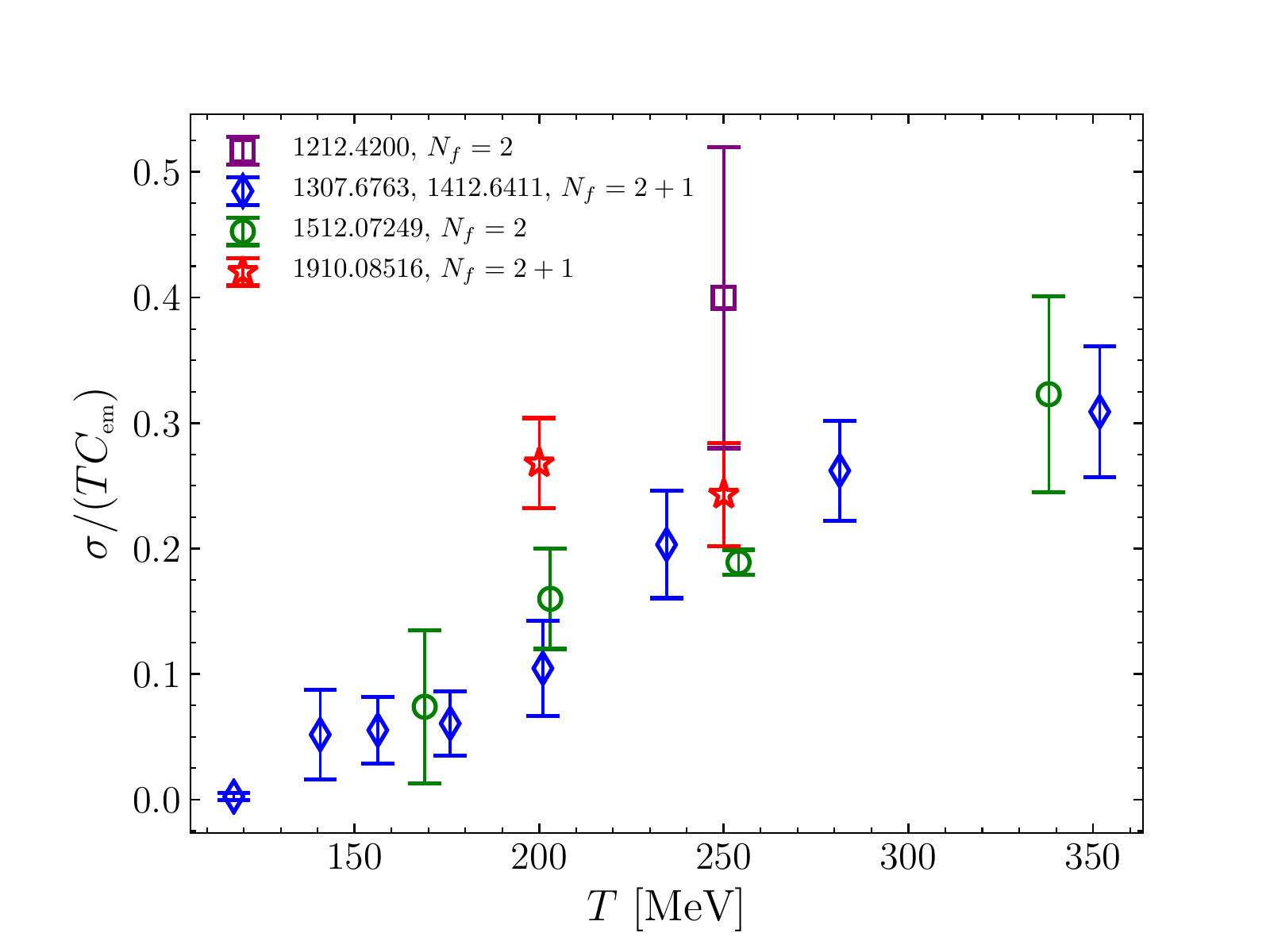}
}
\caption{
Results for the electrical conductivity, normalized as $\sigma/(TC_{\rm em})$, in quenched QCD ($N_f=0$) as a function of $T/T_c$ (left) and
in QCD with $N_f=2$ and $2+1$ dynamical flavours as a function of temperature in MeV (right). Taken from \cite{Aarts:2020dda}.
}
\label{eq:Aarts_conductivity}       
\end{figure}

Fig.~\ref{eq:Aarts_conductivity} shows a collection of available results for the electrical conductivity in quenched (left) and full QCD (right). More details and discussions on this topic can be found in the review \cite{Aarts:2020dda}. Although the full QCD results shown in the right plot are obtained on rather coarse lattices and not continuum extrapolated yet, a qualitative comparison to the quenched results indicates different behavior when approaching small temperatures. While the quenched results show no clear temperature dependence down to the critical temperature\index{critical!temperature}, the conductivity in full QCD shows a decreasing trend towards smaller temperatures. This may be attribute to the change of the transition from first order in quenched to a cross-over in full QCD and the system becoming more non-perturbative and more interacting when dynamical fermion degrees of freedom are included. Final conclusions on this remain for the future when calculations on larger and finer lattices become available allowing for continuum extrapolations also in full QCD.

Taking the limit of zero frequency in Eq.~(\ref{eq:PhotonRata}) allows to estimate
the soft photon rate which can be written in 
terms of the electrical conductivity,
\begin{eqnarray}
	\label{eqn_photonrate_conductivity}
	\lim_{\omega\to0}\omega\frac{\mathrm{d} \Gamma_{\gamma}}{\mathrm{d}^3\vec{k}}
	=\frac{\alpha_{em} C_{em}}{2\pi^2}\left(\frac{\sigma}{C_{em}T}\right)T^2.
\end{eqnarray}
In the following section we will discuss how to obtain the full thermal photon rate from vector meson correlators at non-zero momenta.


\section{Lattice estimate of thermal photon rate}

In the previous section we discussed how to obtain thermal dilepton rates and estimates on the electrical conductivity from continuum extrapolated vector meson correlation functions at zero momentum. We will now extend this discussion to finite momentum, $k$. Extracting the spectral function at finite $k$ allows to use Eq.~(\ref{eq:PhotonRata}) to calculate the thermal photon rate\index{thermal!photon rate}. 
We will follow here \cite{Ghiglieri:2016tvj} where the same lattice setup is used as the one in previous section. 
The calculation of the photon requires the extraction of the spectral function at the photon point ($\omega = \vert \vec{k} \vert$), meaning that the momentum dependence of the spectral function has to be carefully studied. Compared to the extraction of the electrical conductivity we are now interested in frequencies of the order of $k$. At larger $k$ it is expected that the spectral function becomes more perturbative and perturbatively constraining the spectral function in the large frequency region becomes more effective and provides a cleaner extraction of the spectral information.

\begin{figure}[t]
\centering{
\includegraphics[width=0.49\textwidth]{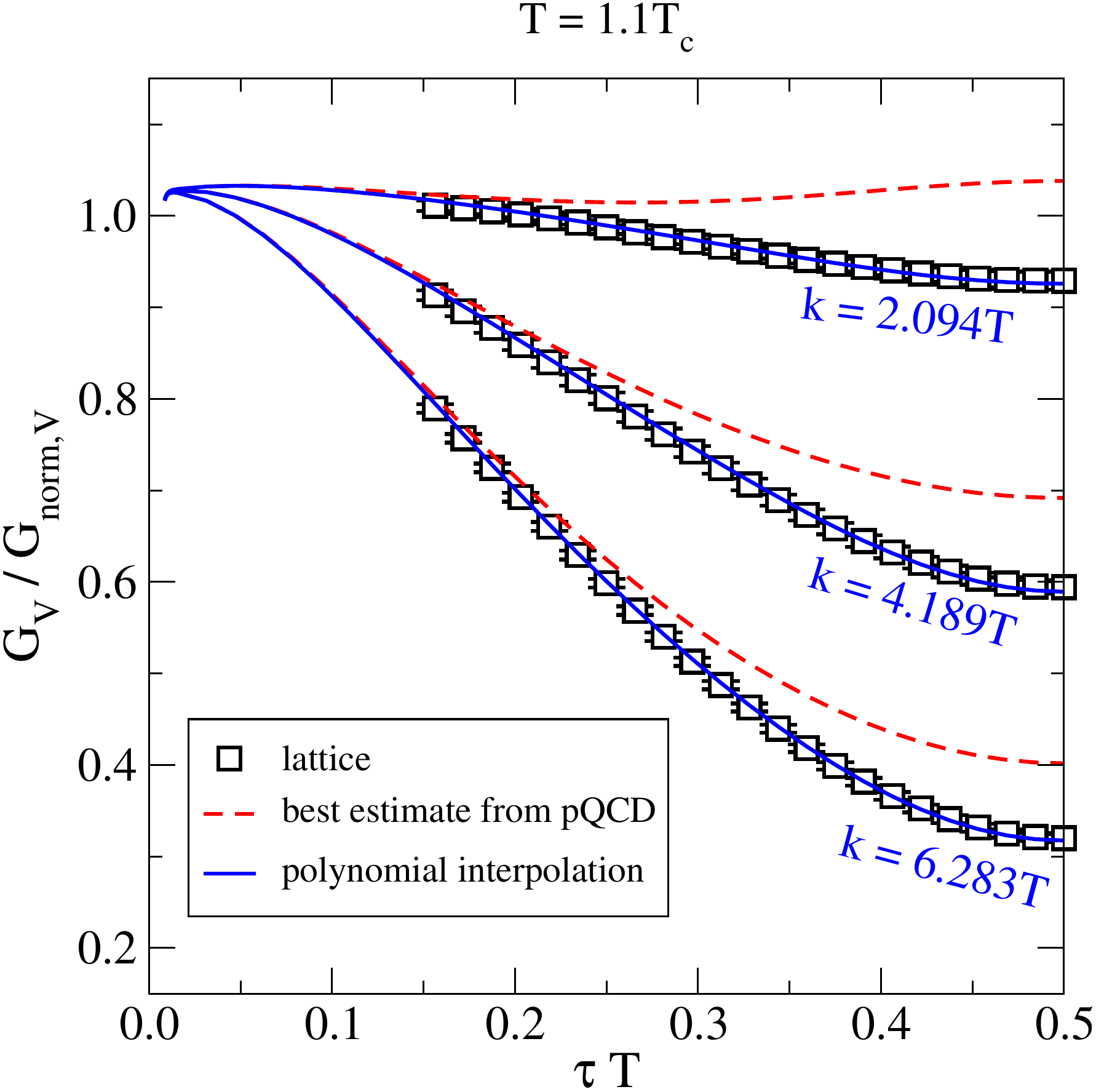}
\includegraphics[width=0.49\textwidth]{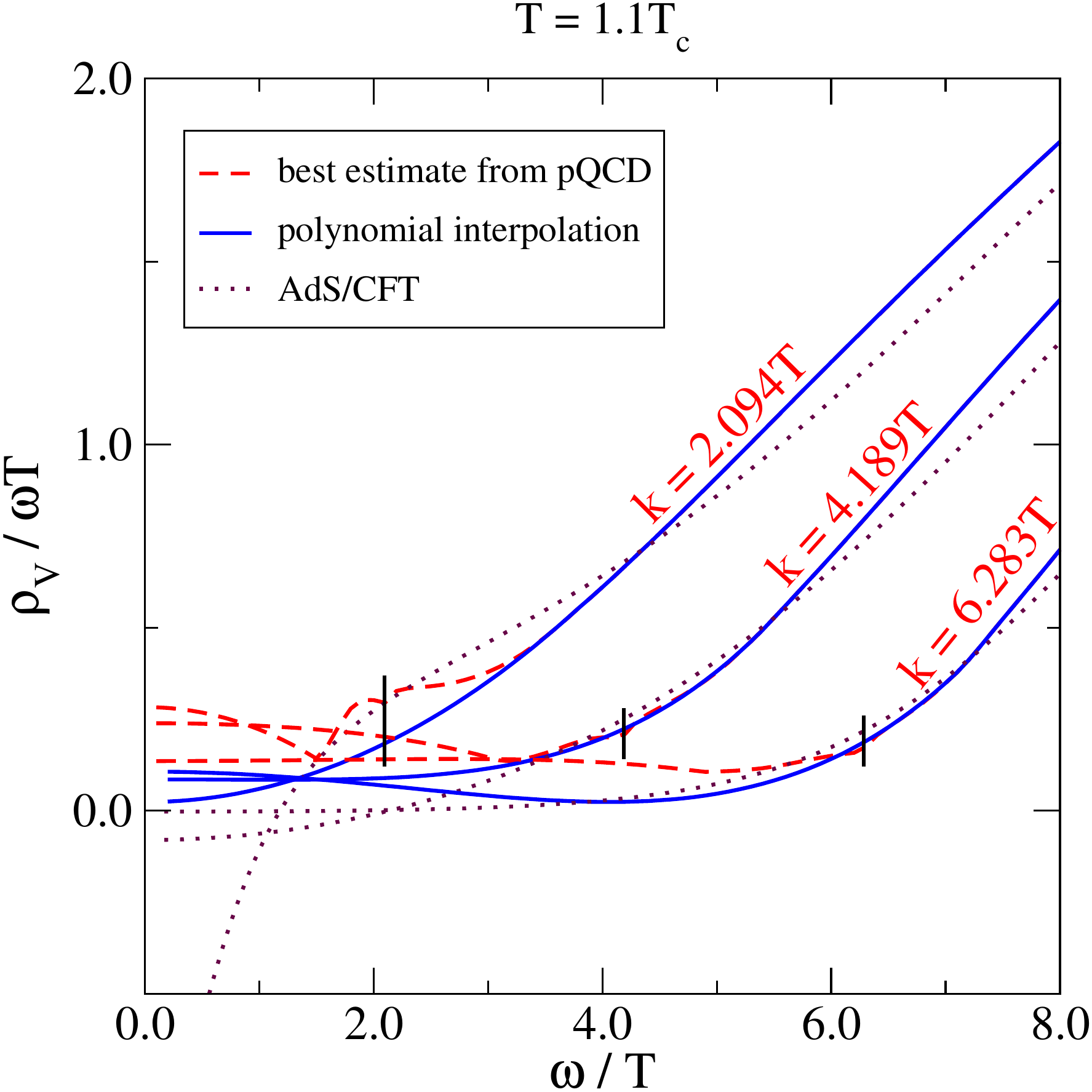}}
\caption{Vector correlators (left panel) obtained by the fit to (\ref{eq:rhoFit}) and resulting spectral functions (right panel) at finite momenta $(2.094T, ~ 4.189T, ~ 6.283T)$. Taken from \cite{Ghiglieri:2016tvj}.}
\label{fig15}       
\end{figure}

We start with a discussion about the perturbative knowledge on the behavior of the spectral function in different regimes.
In the high frequency limit where in the limit of (asymtocically) large invariant mass,  
$M \gg\pi T$,
with $M^2=\omega^2-k^2$
the non-interacting spectral function is known analytically 
\cite{Aarts:2005hg} as
\begin{equation}
\rho_V(\omega, \vec{k} ) = \frac{N_c T M^2 }{2\pi k} \left\{\log\left[\frac{\cosh(\frac{\omega+k}{4T})}{\cosh(\frac{\omega-k}{4T})}\right] - \frac{\omega\theta(k-\omega)}{2T} \right\}\;.
\label{eq:rhoV1}
\end{equation}
Perturbatively, at finite temperature a
logarithmically
enhanced term has been worked out analytically
\cite{Kapusta:1991qp,Baier:1991em}
for the photon rate,
\begin{equation}
\rho_V(k, \vec{k}) = \frac{\alpha_sN_cC_FT^2}{4}\log\left(\frac{1}{\alpha_s}\right)\left[1-2n_F(k)\right] + O(\alpha_s T^2)\;.
\label{eq:rhoV2}
\end{equation}
Non-logarithmic terms
are only known in numerical form~\cite{Arnold:2001ba,Arnold:2001ms}. Recently, 
these results have been extended to 
$
 {\mathcal{O}}(\alpha_{\rm s}^{3/2} T^2)
$ 
both at vanishing~\cite{Ghiglieri:2013gia} and
non-vanishing photon masses ($|M| \lesssim gT$, 
where $g \equiv \sqrt{4\pi\alpha_{\rm s}}$)~\cite{Ghiglieri:2014kma}, which we will use in the following for the large frequency part of the spectral function.

For $M \gg \pi T$, the spectral function goes over into a vacuum
result~\cite{CaronHuot:2009ns} which is known to relative accuracy 
${\mathcal{O}}(\alpha_{\rm s}^4)$~\cite{Baikov:2008jh,Baikov:2012zn} and can directly be taken 
over for a thermal analysis~\cite{Burnier:2012ts,Laine:2013vma}.
Such precisely determined 
results play an essential role in the investigation as they constrain the UV behavior and lead to more reliable extractions of the small and intermediate frequency parts.

\begin{figure}[thbp]
\centering{
\includegraphics[width=0.76\textwidth]{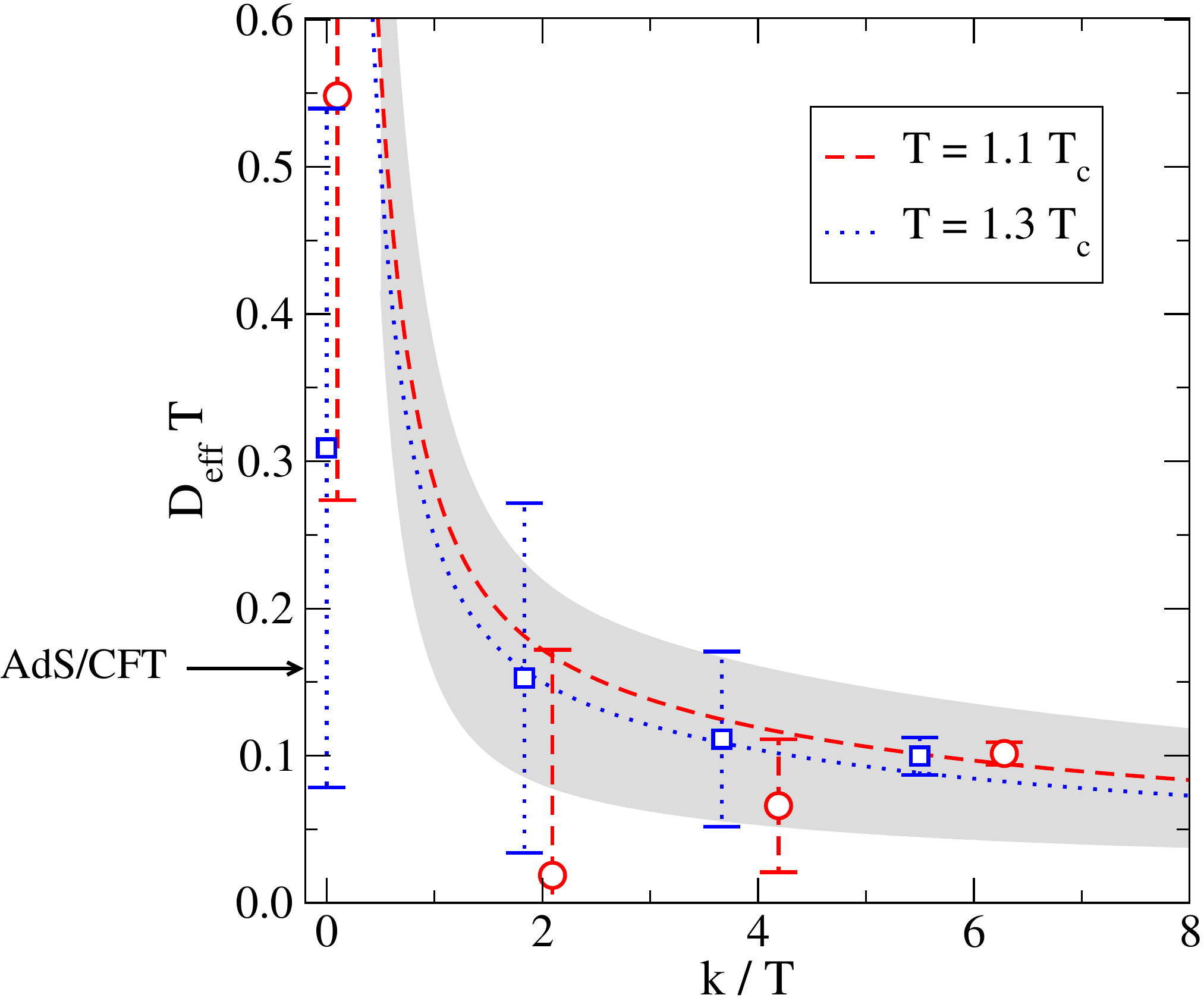}}
\caption{Lattice results for $D_{\rm eff}$ defined in (\ref{eq:Deff}) which can be used to calculate the thermal photon rate using (\ref{eq:dd32}). Also shown is the NLO perturbative prediction from \cite{Braaten:1989mz} and the AdS/CFT value of $DT=1/2\pi$ from \cite{Policastro:2002se} for $k=0$.
Taken from \cite{Ghiglieri:2016tvj}.}
\label{fig16}       
\end{figure}

In the low frequency limit the form of the spectral function can be deduced from the so-called hydrodynamic regime, where the general theory of statistical fluctuations applies and for
$k\lesssim\alpha_s^2 T$,
the spectral function has the form,
\begin{equation}
\frac{\rho_V(\omega,\vec{k})}{\omega} = \left(\frac{\omega^2-k^2}{\omega^2+D^2k^4}+2\right)\chi_q D\;,
\label{eq:rhoVomega}
\end{equation}
with the diffusion coefficient, 
\begin{equation}
D = \frac{1}{3\chi_q}\lim_{\omega\rightarrow0^+}\sum_{i=1}^3\frac{\rho^{ii}(\omega,\vec{0})}{\omega}\;,
\label{eq:D22}
\end{equation}
which is related to the electrical conductivity and the soft photon rates as already discussed in the previous section.

Knowing the form of the spectral function in the two regimes of low frequencies from hydrodynamics and and high frequencies from perturbation theory, one still has to model the spectral function in the intermediate frequency regime incorporating this knowledge.
In \cite{Ghiglieri:2016tvj} a high order polynomial Ans\"{a}tz,
\begin{equation}
\rho_{\rm fit}(\omega) = \frac{\beta\omega^3}{2\omega_0^3}\left(5 - \frac{3\omega^2}{\omega_0^2}\right) - \frac{\gamma\omega^3}{2\omega_0^2}\left(1-\frac{\omega^2}{\omega_0^2}\right)
+\sum_{n=1}^{n_{\rm max}}\frac{\delta_n\omega^{1+2n}}{\omega_0^{1+2n}}\left(1-\frac{\omega^2}{\omega_0^2}\right)\;,
\label{eq:rhoFit}
\end{equation}
was used with the constraint to smoothly match the perturbative result at $\omega=\omega_0$
\begin{equation}
\rho_V(\omega_0) = \beta, \hspace{40pt} \rho'_V(\omega_0) = \gamma\;, 
\label{eq:rhoV4}
\end{equation}
and $n_{\rm max}+1$ free parameters starting with a linear 
behavior for $\omega\ll T$.
The matching frequency is defined as $\omega_0\simeq\sqrt{k^2+(\pi T)^2}$
and $n_{\rm max}=0$ and $n_{\rm max}=1$ was used.
More details on the Ans\"{a}tz and the fitting strategy can be found in 
\cite{Ghiglieri:2016tvj}.

Results from this procedure are shown in Fig.~\ref{fig15} for a temperature of $1.1~T_c$ and three momenta $k$ in comparison to the best estimate from perturbation theory. 
While at $\tau \rightarrow 0$, all correlators approach the perturbative result, at larger values of $\tau$ non-perturbative effects are clearly visible.
This deviations become smaller for larger values of $k$, i.e. the lattice results approach the perturbative behavior for increasing momenta.


\begin{figure}[t]
\centering{
\includegraphics[width=0.76\textwidth]{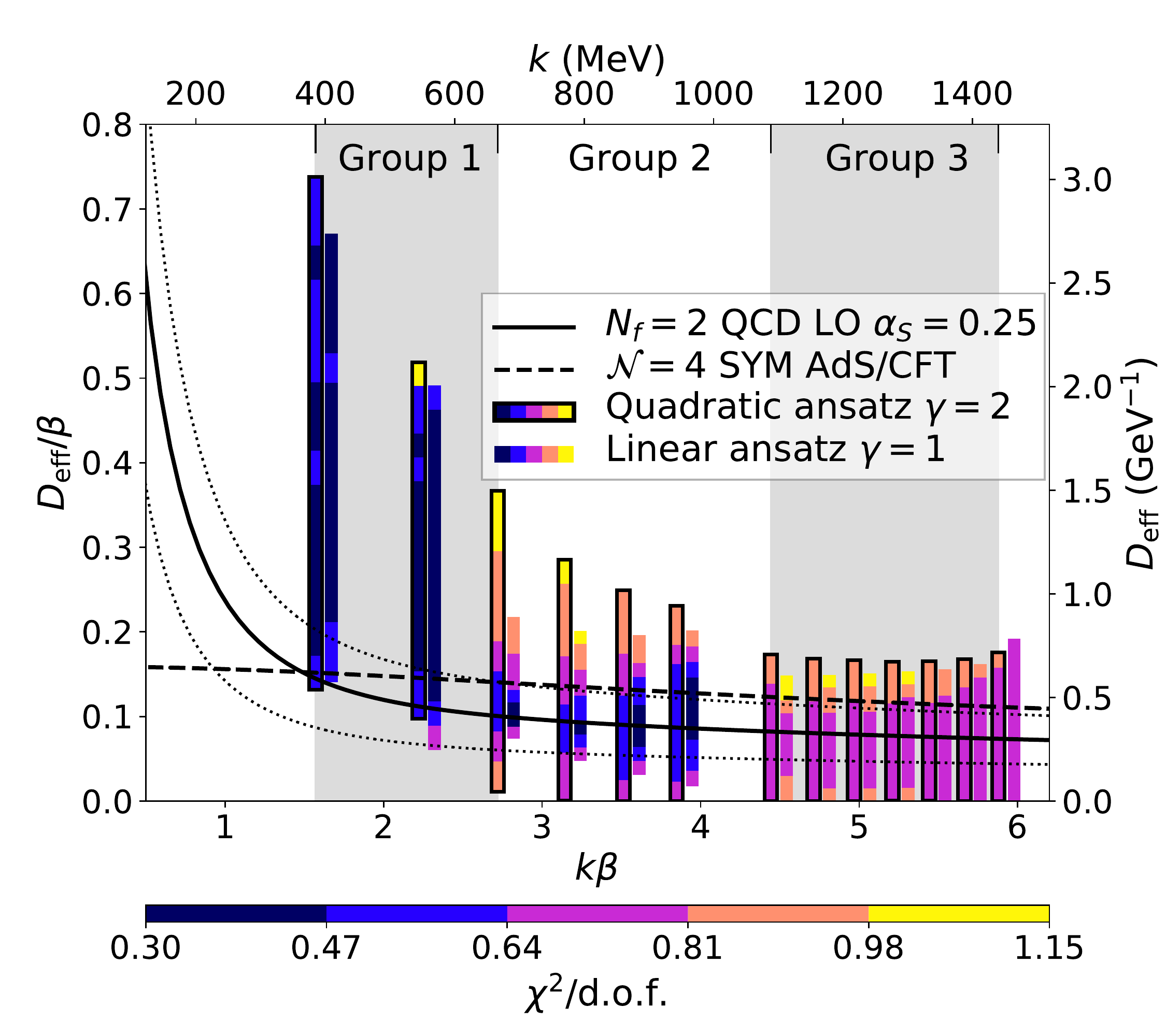}
}
\caption{Lattice result for the effective diffusion coefficient $D_{\rm eff}$ from the 2-flavor lattice QCD calculation from \cite{Ce:2020tmx}.}
\label{figMeyer}       
\end{figure}


The spectral function at the photon point $\omega=k$ 
\begin{equation}
D_{\rm eff}(k) =
\begin{cases}
    \frac{\rho_V(k,\vec{k})}{2\chi_q k},& \text{for}   \ \ k> 0\\
    \lim_{\omega\rightarrow0^+}\frac{\rho_{ii}(\omega,\vec{0})}{3\chi_q\omega}, & \text{for} \ \  k=0      
\end{cases}
\label{eq:Deff}
\end{equation}
can be used to compute the thermal photon rate\index{thermal!photon rate}
\begin{equation}
\frac{d\Gamma_\gamma(k)}{d^3 \vec{k} } = \frac{2\alpha_{em}\chi_q}{3\pi^2}n_B(k)D_{\rm eff}(k) + O(\alpha_{em}^2)\;.
\label{eq:dd32}
\end{equation}
While this equation is perturbative with respect to electromagnetic
interactions, the function $D_{\rm eff}$ contains all non-perturbative
information from strong interactions. 
The results are shown in Fig.~\ref{fig16}.
The results approach the NLO 
perturbative prediction (valid for $k\gg gT$) for large momenta, while for momenta $k/T<3$ 
non-perturbative effects become visible. 
The electrical conductivity could also be obtained in the $k\rightarrow0$ limit, but the available momenta are not small enough for such a determination from $D_{\rm eff}$.
The value shown in the plot at $k=0$ was calculated from the result
obtained at zero $k$ from \cite{Ding:2016hua},
see previous section. The AdS/CFT value is $DT=1/(2\pi)$.

The results of this study show the potential for obtaining results for the photon rates obtained from continuum extrapolated lattice QCD correlation functions\index{QCD!correlation functions} incorporating perturbative knowledge in the spectral reconstruction. Obtaining results at even smaller momenta would be very interesting for future studies.
In a finite-size box momenta are given by $k = 2 \pi n /(a N_s)$, where $a$ is the lattice spacing
and $n$ is an integer and $a N_\tau = 1/T$, i.e. in units of temperature,
\begin{equation}
 k/T = 2\pi n \times \frac{N_\tau}{N_s}
 \;,
\end{equation}
where $N_\tau$ and $N_s$ are the temporal and spatial lattice extents, respectively. Therefore going to small momenta requires to do the calculations at larger momenta and therefore become more expensive in terms of numerical calculations. 

First results on the photon rate obtained in full QCD using a different method to extract the photon rates can be found in \cite{Ce:2020tmx} and are shown in Fig.~\ref{figMeyer}.
This study is based on the difference between the spatially
transverse and longitudinal parts of the vector meson spectral function, which has the advantage that UV contributions are exponentially suppressed. 

In addition to improving the lattice QCD results, e.g. obtaining continuum results also in full QCD including realistic dynamical fermion degrees of freedom, 
combining different methods for the spectral reconstruction, including perturbative models, special combinations of different operators, reconstruction methods discussed in Sec.~\ref{Sec:inverse} or machine learning techniques may improve the determination of spectral properties from lattice QCD correlation functions\index{QCD!correlation functions} in the future. 

\section{Charmonia and bottomonia spectral function from lattice QCD}\label{sec:charm}

So far we have discussed the light quark correlation and spectral functions and related observables.
In the following sections we will discuss meson correlation functions in the heavy flavor sector. We focus on charmonium and bottomonium in the vector channel, namely $J/\psi$ and $\Upsilon$. We perform lattice calculations on very large and fine lattices so that charm and bottom quarks can be treated relativistically. The results given below are based on \cite{Ding:2018uhl}. As preparation we first give a brief introduction to free spectral functions that will be used in the spectral analysis.

\subsection{Free spectral function}
\label{free-spf-sec}

\begin{figure}[tbh]
\centering{
\includegraphics[width=0.98\textwidth]{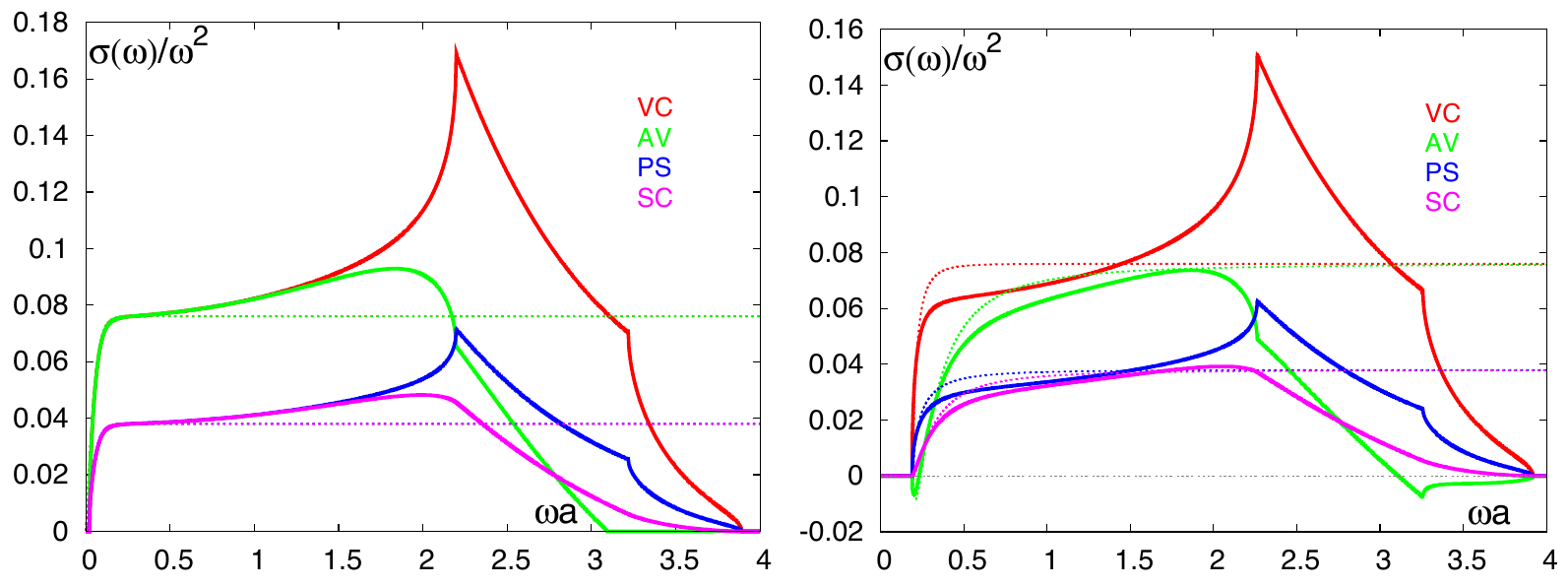}}
\caption{Free meson spectral functions in the continuum (\textit{dotted}) and on the lattice (\textit{solid}) in different channels calculated according to \cite{Aarts:2005hg} taken from \cite{Ding:2010smi}. In left panel the quark mass is $am=0$ and in the right panel the quark mass $am=0.01$. $N_{\tau}=64$ in both panels.}
\label{figfreespf}      
\end{figure}

The non-interacting spectral functions for massive free quarks, including the one in the continuum and on the lattice, can be analytically calculated in the high temperature limit \cite{Karsch:2003wy,Aarts:2005hg}. At zero momentum, the continuum free meson spectral function reads
\begin{align}
 \rho_H(\omega) =&
 \Theta(\omega^2-4m^2)\frac{N_c}{8\pi \omega}
 \sqrt{\omega^2-4m^2}\left[1-2n_F(\omega/2)\right] 
 \Big[ \omega^2\left( a_H^{(1)} -a_H^{(2)} \right) \nonumber \\
&+4m^2\left( a_H^{(2)} -a_H^{(3)} \right) \Big]+2\pi \omega\delta(\omega) N_c
\Big[ \left(a_H^{(1)} + a_H^{(2)}\right) I_1 \nonumber \\
&+\left( a_H^{(2)} -a_H^{(3)}\right) I_2 \Big],
\label{freespf_cont}
\end{align}
with
\begin{align}
I_1 = -2\int_{\mathbf k} n'_F(\omega_{\mathbf k}), 
\;\;\;\;\;\;
I_2 = -2\int_{\mathbf k} \frac{k^2}{\omega_{\mathbf k}^2 }n'_F(\omega_{\mathbf k}).
\end{align}
Here $n_F$ is the fermi distribution and $n'_F$ the partial derivative with respect to $\omega_{{\mathbf k}}$. The coefficients $a^{(1)}_H$, $a^{(2)}_H$ and $a^{(3)}_H$ can be found in Tab.\ref{tablecoeff}. We could see that in the pseudo-scalar, the $\delta$ function sitting at zero frequency vanishes because of the coefficients. So there will be no transport peak in this channel. While for vector channel it does not. 
\begin{table}[h]
\begin{center}
\begin{tabular}{|c|c|c|c|c|c|}
\hline
& $\Gamma_H$    & $a_H^{(1)}$ & $a_H^{(2)}$ & $a_H^{(3)}$  \\
\hline
$\rho_{\rm S}$  &${1\!\!1}$      & $1$   & $-1$  & $1$   \\
$\rho_{\rm PS}$ & $\gamma_5$    & $1$   & $-1$  & $-1$  \\
\hline
$\rho^{00}$ &$\gamma^0$ & $1$   & $1$   & $1$   \\
$\rho^{ii}$ &$\gamma^i$ & $3$   & $-1$  & $-3$  \\
$\rho_{\rm V}$  &$\gamma^\mu$   & $2$   & $-2$  & $-4$  \\
\hline
$\rho^{00}_5$   &$\gamma^0\gamma_5$ & $1$   & $1$   & $-1$  \\
$\rho^{ii}_5$   &$\gamma^i\gamma_5$ & $3$   & $-1$  & $3$   \\
$\rho_{\rm A}$  &$\gamma^\mu\gamma_5$   & $2$   & $-2$  & $4$   \\
\hline
\end{tabular}
 \caption{Coefficients $a_H^{(i)}$ for free spectral functions in 
 different channels $H$ taken from \cite{Aarts:2005hg}.
}
\label{tablecoeff}
\end{center}
\end{table}

The free spectral function can also be calculated on the lattice using the lattice propagtors in momentum space as a sum over lattice momenta. For the Wilson fermion discretization of the fermionic action, it is given by

\begin{eqnarray}
&&\hspace*{-0.6cm}
\rho^{\rm Wilson}_H(P) = \frac{4\pi N_c}{L^3}\sum_{\mathbf k} \sinh\left(\frac{\omega}{2T}\right) 
\bigg\{\nonumber \\ 
&&\hspace*{-0.6cm} \;\;\;\;\;\;\;\;
\bigg[ 
a_H^{(1)} S_4({\mathbf k})S^\dagger_4({\mathbf r}) 
+ a_H^{(2)} \sum_i S_i({\mathbf k})S^\dagger_i({\mathbf r})
+ a_H^{(3)} S_u({\mathbf k})S^\dagger_u({\mathbf r}) \bigg] 
\delta(\omega +E_{\mathbf k}-E_{\mathbf r})
\nonumber \\&&\hspace*{-0.6cm} \;\;\;\;\;\;\;\;
+ \bigg[
a_H^{(1)} S_4({\mathbf k})S^\dagger_4({\mathbf r}) 
- a_H^{(2)} \sum_i S_i({\mathbf k})S^\dagger_i({\mathbf r})
- a_H^{(3)} S_u({\mathbf k})S^\dagger_u({\mathbf r}) \bigg]
\delta(\omega -E_{\mathbf k}-E_{\mathbf r})
\nonumber \\ &&\hspace*{-0.6cm} \;\;\;\;\;\;\;\;
+ (\omega\to-\omega)
\bigg\}.
\label{freespfwilson}
\end{eqnarray}
Here $L=aN_{\sigma}$ is the spatial extent of the lattice. And the lattice propagators are given by
\begin{eqnarray}
\nonumber
&S_4({\mathbf k}) =\frac{\sinh \left(E_{\mathbf k}/\xi\right)}{2{\cal E}_{\mathbf k}\cosh(E_{\mathbf k}/2T)}, \nonumber \\
&S_i({\mathbf k}) =\frac{1}{\xi} \frac{i\sin k_i}{2{\cal E}_{\mathbf k}\cosh(E_{\mathbf k}/2T)}, \nonumber \\
&S_u({\mathbf k}) = -\frac{1-\cosh \left(E_{\mathbf k}/\xi\right) +{\cal M}_{\mathbf k}}{2{\cal E}_{\mathbf k}\cosh(E_{\mathbf k}/2T)},
\label{eqsss}
\end{eqnarray}
where $\epsilon=a/a_{\tau}$ is the anisotropy parameter, i.e. allows for different lattice spacings in temporal and spatial direction and $\epsilon=1$ for an isotropic lattice. Relevant quantities are defined as
\begin{eqnarray}
&{\cal K}_{\mathbf k} = \frac{1}{\xi}\sum_{i=1}^3 \gamma_i\sin k_i,\nonumber \\
&{\cal M}_{\mathbf k} = \frac{1}{\xi}\left[ r \sum_{i=1}^3 \left(1-\cos
k_i\right) + m\right]\nonumber \\
&{\cal E}_{\mathbf k} = (1+{\cal M}_{\mathbf k}) \sinh \left(E_{\mathbf k}/\xi\right),
\end{eqnarray}
and the energy of a single partical $E_{{\mathbf k}}$ is determined by the following equation
\begin{equation}
\cosh \left(E_{\mathbf k}/\xi\right) = \sqrt{ 1 + {\cal K}_{\mathbf k}^2+(m/\xi)^2}.
\end{equation}

\begin{figure}[thbp]
    \centering
        \includegraphics[clip, trim=4cm 11cm 2.0cm 5.5cm, width=0.98\textwidth]{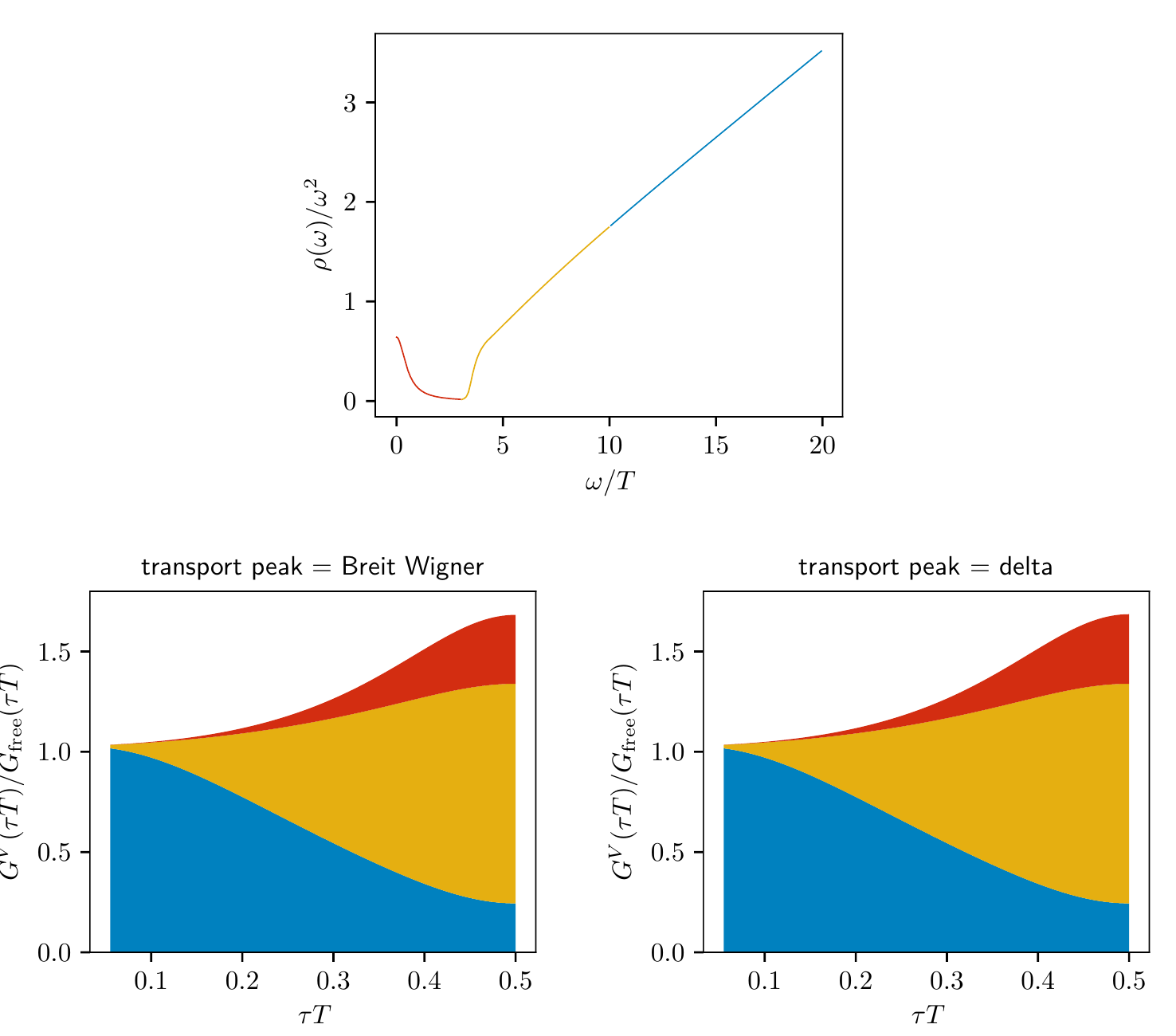}
    \caption{Upper: a sketch showing different parts of the spectral function. Lower: the contributions of different parts of the spectral function to the correlators. All three figures here are taken from \cite{Sandmeyer:2019gol}.}
    \label{fig:differentparts_spf_corr}
\end{figure}

Fig.~\ref{figfreespf} shows the free particle spectral function in the continuum and on the lattice (Wilson fermion discretization) calculated according to Eq.~(\ref{freespf_cont}) and Eq.~(\ref{freespfwilson}). The lattice spectral functions approach the continuum one at small frequencies but differ at large frequencies and even vanish above a cut-off around $a\omega \simeq 4$, which is dictated by the lattice spacing. The peaks and cusps in the lattice spectral functions correspond to effect of the so called Wilson doublers. If one would scale the x-axis to physical units or units of temperature, i.e. $\omega/T=a\omega N_{\tau}$, it becomes clear that these lattice discretization and cut-off effects move to larger frequencies with increasing $N_{\tau}$ and disappear in the $N_{\tau}\rightarrow\infty$ limit and the lattice spectral functions approach their continuum values in this limit.

In the interacting theory these discretization effects are also present and can lead to cut-off effects in the spectral function in the physically interesting frequency regions when the lattice spacings are not small enough, especially for heavy quarks, e.g. for a typical inverse lattice spacing of $a^{-1}=3.6$~GeV corresponding to a lattice spacing of $a=0.055$~fm the peak in the lattice spectral function will be of the order of the bottomonium ground state and would influence any studies of this state. Note that according to Eq.~(\ref{freespfwilson}) these discretization effects stem from finite spacial lattice spacing effects. Therefore on anisotropic lattices, where the temporal extend is chosen to be smaller as the spatial one, e.g. to improve the spectral reconstruction at fixed temperatures using larger $N_{\tau}$, these effects are given by the coarser spatial lattice spacing.

\begin{figure}[thbp]
    \centering
        \includegraphics[ width=0.65\textwidth]{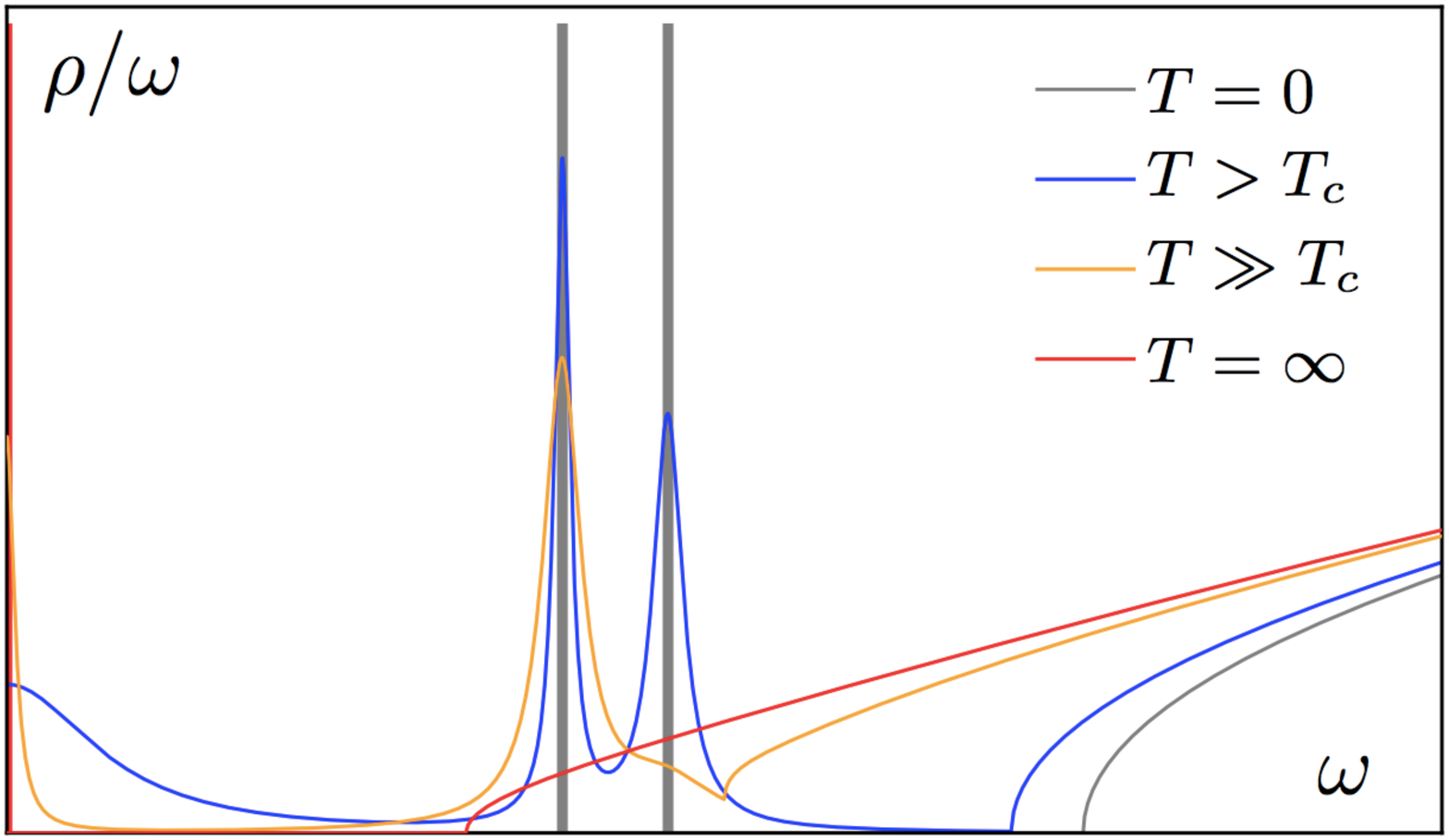}
    \caption{A sketch of the spectral function in the vector channel at different temperature~\cite{Sandmeyer:2019gol}.}
    \label{fig:vecotr_spf_full}
\end{figure}

Next we will consider the small frequency part of the vector meson spectral function, i.e. the transport peak. In the non-interacting limit (infinite temperature limit), in the zero frequency limit a $\delta$-function appears, \cite{Karsch:2003wy,Aarts:2005hg}
\begin{equation}
\rho_{ii}(\omega \ll T)\sim2\chi_{00}\frac{T}{M}\omega\delta(\omega).
\label{eq:TransportRhodelta}
\end{equation}
When the interaction is turned on, this $\delta$ peak in the spatially polarized vector channel gets smeared into a Breit-Wigner distribution \cite{Petreczky:2005nh}
\begin{equation}
\rho_{ii}(\omega \ll T)\sim2\chi_{00}\frac{T}{M}\frac{\omega\eta}{\omega^2+\eta^2}\; ,
\label{eq:TransportRho}
\end{equation}
with $\eta=T/MD$ with the quark mass $M$ and diffusion coefficient $D$. Note that the temporal vector channel, $\rho_{00}$ is protected by the conservation of particle number to remain a delta function at zero momentum.
For the moment we ignore possible bound states or resonance peaks in the intermediate frequency range and construct a spectral function with only a transport peak and a free continuum spectral function. 
This would correspond to a system at very high temperatures where any bound states are already melted.
An example of such a spectral function is shown in the top panel of Fig.~\ref{fig:differentparts_spf_corr}. 
In the lower plots 
the contributions of the spectral function, including a transport peak (left) and delta peak (right), to the correlators according to Eq.~(\ref{eq:IntRelation}) in parts is shown. Here one can observe that the correlation function at different distance $\tau T$ are sensitive to different parts of the spectral function due to the existence of the kernel function Eq.~(\ref{eq:Kernel}). At small $\tau T$ the correlator is dominated by the high frequency part of the spectral function while the transport region mainly contributes at large $\tau T$.

At smaller temperatures additional bound state or resonance contributions will appear. At zero temperature bound states, i.e. ground and excited states for charmonium and bottomonium, are known and can be well approximated as series of $\delta$-like functions. At finite temperature, these resonance peaks will probably broaden and their peak locations may change. An important question is at which temperatures which states dissociate and 
if at sufficient high temperatures all bound states are melt and merge into the threshold to a continuum region in the spectral function. Fig.~\ref{fig:vecotr_spf_full} shows a sketch of the spectral function in the vector channel at different temperatures.
From the previous discussions we already learned that disentangling all these contributions to the spectral function is difficult as they contribute to the correlation function at all distances and a spectral reconstruction is required. Therefore it is important to remove the lattice discretization effects by performing the continuum limit of the correlation functions and implement as much prior information as possible into model Ans\"{a}tz for the spectral function, e.g. from perturbation theory in the high frequency part. We will follow this strategy in the next sections for charmonium and bottomonium correlation and spectral functions, first in the pseudo-scalar channel where not transport contribution appears and then in the vector channel including a discussion on the transport contribution.


\subsection{Spectral functions of charmonia and bottomonia in the pseudo-scalar channel}
\label{sec-ps}
In this section we will discuss the determination of 
charmonium and bottomonium spectral functions in the pseudo-scalar channel 
based on comparing continuum extrapolated lattice correlation functions to
perturbative spectral function, where resummed thermal effects around the threshold and vacuum asymptotics above the threshold are incorporated \cite{Burnier:2017bod}. The pseudo-scalar channel has the feature that the transport peak is absent, making the spectral analysis much easier compared to the vector channel.

The strategy is to fit a perturbatively inspired model to the lattice data. The model spectral function consists two parts: the thermal contributions around the threshold and the vacuum contribution above the threshold. The former can be obtained by pNRQCD calculations \cite{Laine:2007gj}:
 \begin{equation}
 \rho^\mathrm{NRQCD}_\mathrm{P} = 
 \frac{M^2}{3}
 \rho^\mathrm{NRQCD}_\mathrm{V} 
 \;,
\end{equation}
where
 \begin{align}
\rho^{\text{pNRQCD}}_V(\omega)=\frac{1}{2}\left( 1-e^{-\frac{\omega}{T}} \right) \int\limits_{-\infty}^{\infty}\text{d}t\ e^{i\omega T}C_>(t,\vec{0},\vec{0}).
\end{align}
 $C_>$ is a Wightman function solvable for a real-time static potential from hard thermal loop\index{thermal!loop} resummation. And at frequencies well above the threshold, ultraviolet asymptotics says that the spectral functions are
\begin{equation}
 \left. \frac{ \rho^{ }_\mathrm{P}(\omega) }
 {\omega^2 m^2(\bar{\mu})} \right|^\mathrm{vac}
 \; \equiv \;
 \frac{N_c}{8\pi}\, \tilde{R}_c^{p} (\omega,\bar{\mu}),
 \label{tilde_def_v}
\end{equation}
where $\tilde{R}_c^{p} (\omega,\bar{\mu})$ and the running mass $m^2(\bar{\mu})$ with scale $\bar{\mu}$ can be found in \cite{Burnier:2017bod}. These two parts are combined by matching $\rho^{\text{pNRQCD}}_V$ smoothly to the vacuum asymptotics. More discussion on the resulting perturbative spectral function, $\rho_\mathrm{P}^\mathrm{pert}$ can be found in \cite{Burnier:2017bod}.

With this perturbative knowledge we can form a model spectral function,
\begin{equation}
 \rho_\mathrm{P}^\mathrm{model}(\omega) \; \equiv \;
 A \, \rho_\mathrm{P}^\mathrm{pert}(\omega - B)
 \; ,
 \label{model-ps}
\end{equation}
that contains two fit parameters $A$ and $B$. $A$ accounts for the uncertainties in the renormalization and $B$ for the uncertainties in the determination of the pole mass.

\begin{figure}[htbp]
    \centering{
     \includegraphics[width=0.48\textwidth]{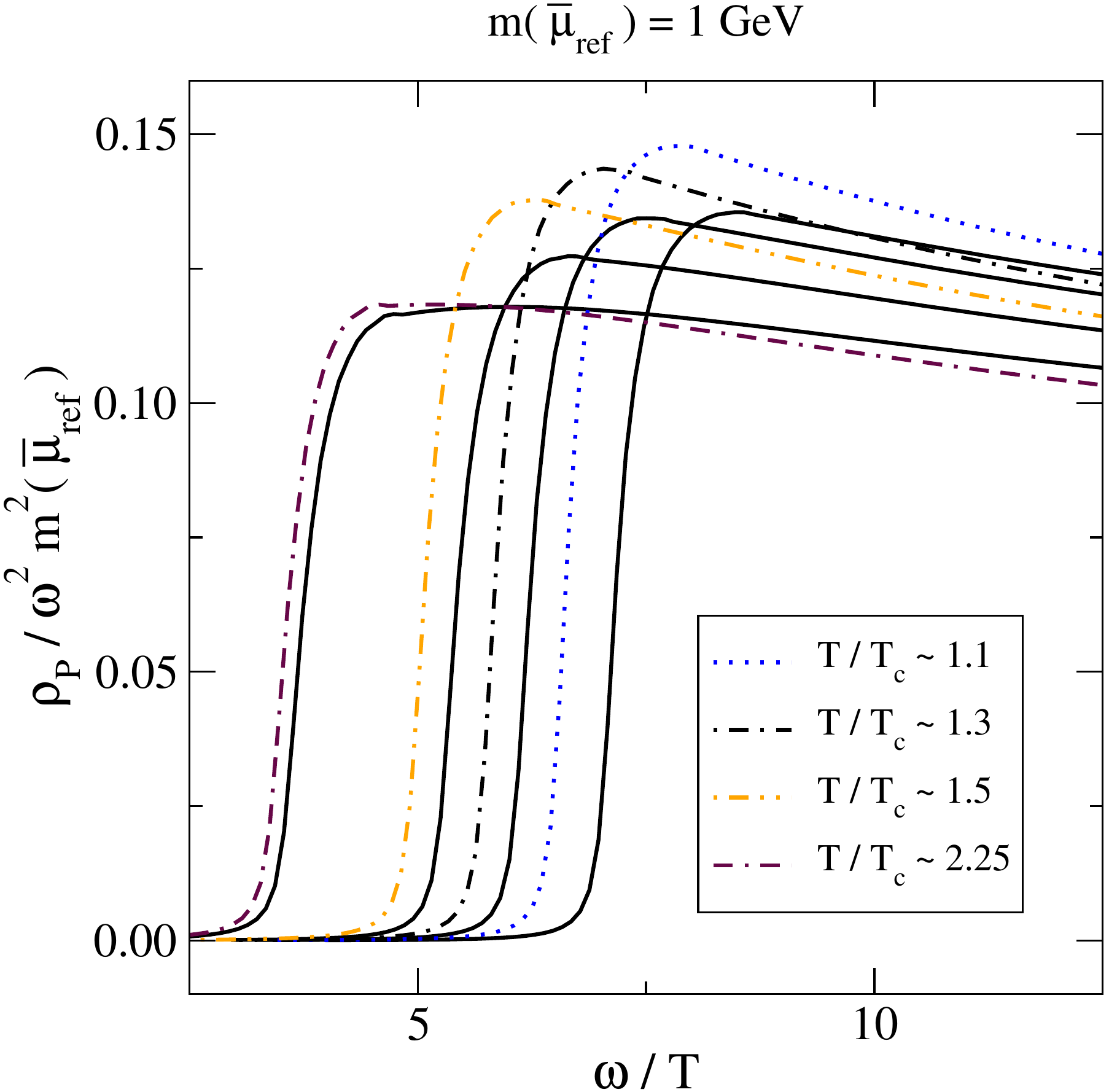}\includegraphics[width=0.48\textwidth]{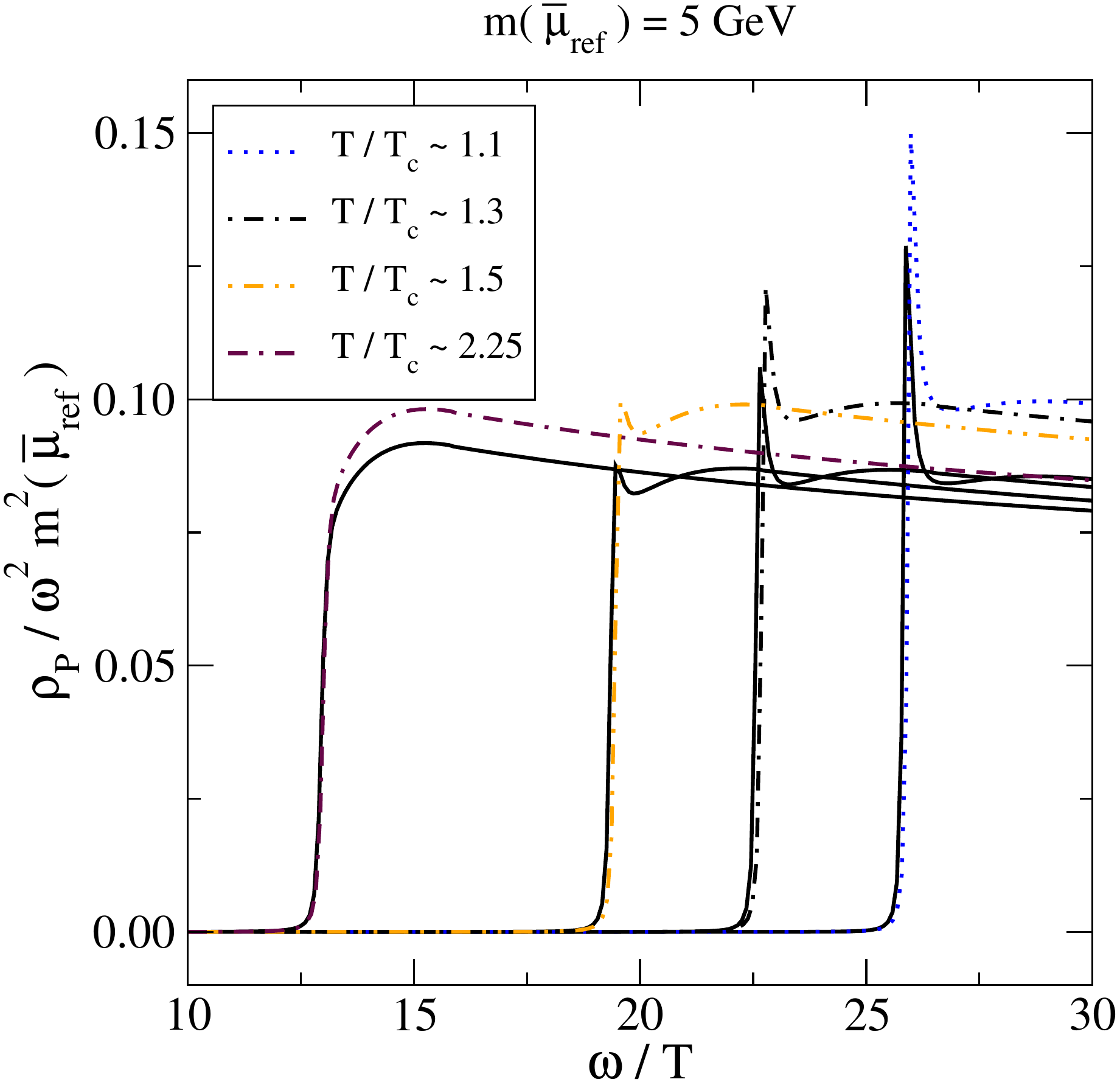}}
 \caption{A comparison of the perturbative pseudo-scalar spectral functions (dash-dotted curves) and their modifications (solid curves) for chamonium (left) and bottomonium (right) \cite{Burnier:2017bod}.}
 \label{spf-ps-comp-cb}    
\end{figure}

\begin{figure}[thbp]
    \centering{
     \includegraphics[width=0.48\textwidth]{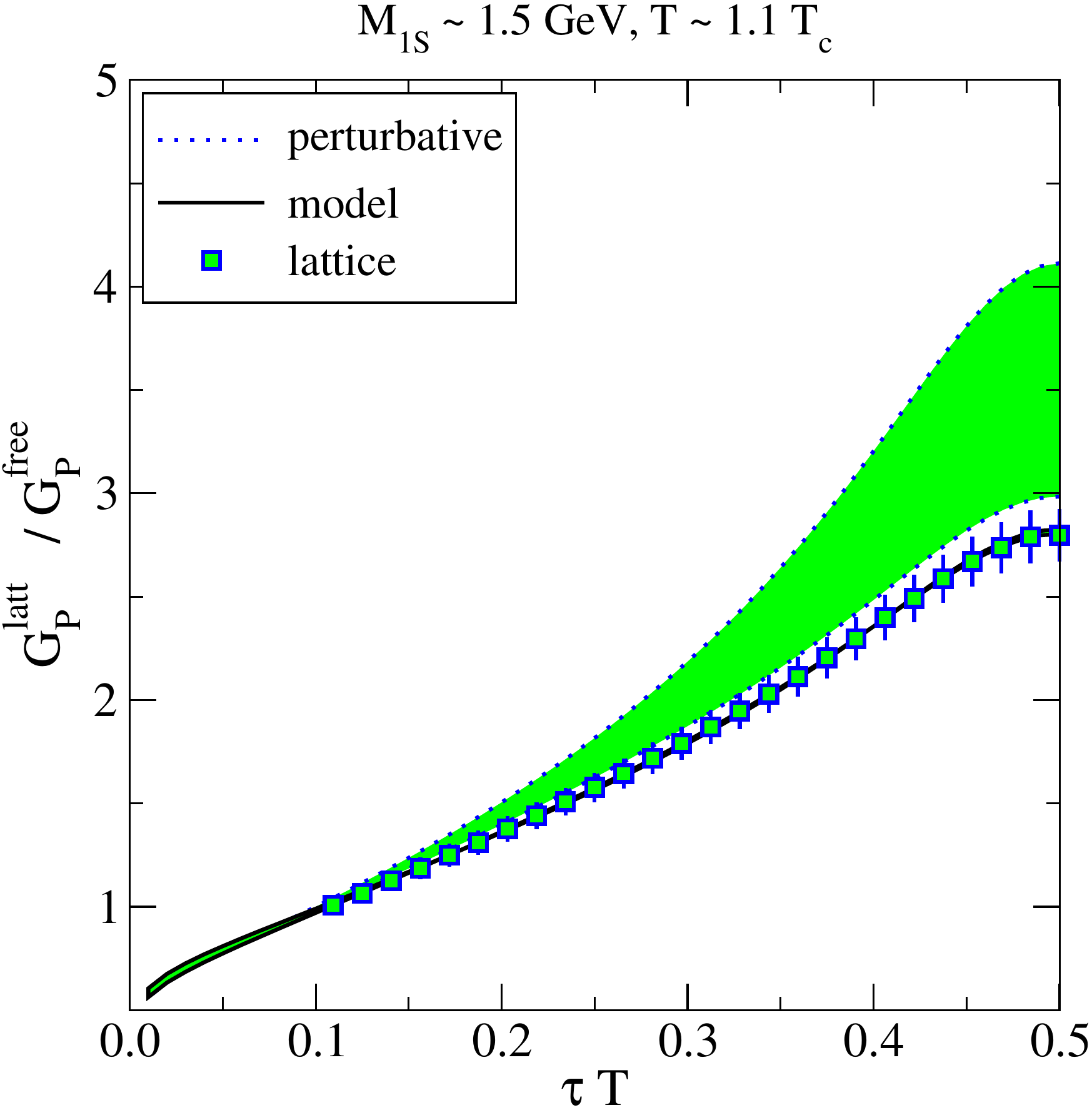}\includegraphics[width=0.48\textwidth]{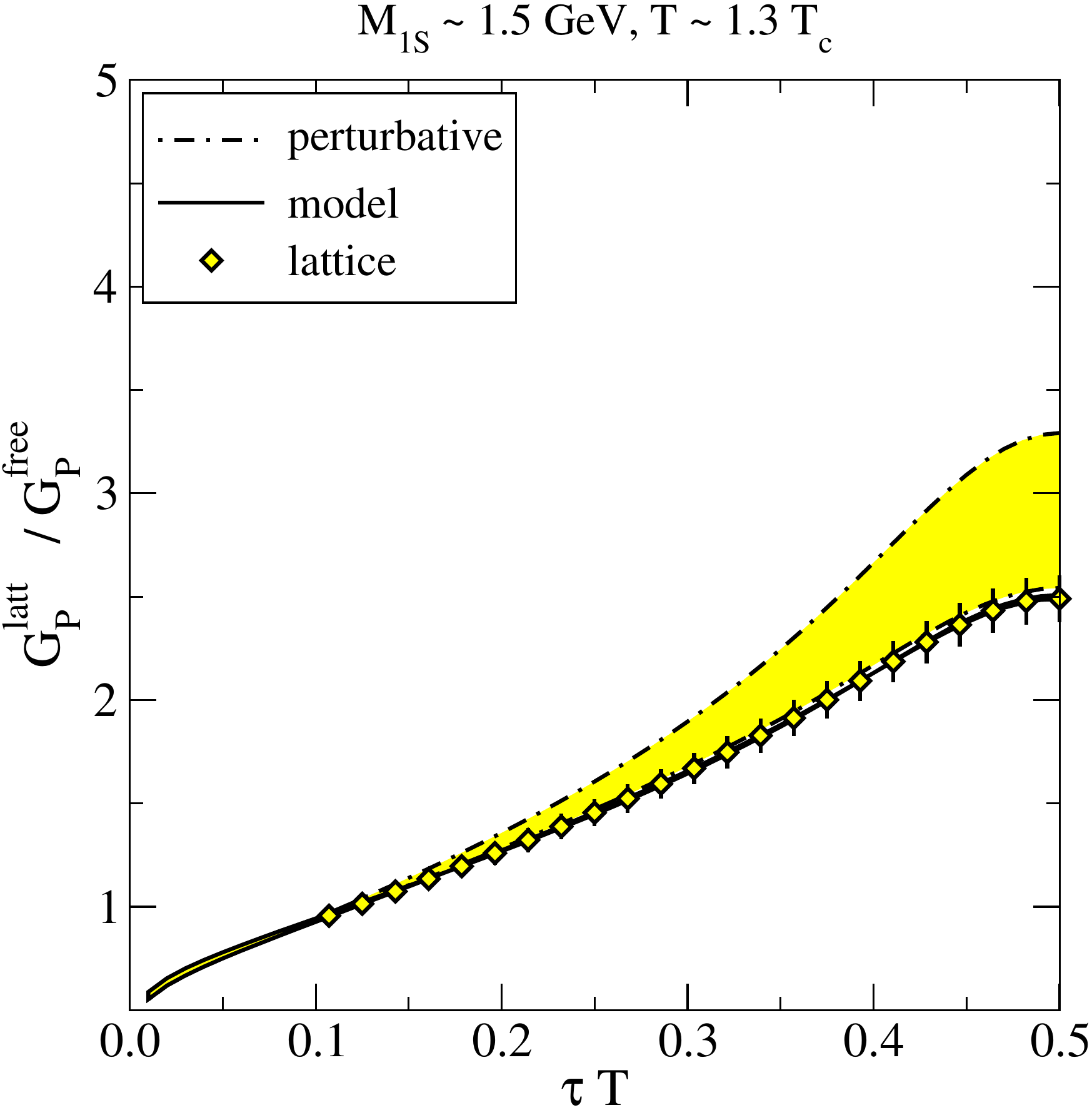}}
         \centering{
     \includegraphics[width=0.48\textwidth]{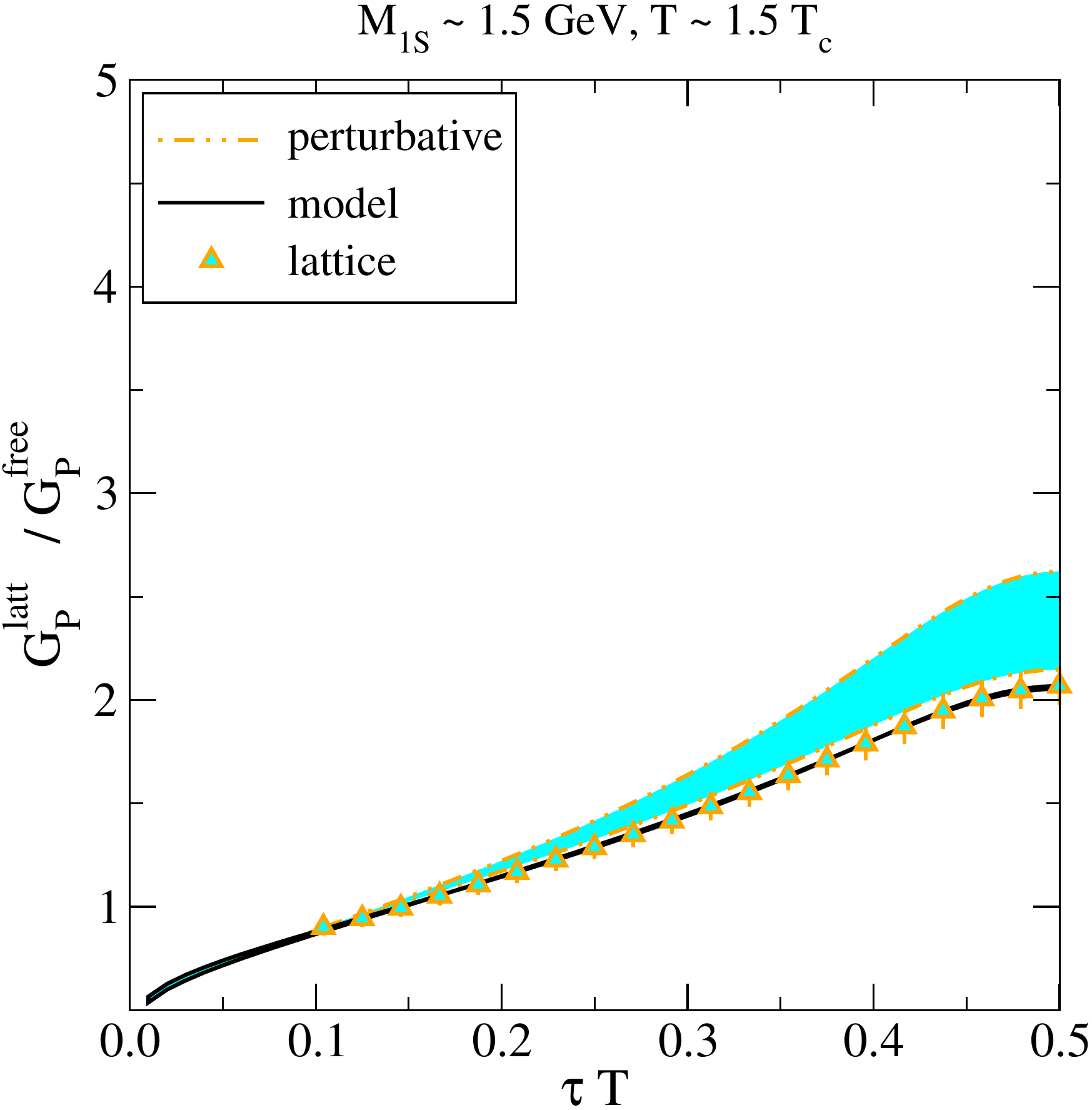}\includegraphics[width=0.48\textwidth]{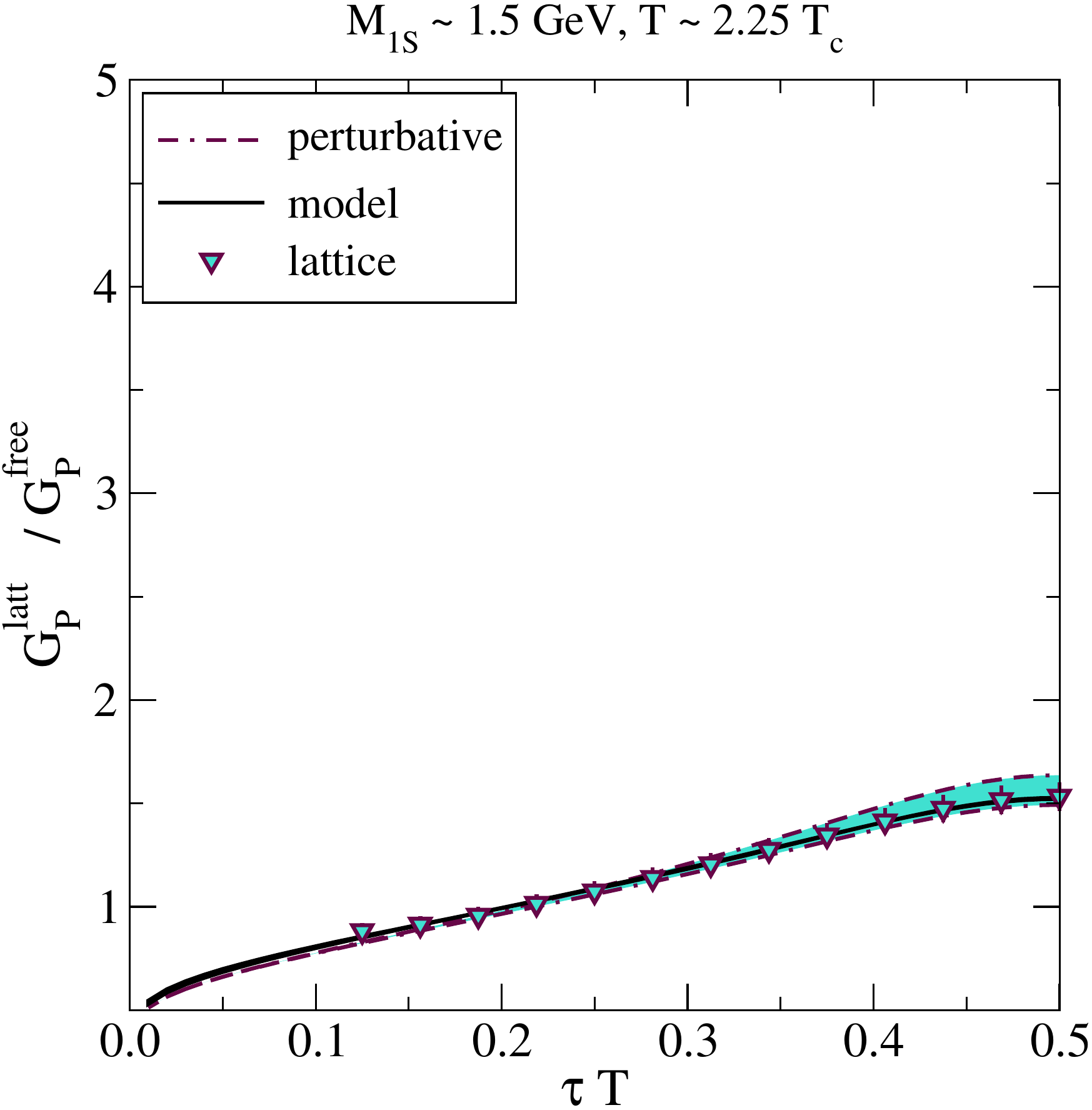}
     }
 \caption{A comoparison of lattice correlators, correlators integrated from the model spectral functions and the resummed perturbative predictions for charmonium in the pseudo-scalar channel \cite{Burnier:2017bod}.}
 \label{corr-ps-comp}    
\end{figure}

This model spectral function can be used to fit the lattice correlators, which have been first interpolated to the physical meson masses and then extrapolated to the continuum. We show the fit results in Fig.~\ref{corr-ps-comp}. We could see that continuum lattice correlators are well described by this model. The corrections to the fit parameters turn out to be small and $\chi^2/$d.o.f$<$1, see Tab.~\ref{table:fits-A-B}. We show the resulting spectral functions in Fig.~\ref{spf-ps-comp-cb}. From our analysis we conclude that for charmonium no resonance peaks are needed for representing the lattice data even at $1.1~T_c$. Only modest threshold enhancement is sufficient in the analyzed temperature region. For bottomonium the thermally broadened resonance peak could survive up to temperatures around $1.5T_c$.

\begin{table}[t]
\small{
\begin{center}
\begin{tabular}{c|ccc|ccc}
 & \multicolumn{3}{c|}{charmonium} 
 & \multicolumn{3}{c}{bottomonium} \\[2mm] 
 $\displaystyle {T } / { T_c^{ }} $ &
 $\displaystyle A $ &
 $\displaystyle { B } / { T } $ & 
 $\displaystyle {\chi^2} / {\mbox{d.o.f.}} $ &
 $\displaystyle A $ &
 $\displaystyle { B } / { T } $ &  
 $\displaystyle {\chi^2} / {\mbox{d.o.f.}} $ 
 \\[3mm]
 \hline 
  $1.1$ & 1.04 & 0.52 & 0.01  & 0.85 & -0.11 & 0.02  \\ 
  $1.3$ & 1.04 & 0.37 & 0.01  & 0.87 & -0.13 & 0.04  \\
  $1.5$ & 1.02 & 0.33 & 0.02  & 0.87 & -0.11 & 0.10  \\ 
 $2.25$ & 1.06 & 0.16 & 0.08  & 0.93 & -0.04 & 0.28  \\ \hline
\end{tabular} 
\end{center}
}
\caption[a]{\small
  Best fit parameters of the model spectral function Eq.~(\ref{model-ps}) to our lattice data.
 }
\label{table:fits-A-B}
\end{table}

Note that the results here are obtained in the quenched approximation for a gluonic system without dynamical fermion degrees of freedom. It remains to be seen if such a perturbative model still is able to describe charmonium and bottomonium systems when extending these studies to full QCD or if non-perturbative effects will become stronger in a more realistic QGP\index{QGP} medium.

\subsection{Spectral functions of charmonia and bottomonia in the vector channel}

\begin{figure}[hhbp]
    \centering{
     \includegraphics[width=0.48\textwidth]{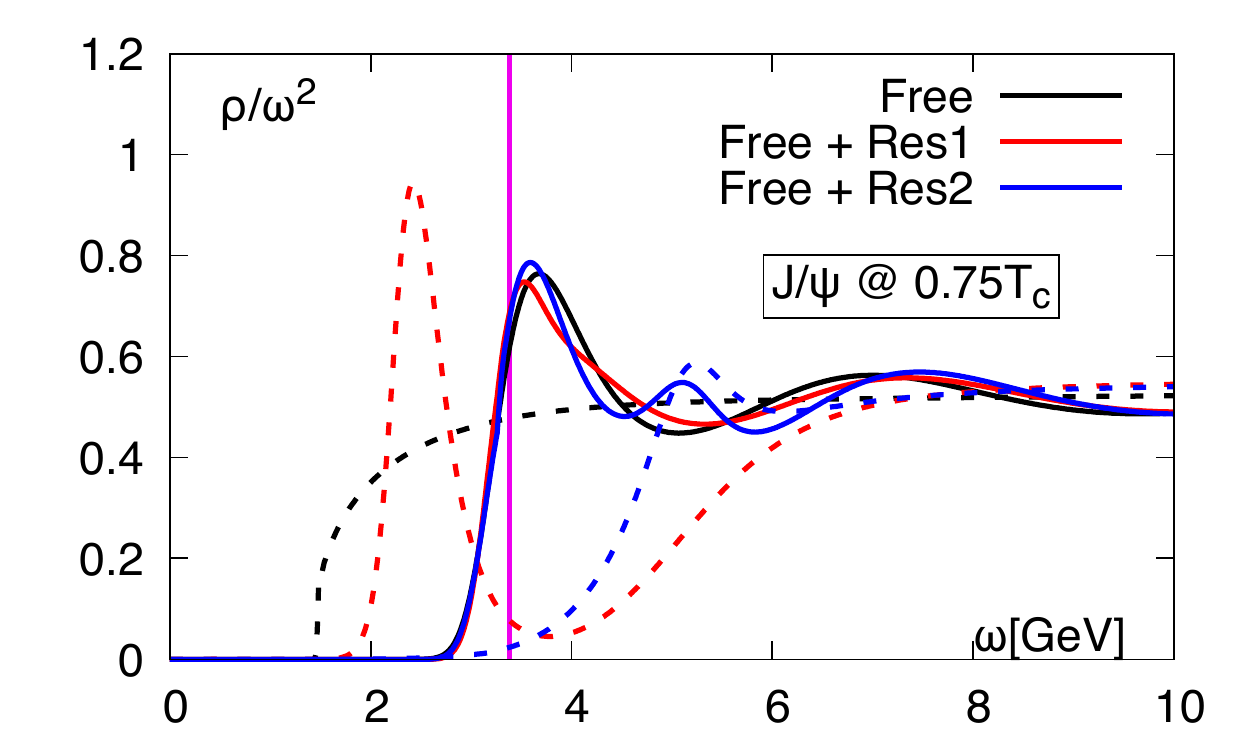}\includegraphics[width=0.48\textwidth]{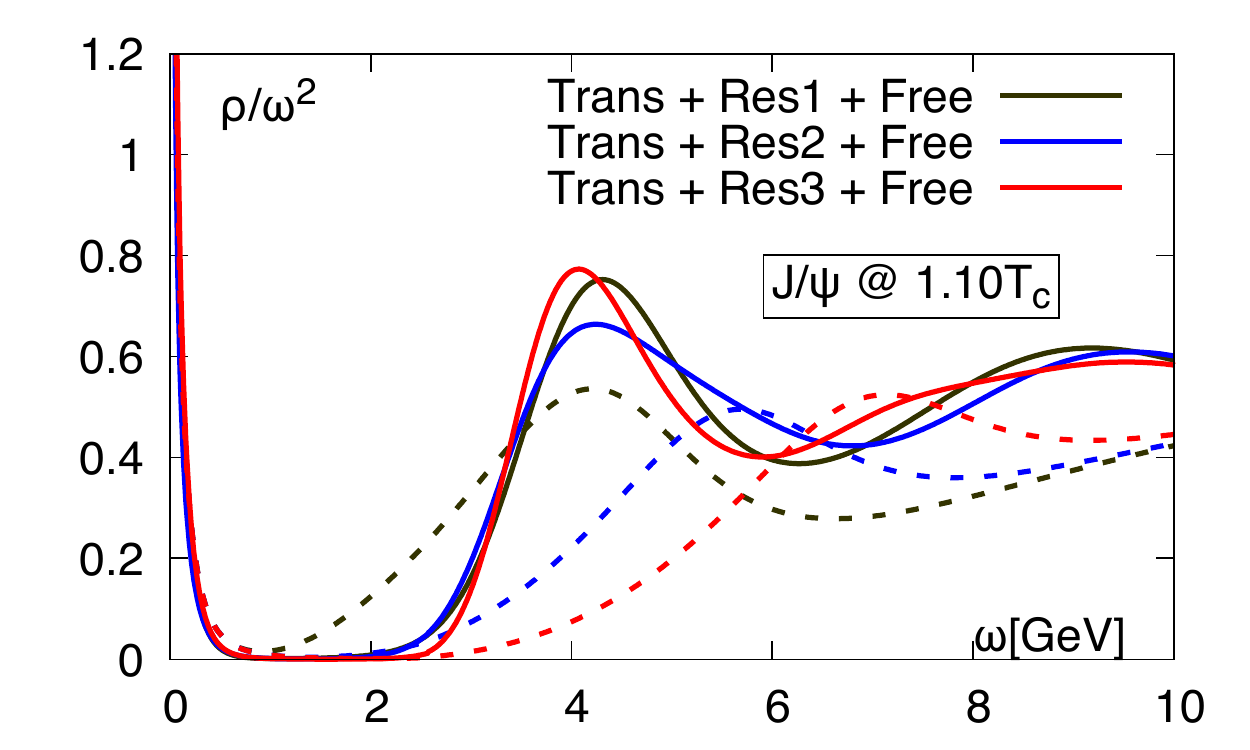}}
         \centering{
     \includegraphics[width=0.48\textwidth]{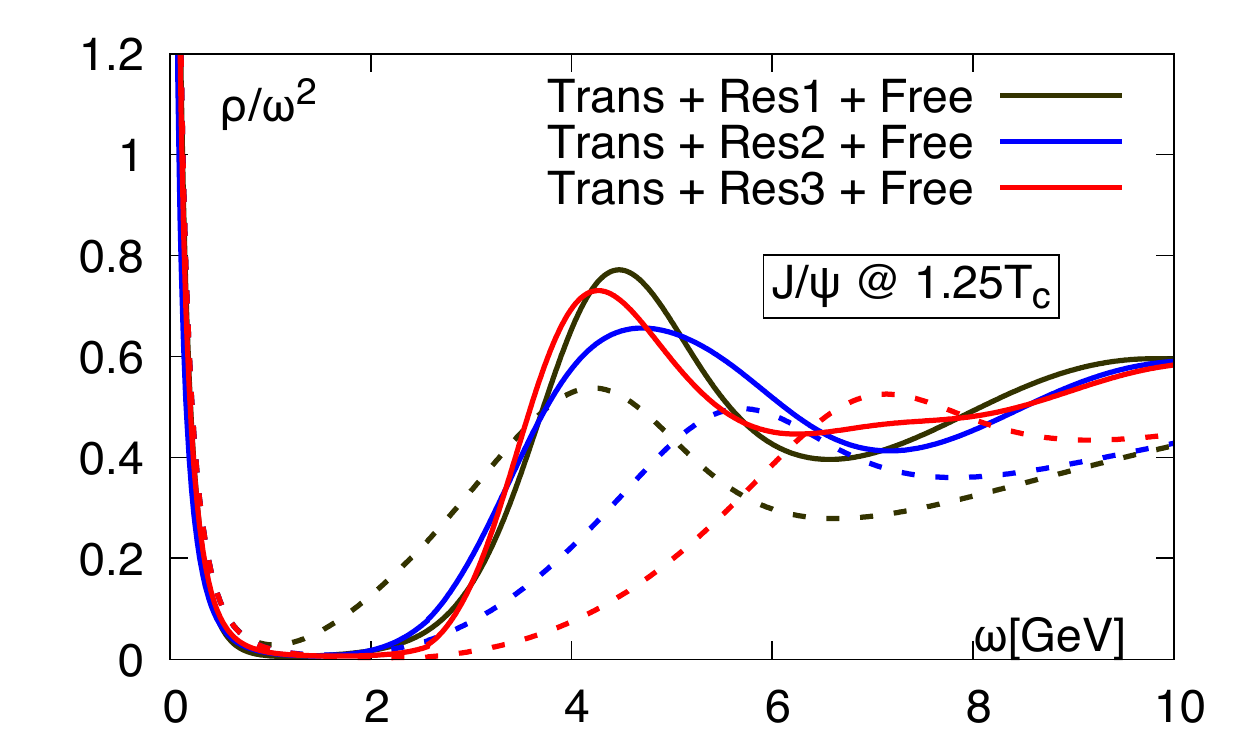}\includegraphics[width=0.48\textwidth]{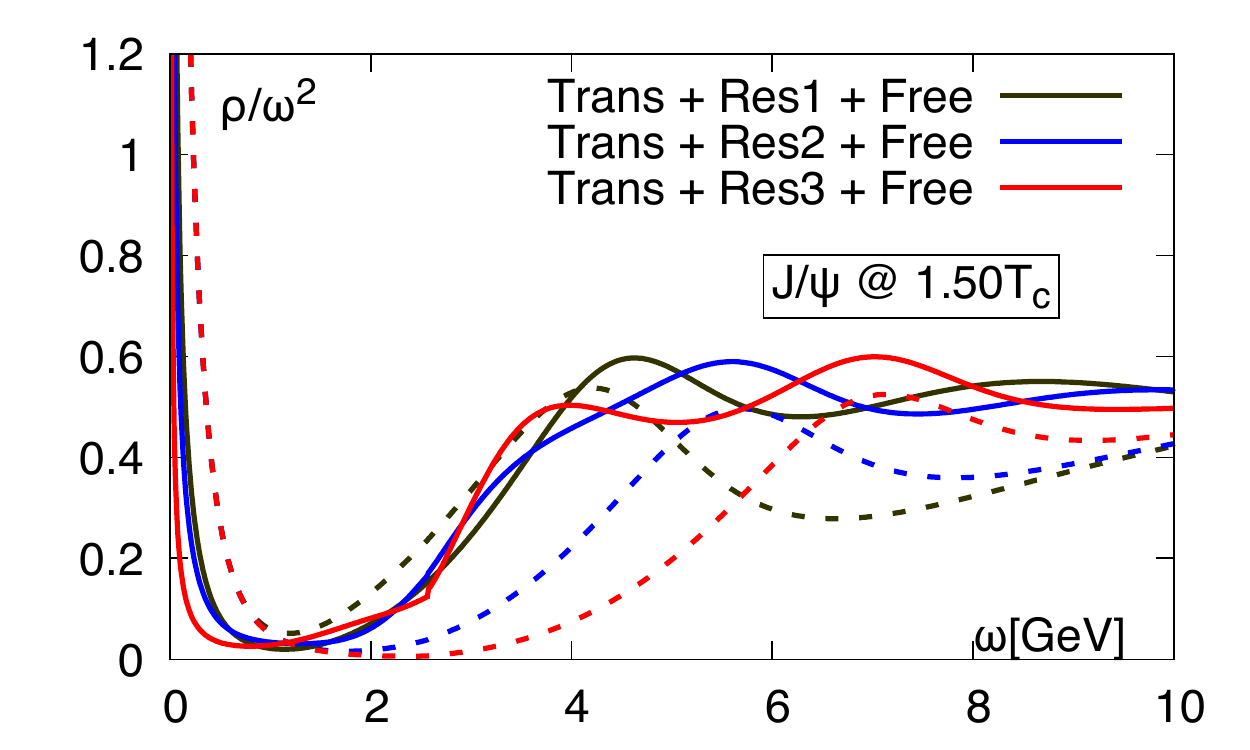}
     }
 \caption{Charmonium spectral functions in the vector channel at different temperatures obtained by MEM with different default models (dashed curves) \cite{Ding:2018uhl}.}
 \label{charm_dm}    
\end{figure}

In this section we discuss lattice results on quarkonium spectral functions in the vector channel obtained from lattice QCD\index{QCD!lattice} simulations at different temperature $T\in [0.75, 2.25]T_c$. Same as previous section, the simulations are performed in quenched approximation on large and fine isotropic lattices. The results in this section are obtained from the finest available lattice with a lattice spacing of $a\simeq 0.009~fm$. 
The spectral functions are reconstructed using the Maximum Entropy Method\index{entropy}, discussed in Sec.~\ref{Sec:inverse}, with default models constructed using the ingredients mentioned in Sec.~\ref{free-spf-sec}. Some reconstructed spectral functions (solid lines) for different default models (dashed lines) are summarized in Fig.~\ref{charm_dm} taken from \cite{Ding:2018uhl}.
In these studies, the determination of the spectral modifications and dissociation temperatures requires carefully examination. For instance the dependence of the results on the default model need to be checked. The deformation of the spectral function with temperature could also come from the reduction of the number of data points of the correlators with increasing temperature, which also needs to be taken into account.
Furthermore the resolution of spectral properties, e.g. the width of spectral peaks, usually is limited using Bayesian spectral reconstruction approaches, even for very large lattice extents and high statistics data. 

\begin{figure}[thbp]
    \centering{
     \includegraphics[width=0.7\textwidth]{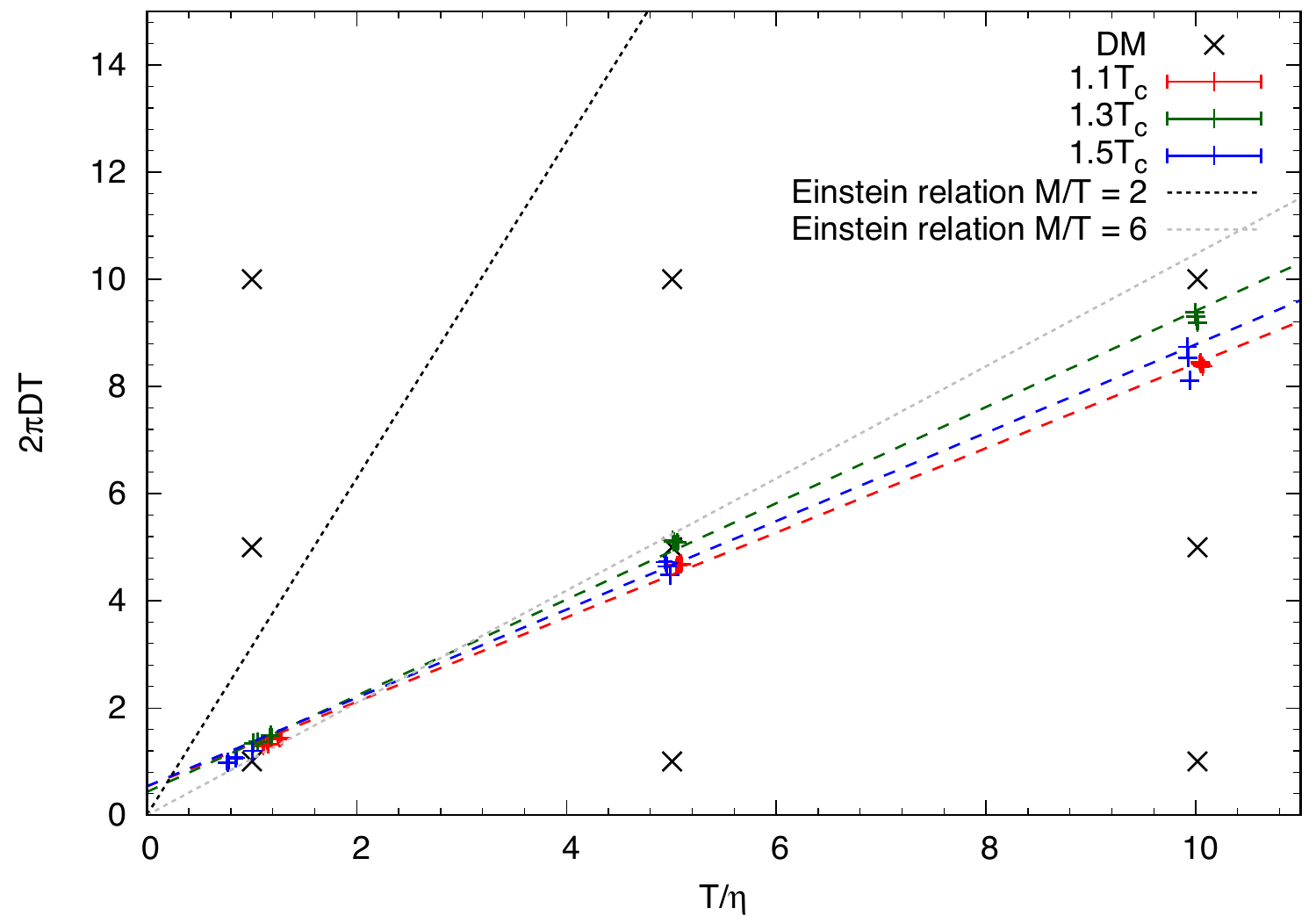}}
 \caption{Linear relation between $2\pi TD$ and $T/\eta$ at 1.10, 1.25 and 1.50$T_c$ for charmonia \cite{Ding:2018uhl}.}
 \label{2piTD}    
\end{figure}

The heavy quark diffusion coefficient for charm quarks, which requires precise reconstruction of the transport peak, still remains unclear. One of the difficulties is that the transport contribution is much smaller than the other contributions to the spectral function leading to only a small curvature in the transport contribution for different distances $\tau T$ in the correlator. A Breit-Wigner distribution with different width all could describe the lattice data equally well, forming a linear relation between $2\pi TD$ and $T/\eta$, see Fig.~\ref{2piTD}.
Although the method seems to be sensitive to the integral over the transport peak, a resolution of the shape of the transport peak is limited.

\begin{figure}[thbp]
    \centering{
     \includegraphics[height=0.35\textwidth, width=0.48\textwidth]{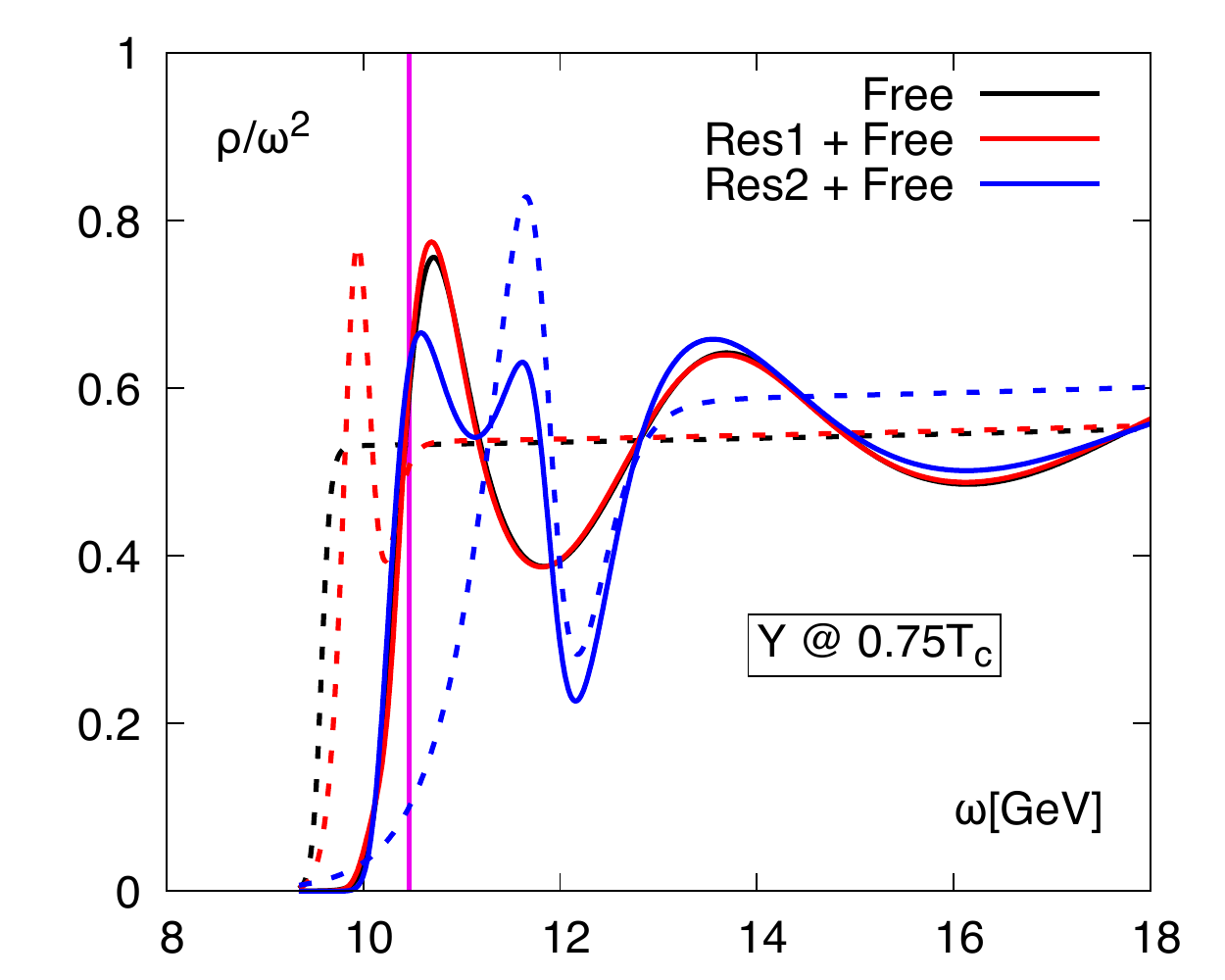}\includegraphics[height=0.35\textwidth,width=0.48\textwidth]{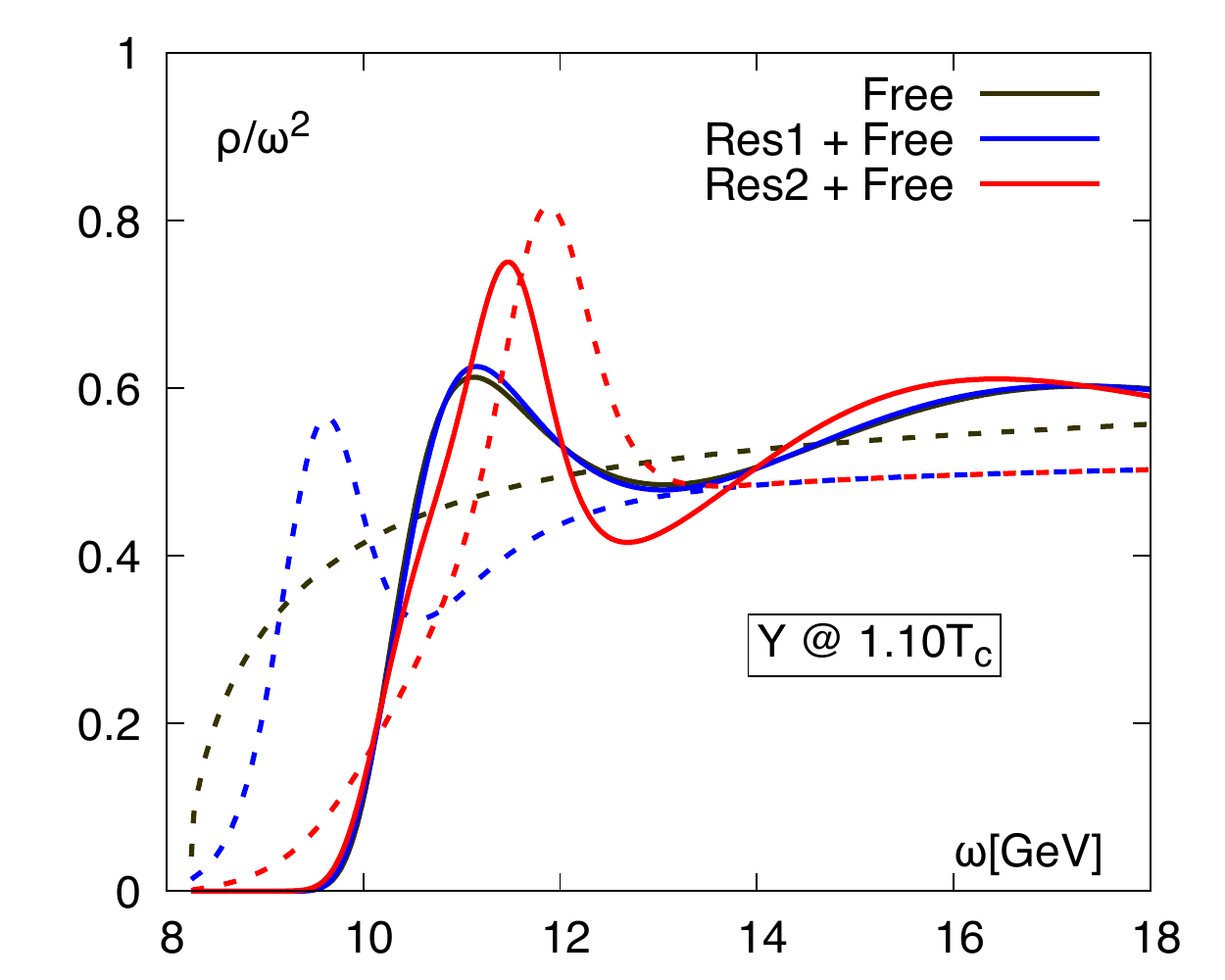}}
         \centering{
     \includegraphics[height=0.35\textwidth,width=0.48\textwidth]{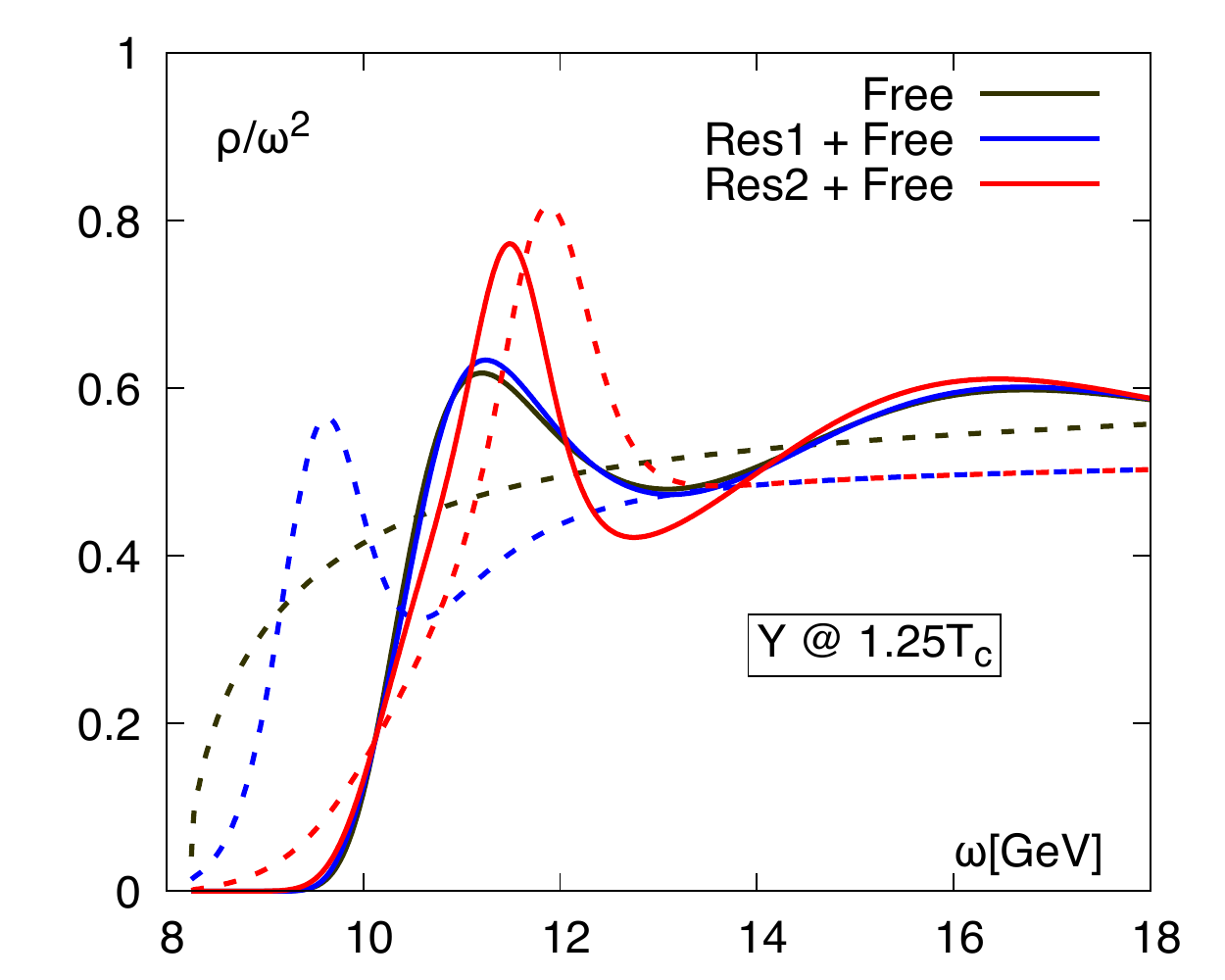}\includegraphics[height=0.35\textwidth,width=0.48\textwidth]{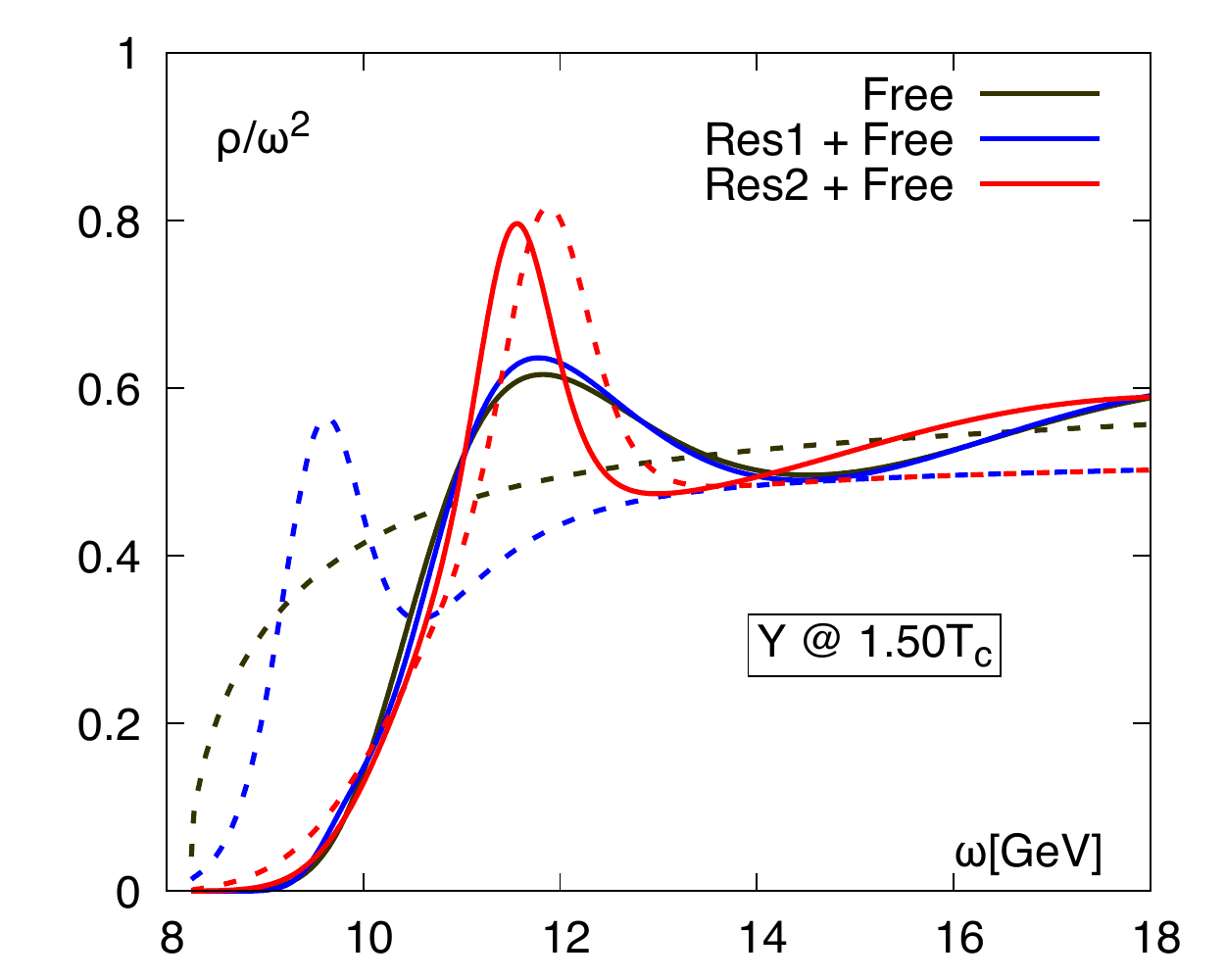}
     }
 \caption{Bottomonia spectral functions at different temperatures obtained by MEM with different default models \cite{Ding:2018uhl}.}
 \label{bottom_dm}    
\end{figure}

In the bottomonium channel, due to the large scale separation, it is always difficult to directly reconstruct the full spectral functions with methods like MEM. A common work around is to use the difference correlator $G_{\mathrm{diff}}(\tau/a)\equiv G(\tau/a)-G(\tau/a+1)$, which is believed to be able to effectively remove the tiny transport contribution. With this method the intermediate and large frequency part of the spectral function can be obtained, see Fig.~\ref{bottom_dm}.
Also here some default model dependencies are visible and the width of the spectral peaks, where the ground state may correspond to the $\Upsilon$ state, can not reliably extracted. This is furthermore hindered by the fact, that estimating the statistical\index{statistical!error} and systematic uncertainties in such Bayesian analyses are difficult. 

Recently a similar study in the vector channel on the same data set as in the pseudo-scalar channel (Sec.~\ref{sec-ps}) using same strategy, i.e. $\chi^2$-fitting based on perturbatively inspired models, has been carried out. Some preliminary results are reported in \cite{Lorenz:2020uik}. But in the vector channel one has to solve the transport peak and will have similar difficulties resolving details of the shape of it as seen in the MEM analysis discussed above, also shown in Fig.~\ref{2piTD}. 
Possible ways to improve on this problem are making use of estimating the heavy quark mass and using the Einstein relation to eliminate one free parameter or analyzing so called thermal moments of the spectral function. 
Furthermore, combining different spectral reconstruction methods will help to improve the extraction of more reliable spectral functions and also allow to better estimate and reduce systematic uncertainties in the spectral reconstruction.

\section{Heavy quark momentum diffusion coefficient}

As we have learned in the previous sections, the reconstruction of the transport peak from the vector channel of current-current correlation functions is challenging. Especially for cases where this peak is expected to be very narrow and possible bound states contribute to the spectral function, like in the heavy quark sector, the sensitivity of the correlation function to the transport contribution is small, hindering a reliable extraction and spectral reconstruction. In this section we will discuss results for the heavy quark momentum diffusion coefficient which can be determined from an operator which is more sensitive to the transport contribution, as the corresponding spectral function contains no bound states and the small frequency limit is expected to be approached in a smooth way.

\begin{figure}[thbp]
\centering{
     \includegraphics[width=0.6\textwidth]{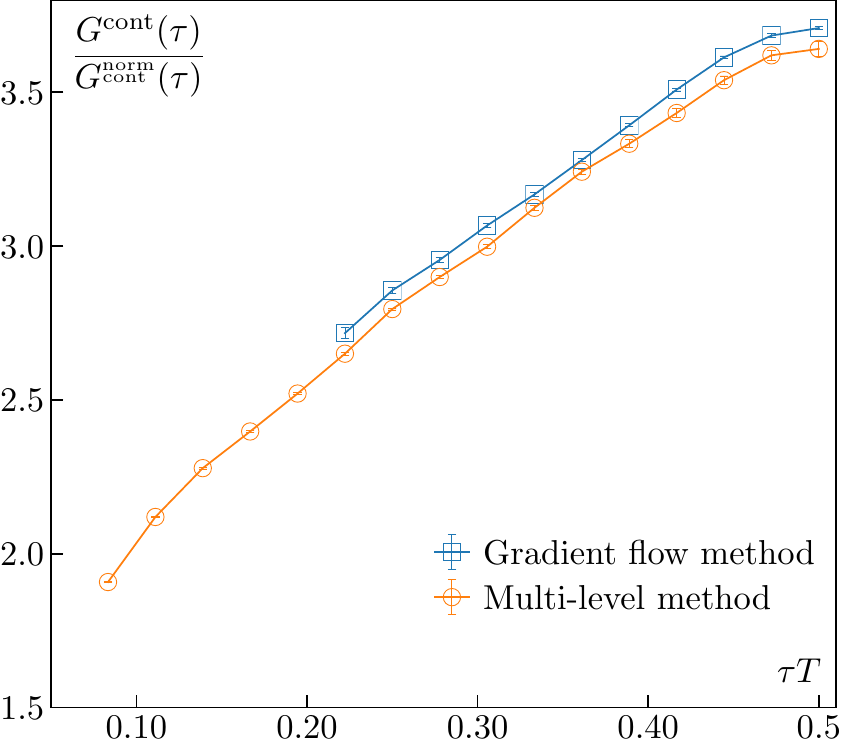}}
 \caption{The comparison of nonperturbatively renormalized continuum color-electric correlator extrapolated to continuum limit and zero flow time obtained using gradient flow method~\cite{Altenkort:2020fgs} and revised continuum correlator from multilevel method taken from~\cite{Francis:2015daa}.}
 \label{EEcorrifg}    
\end{figure}

In 2009 Moore, Laine and Caron-Huot proposed to measure a purely gluonic Euclidean correlator, i.e. color-electric correlators that can be related to the heavy quark diffusion \cite{CaronHuot:2009uh}. The idea is to construct a kinetic mass dependent momentum diffusion coefficient $\kappa^{(M)}$ within the 
heavy quark effective theory that can be related by the heavy quark diffusion coefficient $D$ via Einstein relation $D=2T^2/\kappa^{(M)}$. To get rid of the mass dependence, one carries out the large quark mass limit in the Langevin theory. In this limit the force-force correlator appears in the definition of $\kappa$ could be expanded in inverse powers of 
the heavy quark mass, leading to a gluonic correlation function
of the following form
\begin{equation}
G_E(\tau) = -\frac{1}{3}\frac{\langle{\rm Re}{\rm Tr}\left[U(\frac{1}{T},\tau)gE_i(\tau,\vec{0})U(\tau,0)gE_i(0,\vec{0})\right]\rangle}{\langle{\rm Re}{\rm Tr}U(\frac{1}{T},0)\rangle}\;,
\label{eq:GE}
\end{equation}
where $E_i$ is the color electric field\index{color!electric field}.
Then after analytic continuation a heavy quark momentum diffusion coefficient free of quark mass can be defined by
\begin{equation}
\frac{\kappa}{T^3} = \lim_{\omega\rightarrow0}\frac{2T\rho_E(\omega)}{\omega},
\label{eq:kappa}
\end{equation}
where $\rho_E(\omega)$ is the spectral function encoded in the color-electric correlators\index{color!electric correlators}. 

\begin{figure}[hbtp]
    \centering{
     \includegraphics[width=0.48\textwidth]{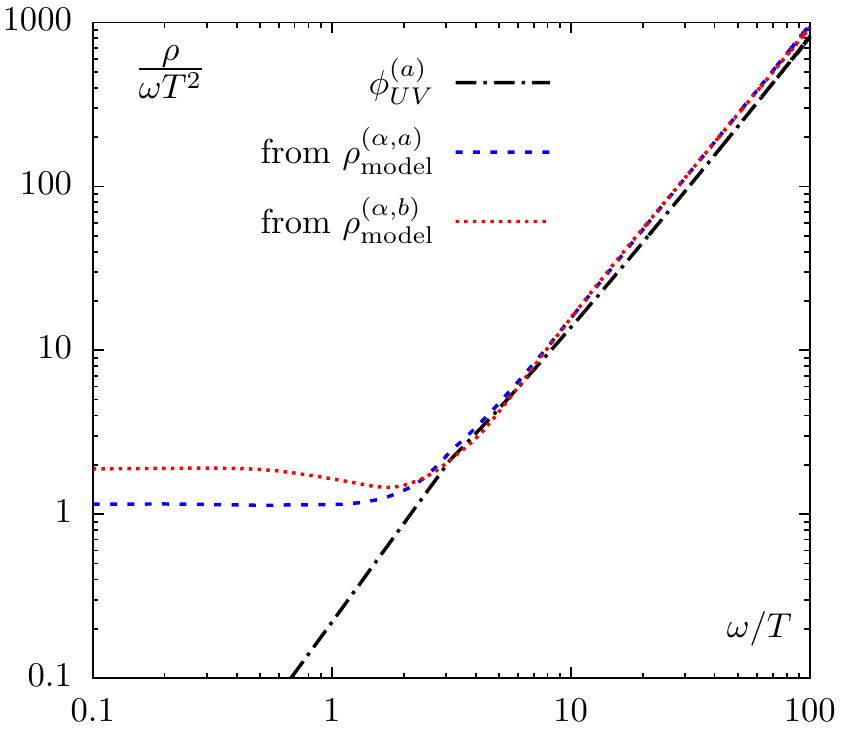} \includegraphics[width=0.48\textwidth]{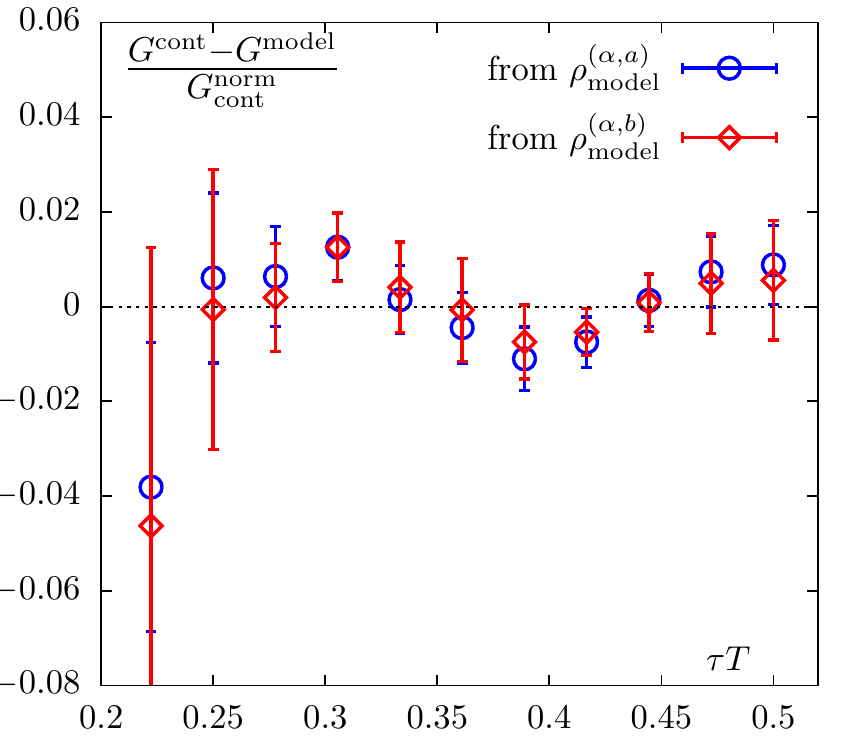}}
 \caption{The fitted spectral functions (\textit{up}) and correlators (\textit{down}) using $\chi^2$ fits based on different perturbatively inspired models and Backus-Gilbert method \cite{Francis:2015daa}.}
 \label{EEcorrifg1}    
\end{figure}

The advantage of the studying the color-electric correlators is that this correlator is much less computation demanding than the current-current correlators as one does not need to calculate the inverse Dirac matrix. Furthermore, the corresponding spectral function related to this correlator has less structures than the one from the current-current correlators and a smooth behavior at small $\omega$ is expected. Perturbative computations of $\kappa/T^3$ show weak convergence in the range of coupling of interest, and hence a non-perturbative approach like lattice QCD is required to make realistic prediction. In the following we show a lattice study of this quantity taken from \cite{Altenkort:2020fgs}, where the gradient flow \cite{Narayanan:2006rf,Luscher:2009eq} method is used to improve the signal of the correlators measured. 

In Yang-Mills gradient flow method the gauge field are  evolved with respect to a fifth dimension which is called flow time ${\tau_{\text{F}}}$. The gauge action will be driven to its saddle point according to a diffusion equation
\begin{align}
    \label{boundary}
    B_{\mu}\Big{|}_{{\tau_{\text{F}}}=0}=A_{\mu}, \quad
    \frac{\partial{B_{\mu}}}{\partial{{\tau_{\text{F}}}}}=D_{\nu}G_{\nu\mu},
\end{align}
where $B_{\mu}({\tau_{\text{F}}},x)$ is the  flowed gauge field and  $A_{\mu}$ the ordinary gauge field. The covariant derivative $D_{\nu}$ respecting the gauge invariance could be easily constructed as
\begin{equation}
    \begin{aligned}
        D_{\mu}&=\partial_{\mu}+[B_{\mu},{\makebox[1ex]{\textbf{$\cdot$}}}]
        \end{aligned}
\end{equation}
and the field strength tensor could also be constructed with the flowed gauge fields. 

Since a fifth dimension is introduced in this method, to obtain the physical quantities, one needs to extrapolate the flow time ${\tau_{\text{F}}}$ to zero. For this one could make use of the small-flow-time-expansion proposed in perturbation theory: the flowed local operator $O(x,{\tau_{\text{F}}})$ can be expanded in ${\tau_{\text{F}}}$-dependent coefficients $c_i$ and the corresponding renormalized, unflowed operator~\cite{Luscher:2011bx}
\begin{equation}
    \label{short-time-expansion}
    O(x,{\tau_{\text{F}}}) \xrightarrow[{\tau_{\text{F}}}\rightarrow 0]{\text{}} \sum_i c_i({\tau_{\text{F}}})O^R_i(x).
\end{equation}
The amount of flow that can be applied to a gauge field can be estimated from perturbative theory. For instance \cite{Eller:2018yje} shows the maximum flow that should be used in the color-electric and energy-momentum tensor correlators at leading order, before the flow influences the correlation function.

\begin{figure}[thbp]
    \centering{
     \includegraphics[width=0.6\textwidth]{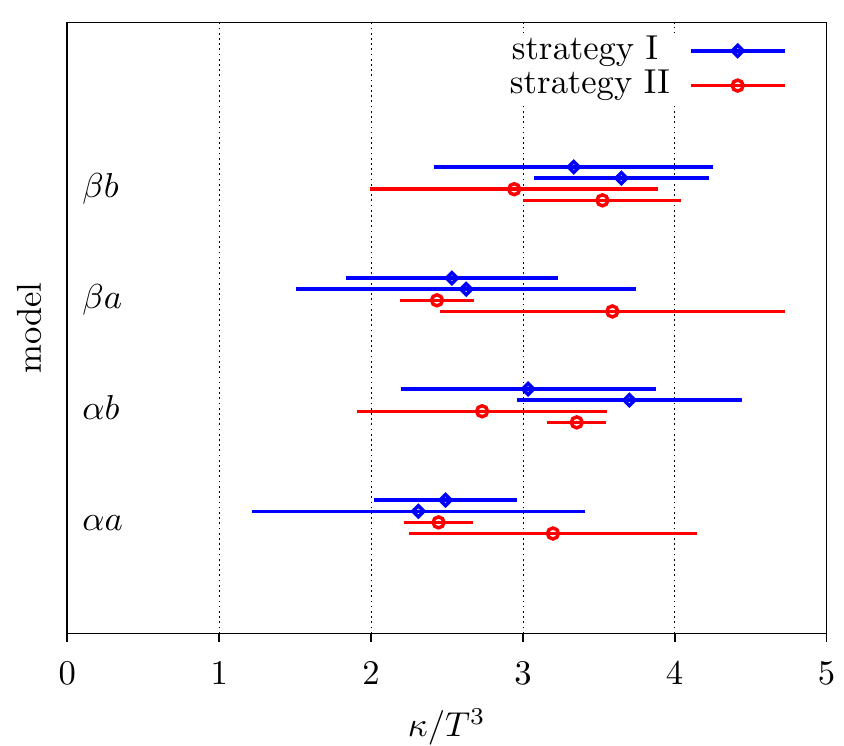}}
 \caption{The heavy quark momentum diffusion coefficient obtained from different fit models and strategies based on correlators obtained from using gradient flow method~\cite{Altenkort:2020fgs}.}
 \label{kappa_compare}    
\end{figure}

Fig.~\ref{EEcorrifg} shows the color-electric correlators\index{color!electric correlators} obtained from using gradient flow method~\cite{Altenkort:2020fgs} and multilevel algorithm~\cite{Francis:2015daa} respectively. We could see the error sizes in both cases are comparable, which suggests that gradient flow a powerful in improving the signal. Besides, there is an small overall shift between the results obtained from these two methods, which could be explained as the difference in the renormalization: in the frame work of gradient flow, the correlators are automatically renormalized well in the multilevel algorithm, the perturbative renormalization constant has been used.

With the data obtained above one could try to reconstruct the spectral function from it. The 
spectral function could be constrained by using physical arguments in
different regions. For small frequencies, i.e., 
$\omega\ll T$ there is a linear behavior as 
predicted by hydrodynamics
\begin{equation}
\rho_{\rm IR}(\omega) = \frac{\kappa\omega}{2T}\;.
\label{eq:rhoIR}
\end{equation}

For large frequencies, $\omega\gg T$, the behavior 
is constrained by perturbation theory
\begin{equation}
\rho_{\rm UV}(\omega) = \left[\rho_{\rm UV}(\omega)\right]_{T=0} + O\left(\frac{g^4T^4}{\omega}\right)\;.
\label{eq:rhoUV}
\end{equation}
Using a renormalization scale $\bar{\mu}_\omega=\omega$ for $\omega\gg\Lambda_{\overline{\rm MS}}$ 
the leading order expansion gives
\begin{equation}
\rho_{\rm UV}(\omega) = \Phi_{\rm UV}(\omega)\left[1+O\left(\frac{1}{\log(\omega/\Lambda_{\overline{\rm MS}})}\right)\right]\;,
\label{eq:rhoUV2}
\end{equation}
with 
\begin{equation}
\Phi_{\rm UV}(\omega) = \frac{g^2(\bar{\mu}_\omega)C_F\omega^3}{6\pi}\;, 
\label{eq:phiUV}
\end{equation}
where the running coupling\index{running coupling} constant could be determined up to 4-loop order at scale $\bar{\mu}_\omega={\rm max}(\omega,\pi T)$. Combining these two parts with interpolation in the intermediate frequencies one could have various Ans\"{a}tz to fit the correlator via 

\begin{equation}
G_{\rm model}(\tau) = \int_0^\infty \frac{d\omega}{\pi}\rho_{\rm model}\left(\omega\right)\frac{\omega}{T}\frac{\cosh(\frac{1}{2} - \tau T)}{\sinh(\frac{\omega}{2T})}\;.
\label{eq:Gmodel}
\end{equation}

\begin{figure}[thbp]
    \centering{
     \includegraphics[width=0.9\textwidth]{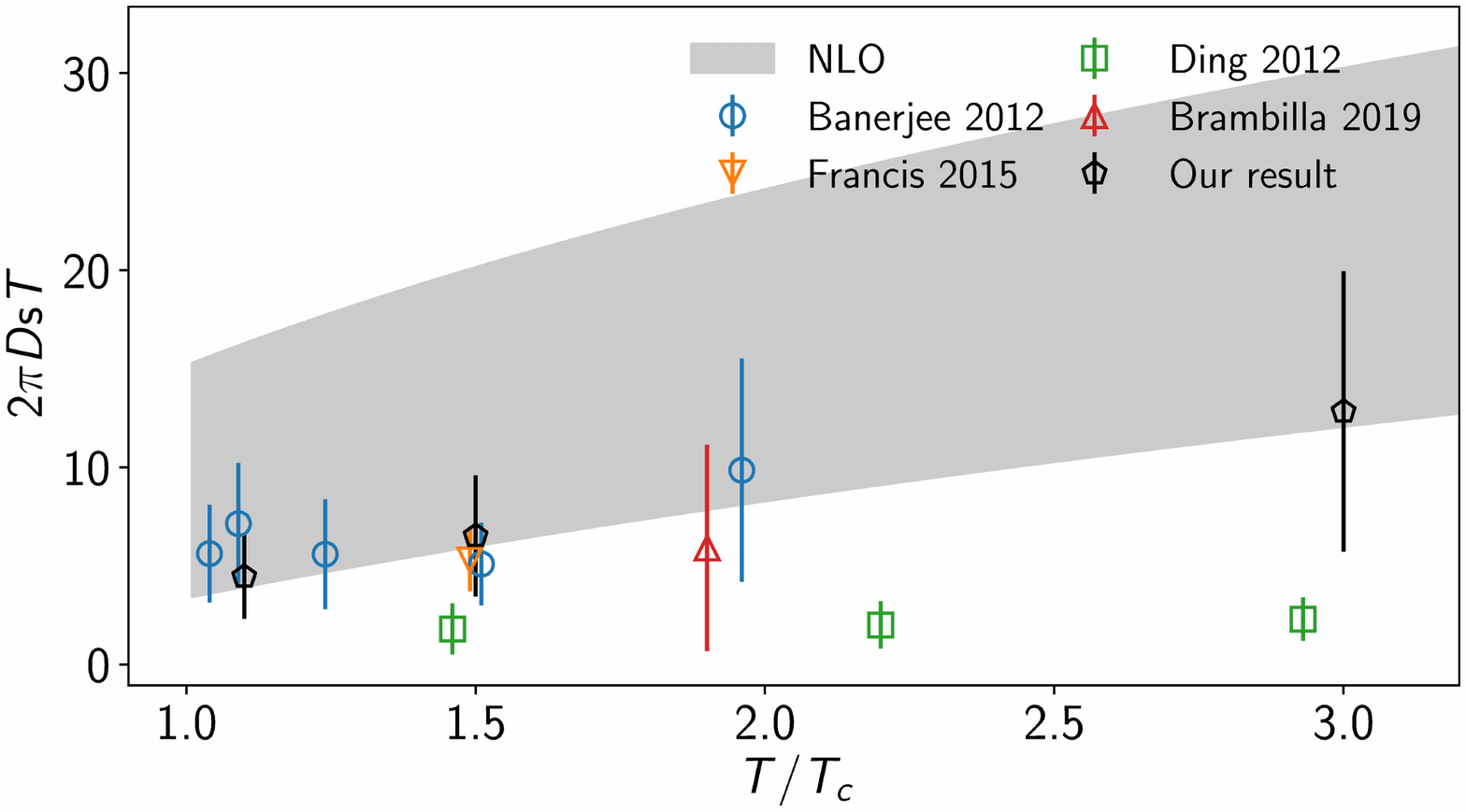}}
 \caption{The temperature dependence of heavy quark  diffusion coefficient extracted from color-electric correlators obtained using multilevel algorithm and comparison with other studies. Taken from \cite{Brambilla:2020siz}.}
 \label{kappa_tum}    
\end{figure}

The fitted spectral function and correlators are shown in Fig.\ref{EEcorrifg1}. The resulting diffusion coefficient is then (see Fig. \ref{kappa_compare})
\begin{equation}
\frac{\kappa}{T^3} = \lim_{\omega\rightarrow0}\frac{2T\rho_E(\omega)}{\omega} = 2.31 ... 3.70 \;.
\label{eq:kappa2}
\end{equation}
This quantity is related in the non-relativistic limit 
to the diffusion coefficient $D$ and the drag coefficient 
$\eta_D$
\begin{equation}
T D = 4\pi \frac{T^3}{\kappa} = 0.54 ... 0.87 \;,
\label{eq:2piTD}
\end{equation}
\begin{equation}
\eta_D = \frac{\kappa}{2 M_{\rm kin}T}\left(1 + O\left(\frac{\alpha_s^{3/2}T}{M_{\rm kin}}\right)\right)\;.
\label{eq:etaD}
\end{equation}
From this one estimates the timescale for
kinetic equilibration of heavy flavors
\begin{equation}
\tau_{\rm kin} = \frac{1}{\eta_D} = (2.31 ... 3.70)\left(\frac{Tc}{T}\right)^2\left(\frac{M}{1.5 \rm GeV}\right)^2\rm fm/c\;.
\label{eq:tau}
\end{equation}

Close to $T_c$ the estimated time is
$\tau_{\rm kin}=1 \rm fm/c$ which is
close to the equilibration time for
light flavors. This may help to explain the flow behavior for the $D$ meson recently observed in experiment.

Recently the TUMQCD collaboration studied the temperature dependence of the the heavy quark momentum diffusion coefficient using multilevel algorithm in a wide temperature range $1.1<T/T_c<10^4$~\cite{Brambilla:2020siz}. Fig.~\ref{kappa_tum} shows the results and a comparison with other studies. At $1.5T_c$ the results are consistent with those obtained using gradient flow method given above.


\section{Conclusions}

Recent progress in the determination of spectral and transport properties have been discussed.
Combining continuum extrapolated correlation functions from Lattice QCD with phenomenologically inspired and perturbatively constrained Ans\"{a}tz, allows to extract transport and spectral properties of the QGP medium. In the light quark sector, continuum estimates for the electrical conductivity, thermal dilepton rate and thermal photon rate have been obtained.
Results for charmonium and bottomonium correlation functions in the gluonic medium can be well described by perturbative models. While for charmonium no resonance peaks are needed at temperature above $T_c$ in these spectral functions, thermally broadend resonance peaks persist up to around $1.5~T_c$. While the extraction of heavy flavor diffusion coefficients from hadronic correlation functions is still challenging, in the heavy quark mass limit, based on calculations of the color-electric field\index{color!electric correlator} correlator, estimates for the heavy quark momentum diffusion coefficient can be extracted.

The methodology developed in these studies so far was applied in the quenched approximation for a pure gluonic medium. The extension to full QCD calculations is essential for a realistic description of the QGP medium since the influence of dynamical fermions will become important especially at temperatures close to $T_c$.

\begin{acknowledgement}
O.K. would like to thank Chihiro Sasaki, David Blaschke, Krzysztof Redlich and Ludwik Turko, the organizers of the \emph{53rd Karpacz Winter School of Theoretical Physics}, for very kind hospitality.
The authors acknowledge support by the Deutsche Forschungsgemeinschaft (DFG, German Research Foundation) through the CRC-TR 211 "Strong-interaction matter under extreme conditions" -- project number 315477589 -- TRR 211. 
\end{acknowledgement}
 

 \section*{Exercises}

\paragraph{Exercise 1}
Construct free meson spectral functions using Eq.~(\ref{freespf_cont})
and discuss the spectral functions for different channels and different quark masses.
Discuss the behavior of the kernel function Eq.~(\ref{eq:Kernel}) as a
function of $\tau T$ for different $\omega$ and the contribution
of different frequency regions of the free spectral functions to the
correlation functions in Eq.~\ref{eq:IntRelation}.
  
\paragraph{Exercise 2}
Calculate the free meson correlation functions using a numerical
integration (e.g. using Gaussian integration) of Eq.~(\ref{eq:IntRelation})). 
Check your result for $m=0$ against Eq.~\ref{eq:Gfree}.
Discuss the mass dependence of the vector meson correlation functions
and the contribution of different frequency regions of the spectral
function to the correlators.

\paragraph{Exercise 3}
Construct model spectral functions using the free spectral functions at vanishing 
as well as charm quark masses and add a transport peak as discussed in sections \ref{sec:dilep}
and \ref{sec:charm}. Discuss the influence of the transport
contribution to the correlation functions. In which cases are the
correlation functions sensitive to this contribution, assuming that a
typical statistical error\index{statistical!error} of lattice correlation functions is of the
order of one percent.

\paragraph{Exercise 4}
Repeat the previous exercise by adding a resonance peak and discuss
the different contributions of the transport and resonance peak and the free
continuum contributions.

%
%






\end{document}